\documentclass{aa}  
\usepackage{graphicx}
\usepackage{txfonts}
\usepackage[utf8]{inputenc}
\usepackage{amssymb,amsmath}
\usepackage{lscape}
\usepackage{longtable}
\usepackage{float}
\usepackage{natbib}
\usepackage{color}
\begin{document} 
%
%
\title{Stellar mass distribution of S$^{4}$G disk galaxies and signatures of bar-induced secular evolution
\thanks{The FITS files of the synthetic images and the tabulated radial profiles of the mean (and dispersion of) stellar mass density, 3.6~$\mu$m surface brightness, Fourier amplitudes, gravitational force, 
and the stellar contribution to the circular velocity are only available in electronic form at the CDS via anonymous ftp to cdsarc.u-strasbg.fr (130.79.128.5) or via http://cdsweb.u-strasbg.fr/cgi-bin/qcat?J/A+A/}
}
  \author{S. Díaz-García
          \and        
          H. Salo
          \and        
          E. Laurikainen
          }
  \institute{Astronomy Research Group, University of Oulu, FI-90014 Finland\\
              \email{simon.diazgarcia@oulu.fi}
             }
  \date{Received April 11, 2016; accepted July 24, 2016}
  \abstract
  {
Models of galaxy formation in a cosmological framework need to be tested against observational constraints, such as the average stellar density profiles (and their dispersion) as a function of fundamental galaxy properties (e.g. the total stellar mass). 
Simulation models predict that the torques produced by stellar bars efficiently redistribute the stellar and gaseous material inside the disk, 
pushing it outwards or inwards depending on whether it is beyond or inside the bar corotation resonance radius.
Bars themselves are expected to evolve, getting longer and narrower as they trap particles from the disk and slow down their rotation speed.
}
   {We use 3.6~$\mu$m photometry from the Spitzer Survey of Stellar Structure in Galaxies (S$^{4}$G) to trace the stellar distribution in nearby disk galaxies ($z\approx0$) with total stellar masses $10^{8.5}\lesssim M_{\ast}/M_{\odot}\lesssim10^{11}$ 
   and mid-IR Hubble types $-3 \le T \le 10$. We characterize the stellar density profiles ($\Sigma_{\ast}$), the stellar contribution to the rotation curves ($V_{3.6 \mu \rm m}$), and the $m=2$ Fourier amplitudes ($A_2$) as a function of $M_{\ast}$ and $T$. 
   We also describe the typical shapes and strengths of stellar bars in the S$^4$G sample and link their properties to the total stellar mass and morphology of their host galaxy.
}
   {
For 1154 S$^4$G galaxies with disk inclinations lower than $65^{\circ}$, we perform a Fourier decomposition and rescale their images to a common frame determined  by the size in physical units,  by their disk scalelength, 
and for 748 barred galaxies  by both the length and orientation of their bars. We stack the resized density profiles and images to obtain statistically representative average stellar disks and bars in bins of $M_{\ast}$ and $T$. 
Based on the radial force profiles of individual galaxies we calculate the mean stellar contribution to the circular velocity.
We also calculate average $A_{2}$ profiles, where the radius is normalized to $R_{25.5}$. 
Furthermore, we infer the gravitational potentials from the synthetic bars to obtain the tangential-to-radial force ratio ($Q_{\rm T}$) and $A_2$ profiles in the different bins. 
We also apply ellipse fitting to quantitatively characterize the shape of the bar stacks.
}
{
For $M_{\ast} \ge 10^{9}M_{\odot}$, we find a significant difference in the stellar density profiles of barred and non-barred systems: 
(i) disks in barred galaxies show larger scalelengths ($h_{\rm R}$) and fainter extrapolated central surface brightnesses ($\Sigma_{\circ}$), 
(ii) the mean surface brightness profiles ($\Sigma_{\ast}$) of barred and non-barred galaxies intersect each other slightly beyond the mean bar length, most likely at the bar corotation, and 
(iii) the central mass concentration of barred galaxies is higher (by almost a factor~2 when $T\le5$) than in their non-barred counterparts. 
The averaged $\Sigma_{\ast}$ profiles follow an exponential slope down to at least $\sim 10 M_{\odot} \rm pc^{-2}$, which is the typical depth beyond which the sample coverage in the radial direction starts to drop. 
Central mass concentrations in massive systems ($\ge 10^{10}M_{\odot}$) are substantially larger than in fainter galaxies, and their prominence scales with $T$. 
This segregation also manifests in the inner slope of the mean stellar component of the circular velocity: lenticular (S0) galaxies present the most sharply rising $V_{3.6 \mu \rm m}$. 
Based on the analysis of bar stacks, we show that early- and intermediate-type spirals ($0 \le T < 5$) have intrinsically narrower bars than later types and S0s, whose bars are oval-shaped. 
We show a clear agreement between galaxy family and quantitative estimates of bar strength. 
In early- and intermediate-type spirals, $A_{2}$ is larger within and beyond the typical bar region among barred galaxies than in the non-barred subsample. 
Strongly barred systems also tend to have larger $A_{2}$ amplitudes at all radii than their weakly barred counterparts.
}
{
Using near-IR wavelengths (S$^{4}$G 3.6~$\mu$m), we provide observational constraints that galaxy formation models can  be checked against. 
In particular, we calculate the mean stellar density profiles, and the disk(+bulge) component of the rotation curve (and their dispersion) in bins of $M_{\ast}$ and $T$. 
We find evidence for bar-induced secular evolution of disk galaxies in terms of disk spreading and enhanced central mass concentration. 
We also obtain average bars (2-D), and we show that bars hosted by early-type galaxies are more centrally concentrated and have larger density amplitudes than their late-type counterparts.
}
\maketitle
%
%
\section{Introduction}
In the lambda cold dark matter ($\Lambda$CDM) model, the seeds for dark matter halos arise from quantum fluctuations amplified by cosmic inflation. 
The halos gain angular momentum from cosmological torques as they grow. Galaxies are formed from the cooling and condensation of gas in their centres \citep[][]{1978MNRAS.183..341W,1980MNRAS.193..189F}, 
and the baryons inherit the angular momemtum from their host halos. 

In the early universe galaxies suffered recurrent mergers, which became less frequent with time. 
At $z\sim 2-2.5$ galaxies were still thick, gas-rich, and clumpy, and actively formed stars \citep[e.g.][]{2007ApJ...670..237B,2008ApJ...687...59G,2009ApJ...703..785D,2011ApJ...739...45F}. 
\citet[][]{2013ApJ...771L..35V} found that in galaxies with present-day stellar masses similar to that of the Milky Way (MW, log$_{10} (M_{\ast}/M_{\odot}) \approx 10.7$) 
most of the star formation had already taken place before $z \sim 1$, at which time these systems typically show an almost fully assembled backbone with a quiescent 
bulge\footnote{Throughout this paper, and unless stated otherwise, the term `bulge'  refers to the excess of flux above the disk in the central regions of galaxies, independent of the stellar structures that emit this light.} and a slowly star-forming disk. 
This is consistent with studies showing a lack of strong evolution in the stellar mass-size relation in disk galaxies over the last 8 billion years \citep[e.g.][]{1998ApJ...500...75L,1999ApJ...519..563S,2005ApJ...635..959B}. 
In that period the bar fraction \citep[e.g.][]{2004ApJ...615L.105J,2004ApJ...612..191E} or the locii of outer rings relative to bars \citep[e.g.][]{2012A&A...540A.103P} have been found to be constant. 
However, \citet[][]{2008ApJ...675.1141S} found a constant bar fraction up to $z\sim 0.84$ only for the most massive spirals ($M_{\ast}\gtrsim10^{11}M_{\odot}$), whereas for fainter and bluer systems it declined substantially beyond $z\sim 0.3$. 
Recent work by \citet[][]{2015MNRAS.451....2S} concludes that disk galaxies have experienced a substantial peripheral growth since $z\sim1$ \citep[based on measurements of the Petrosian radius;][]{1976ApJ...209L...1P}. 

Disks in spiral galaxies exhibit luminosity profiles that tend to decay exponentially in the radial direction \citep[][]{1970ApJ...160..811F}.
Exponential or quasi-exponential stellar disks have been produced in a $\Lambda$CDM framework in simulation models \citep[e.g.][]{2003ApJ...597...21A,2004ApJ...606...32R,2009MNRAS.396..121D}. 
Recent observations in optical \citep[e.g.][]{2005ApJ...626L..81E,2006A&A...454..759P,2008AJ....135...20E} 
and infrared wavelengths \citep[e.g.][]{2013ApJ...771...59M,2014MNRAS.441.1992L,2014ApJ...782...64K} show that disks can have breaks in their radial surface brightness profiles.
Specifically, disks are classified as Type I if they are purely exponential, and Type II and Type III respectively if they become steeper or shallower after the break.

Roughly two-thirds of the local galaxies have a bar \citep[e.g.][]{1991rc3..book.....D,2000ApJ...529...93K,2002MNRAS.336.1281W,2004ApJ...607..103L}. 
Stellar bars are known to conduct angular momentum of the baryonic and dark matter components throughout the disks of spiral galaxies \citep[e.g.][]{1985MNRAS.213..451W, 2002MNRAS.330...35A,2007ApJ...659.1176M}. 
Namely, angular momentum is emitted from the material in the surroundings of the inner Lindblad resonance of the bar, 
and absorbed by the material near the resonances  associated with the spheroidal components (dark matter halo and bulge, when present) and with the outer disk. 
According to different dynamical models \citep[e.g.][]{1991MNRAS.250..161L,1998ApJ...493L...5D,2002MNRAS.330...35A,2003MNRAS.341.1179A,2006ApJ...637..214M}, 
while bars exchange angular momentum they become narrower, longer, stronger, and they slow down their rotation speed. 
Early analytical studies by \citet[][]{1980A&A....81..198C} proved that the orbits making up the bar do not extend beyond the corotation resonance radius ($r_{\rm cr}$). 
The bar slow-down predicted by N-body simulations is not easy to reconcile with bar pattern speed estimates from observations. 
\citet[][]{2005ApJ...631L.129R,2008MNRAS.388.1803R} showed that bars in early-type galaxies have typical ratios $r_{\rm cr}/r_{\rm bar}\le1.4$  (known as fast bars), while later-types have bars which can be both fast and slow. 
Recent measurements using the Tremaine-Weinberg method \citep[][]{1984MNRAS.209..729T} indicate fast bars for all morphological types \citep[][]{2015A&A...576A.102A}.
Models by \citet[][]{1992MNRAS.259..345A} showed that $r_{\rm cr}/r_{\rm bar}$ determines the shape of the offset dust lanes. 
Among fast bars, the curvature of dust lanes was found to inversely scale with the strength of the bar in the recent work by \citet[][]{2015MNRAS.450.2670S}, 
while slow bars had similar values of the mean curvature for all bar strength values. 
This confirmed the theoretical prediction in \citet[][]{1992MNRAS.259..345A} \citep[see also][]{2002MNRAS.337..808K,2009ApJ...706L.256C}. 

Possible observational evidence for the bar evolution was provided in \citet[][]{2007MNRAS.381..401L}, \citet[][]{2007ApJ...670L..97E}, \citet[][]{2011MNRAS.415.3308G}, 
and \citet[][hereafter DG2016]{2016A&A...587A.160D}, by finding a dependence between different proxies of the bar strength and estimates of their length, and by studying the evolution of the bar parameters in the Hubble sequence.
The bar evolution is also manifested  in the buckling instability in the vertical direction that gives birth to boxy/peanut (B/P) bulges 
\citep[][]{1981A&A....96..164C,1990A&A...233...82C,1991Natur.352..411R,2002MNRAS.330...35A,2004ApJ...604L..93D,2004ApJ...613L..29M,2006ApJ...645..209D}. 
The so-called barlenses \citep[][]{2011MNRAS.418.1452L} are thought to be boxy/peanut bulges seen in the face-on view 
\citep[][]{2007MNRAS.381..401L,2014MNRAS.444L..80L,2015MNRAS.454.3843A}. For a review of the properties of B/P bulges, the reader is referred to \citet[][]{2016ASSL..418...77L} and \citet[][]{2016ASSL..418..391A}.

Bars participate in the redistribution of stars and gas inside the disk \citep[][and references therein]{2013seg..book..305A} by pushing them outwards (inwards) beyond (within) the corotation radius (CR).
Simulation models predict that this process increases the disk size \citep[e.g.][]{1971ApJ...168..343H} and causes secular evolution of bulges \citep[e.g.][]{1992MNRAS.259..328A,1992MNRAS.258...82W,1993A&A...268...65F} 
after the bar-funneled cold gas is transformed into stars, and also as a result of old stars being driven inward by the bar torques \citep[][]{2004A&A...423..849G}. 
Actually, signatures of disk-like central components in the kinematics of observed barred galaxies have been found \citep[e.g.][]{2009A&A...495..775P,2015MNRAS.451..936S}. 
In addition, an enhanced star formation in the central parts of barred galaxies has been detected \citep[e.g.][]{2011MNRAS.416.2182E,2012ApJS..198....4O,2015AJ....149....1Z,2015A&A...584A..88F}. 
Based on studies of the bar fraction in the Galaxy Zoo \citep[][]{2011MNRAS.411.2026M} and measurements of bar sizes, \citet{2013ApJ...779..162C} provided empirical evidence for bar-driven secular evolution within the central kpc of disk galaxies.

Furthermore, bars are known to be responsible for the formation of resonance rings \citep[e.g.][]{1981ApJ...247...77S,1986ApJS...61..609B,2000A&A...362..465R}. 
Whether stellar bars drive spiral density waves has been a matter of debate \citep[e.g.][]{1979ApJ...233..539K,2003MNRAS.342....1S,2004AJ....128..183B,2005AJ....130..506B,2009AJ....137.4487B,2009MNRAS.397.1756D}. 
Recent work by \citet{2010ApJ...715L..56S} supports such causality. Using SDSS-DR2 photometry, observational evidence for the bar-induced disk secular evolution was provided by \citet[][]{2013MNRAS.432L..56S}, 
who found that barred galaxies with stellar masses $M_{\ast}>10^{10}M_{\odot}$ at redshifts $0.02 \le z \le 0.07$ typically have fainter extrapolated central surface brightness and larger disk scalelengths than their non-barred counterparts.

Based on IFU observations of three prominent galactic bulges and full spectral fitting methods, recent work by \citet[][]{2015MNRAS.446.2837S} found that at least $50\%$ of the stars in those bulges formed at $z\sim4$. 
They also detected a younger component ($\sim 1-8$ Gyr). Two bulge population families were found in the models of MW-type galaxies in a cosmological framework by \citet[][]{2013ApJ...763...26O}, 
the first forming during an early starburst-collapse (old stars) and the second during the phase driven by processes such as disk instabilities and/or mergers (young stars).

In order to obtain the observational constraints needed to check  galaxy formation models, 
\citet[][]{2009MNRAS.396..121D} proposed  measuring the average deprojected surface brightness profiles as a function of the primary galaxy parameters, such as stellar mass, colour, and size. 
Mid-IR rest-frame wavelengths are well suited to this as they are very sensitive to the old stellar populations and are barely affected by dust absorption.
For this reason, the Spitzer Survey of Stellar Structure in Galaxies \citep[S$^{4}$G;][]{2010PASP..122.1397S}, which includes 2352 nearby galaxies, is an ideal sample to carry out such a study at $z\approx0$. 

Using S$^{4}$G 3.6~$\mu$m imaging, DG2016 provided a first-order estimate of the halo-to-stellar mass ratio ($M_{\rm halo}/M_{\ast}$) from comparisons of the stellar component of the circular velocity 
to kinematic H{\sc\,i} data compilation from \citet[][]{2009AJ....138.1938C,2011MNRAS.414.2005C} and HyperLEDA\footnote{We acknowledge the usage of the database (http://leda.univ-lyon1.fr).} \citep{2003A&A...412...45P}. 
\citet[][]{2016A&A...587A.160D} found a good agreement in the slope of the $M_{\rm halo}/M_{\ast}$-$M_{\ast}$ relation with the best-fit model at $z\approx0$ in $\Lambda$CDM cosmological simulations \citep[e.g.][]{2010ApJ...710..903M}. 
Based on various bar measurements, DG2016 carried out a statistical analysis of the properties of bars and their fraction at $z=0$ as seen in the S$^{4}$G sample, 
providing possible evidence for the growth of galactic bars within a Hubble time. 
In the current paper, the characterization of galactic bars is done with an inverted approach: 
first we stack images of individual galaxies to obtain average bar density distributions as a function of stellar mass, revised Hubble type and galaxy family; and then study the properties of these stacked bar images.

This paper is organized as follows. In Sect.~\ref{data1} we present the  S$^{4}$G data and the criteria used to define the samples. 
In Sect.~\ref{scaling-bars} we describe the methodology for resizing and stacking  the galaxy images and 1-D profiles. 
In Sect.~\ref{disk-char} we study the luminosity profiles of the stellar disks based on the stacks and we study the mean $A_2$ profiles and stellar component of the circular velocity. 
In Sect.~\ref{bars-char} we analyse the shape and strength of the average bars. 
In Sect.~\ref{bar-hubble} we characterize the bars in the Hubble sequence. 
In Sect.~\ref{disk-assemb} and Sect.~\ref{disk-hubble} we discuss on the assembly and secular evolution of disk galaxies; we emphasize the role of bars in this process, whose effect is demonstrated based on the properties of the disk stacks. 
In Sect.~\ref{summarysection} we summarize the main results of this paper.
%
%
\section{Data and sample selection}\label{data1}
%
%
The S$^{4}$G survey \citep[][]{2010PASP..122.1397S} consists of 2352 nearby galaxies (distance $\lesssim 40$ Mpc)
observed in the 3.6~$\mu$m and 4.5~$\mu$m bands with the InfraRed Array Camera \citep[IRAC;][]{2004ApJS..154...10F} installed on board  the \emph{Spitzer} Space Telescope \citep{2004ApJS..154....1W}. 
Data taken from HyperLEDA were used to define the sample which is composed of bright and large galaxies (extinction-corrected total blue magnitude $m_{B_{\rm corr}} \textless 15.5$ mag and 
blue light isophotal angular diameter $D_{25} \textgreater 1^{\prime}$) located far from the Milky Way plane (Galactic latitude $|b|\textgreater 30^{\circ}$).
The galaxies were mapped out to $1.5\cdot D_{25}$ or more. However, for 125 galaxies the mosaics did not reach that far. A wide range of masses ($\sim$~5 orders of magnitude) and all Hubble types ($T$) are included in the S$^{4}$G sample. 
However, there is a bias towards late-type systems because of the distance restriction based on H{\sc\,i} recessional velocities \citep[observations are currently extended to gas-poor early-type galaxies; see][]{2013sptz.prop10043S}. 

Here we use the 3.6~$\mu$m images (FWHM~$\approx2.1\arcsec$, pixel scale of $0.75\arcsec$pixel$^{-1}$)
allowing us to reach stellar mass surface densities of $\sim$~1~$M_{\odot}/$pc$^{2}$ ($\mu_{\rm 3.6 \mu \rm m}$(AB)(1$\sigma$) $\sim$27 mag arcsec$^{-2}$).
For further details on the processing and reduction of the raw data to obtain science-ready imaging, the sky subtraction and masking process, see \citet[][]{2015ApJS..219....3M} and \citet[][]{2015ApJS..219....4S}. 
%
%
We use the morphological classification from \citet[][hereafter B2015]{2015ApJS..217...32B}, who used the revised version of de Vaucouleurs Hubble-Sandage system \citep[][]{1959HDP....53..275D}. 
Measurements of the bar sizes ($r_{\rm bar}$) and position angles ($PA_{\rm bar}$) are taken from the catalogue of structures by \citet[][hereafter HE2015]{2015A&A...582A..86H}. 
We use the disk orientation parameters, bulge-to-total ratios (B/T), 
and disk scalelengths ($h_{\rm R}$) from S$^{4}$G Pipeline 4 \citep[][hereafter P4]{2015ApJS..219....4S}, where a 2-D decomposition of the 3.6~$\mu$m light distribution into different structure components was done 
using GALFIT \citep{2010AJ....139.2097P} and IDL-based visualization tools (GALFIDL). 
Distances and total stellar masses are taken from S$^{4}$G Pipeline 3 \citep[][hereafter P3]{2015ApJS..219....3M}. 
The isophotal radii at the surface brightness 25.5 mag arcsec$^{-2}$ ($R_{25.5}$), calculated from the 3.6~$\mu\rm m$ images, are also taken from P3.

In order convert the 3.6~$\mu$m surface brightness ($\mu$) to stellar mass density ($\Sigma_{\ast}$),
we use the formulae given in \citet{2013ApJ...771...59M}:
\begin{equation}
{\rm log_{10}} (M_{\ast}/M_{\odot})=-0.4 M_{\rm 3.6\mu m} + 2.13,
\label{munoz1}
\end{equation} 
\begin{equation}
{\rm log_{10}} (\Sigma_{\ast})[M_{\odot}\rm \,kpc^{-2}]=16.76-0.4\mu \rm [mag\,arcsec^{-2}],
\label{munoz2}
\end{equation} 
where $M_{\rm 3.6\mu m}$ refers to the total 3.6~$\mu$m absolute magnitude (AB system), assuming a mass-to-light ratio $M/L=\Upsilon_{3.6 \rm \mu m}=0.53$ \citep[][]{2012AJ....143..139E}.

Only 1345 S$^4$G disk galaxies (ellipticals excluded) have P4 inclinations lower than $65^{\circ}$;  
1226 of these non-highly inclined galaxies have \emph{`ok'} quality flags in P4, meaning that the inclination can be reliably estimated from the ellipse fits. 
Our sample is comprised of 1154 galaxies that reach $R_{25.5}$ (see Sect.~\ref{coaddbars} for further details). 
About $2/3$ of these galaxies (748) are barred according to B2015, and have reliable measurements of bar lengths in HE2015, constituting our subsample of barred galaxies.
%
%
\begin{figure}
\centering   
\begin{tabular}{c c}
   \includegraphics[width=0.41\textwidth]{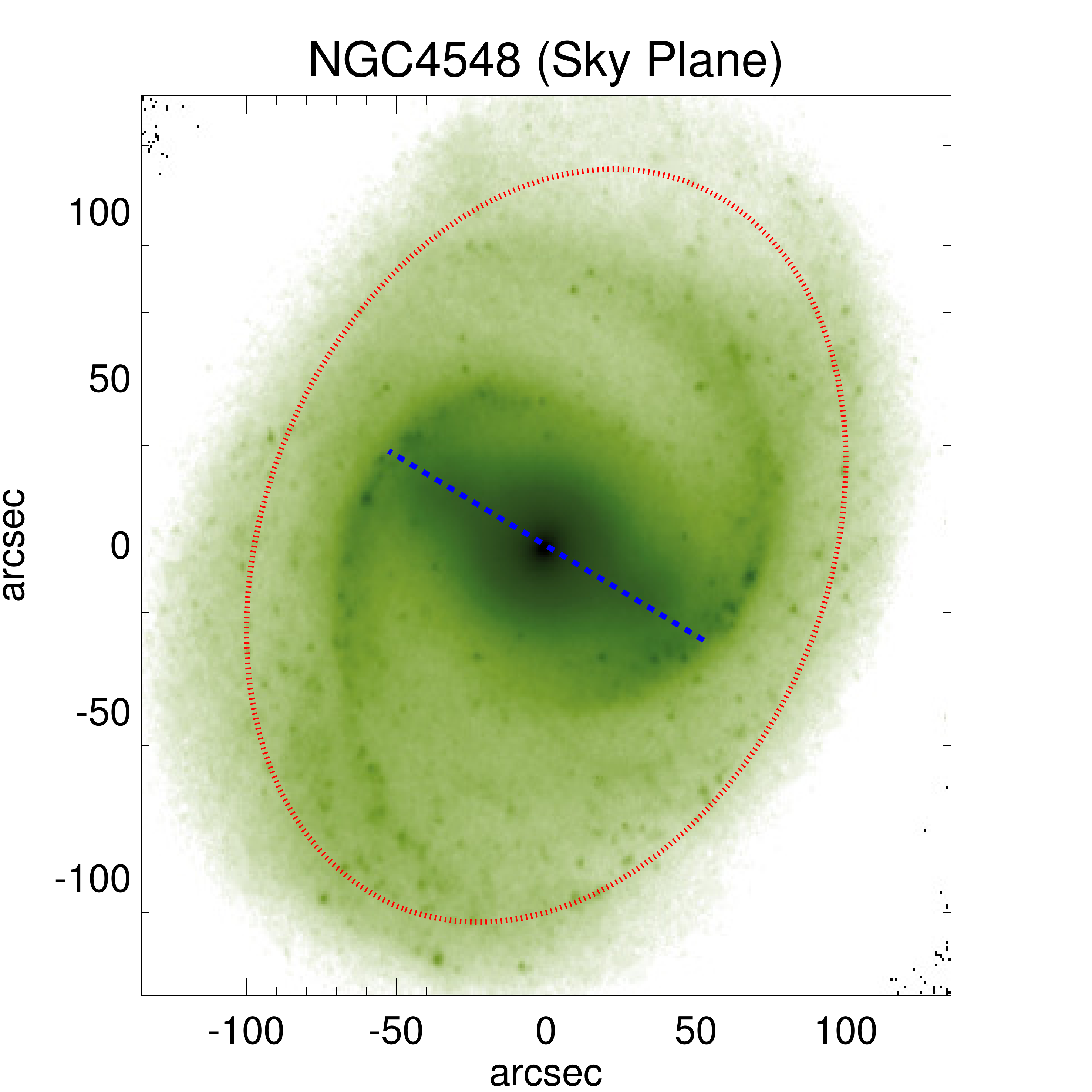}\\
   \includegraphics[width=0.4\textwidth]{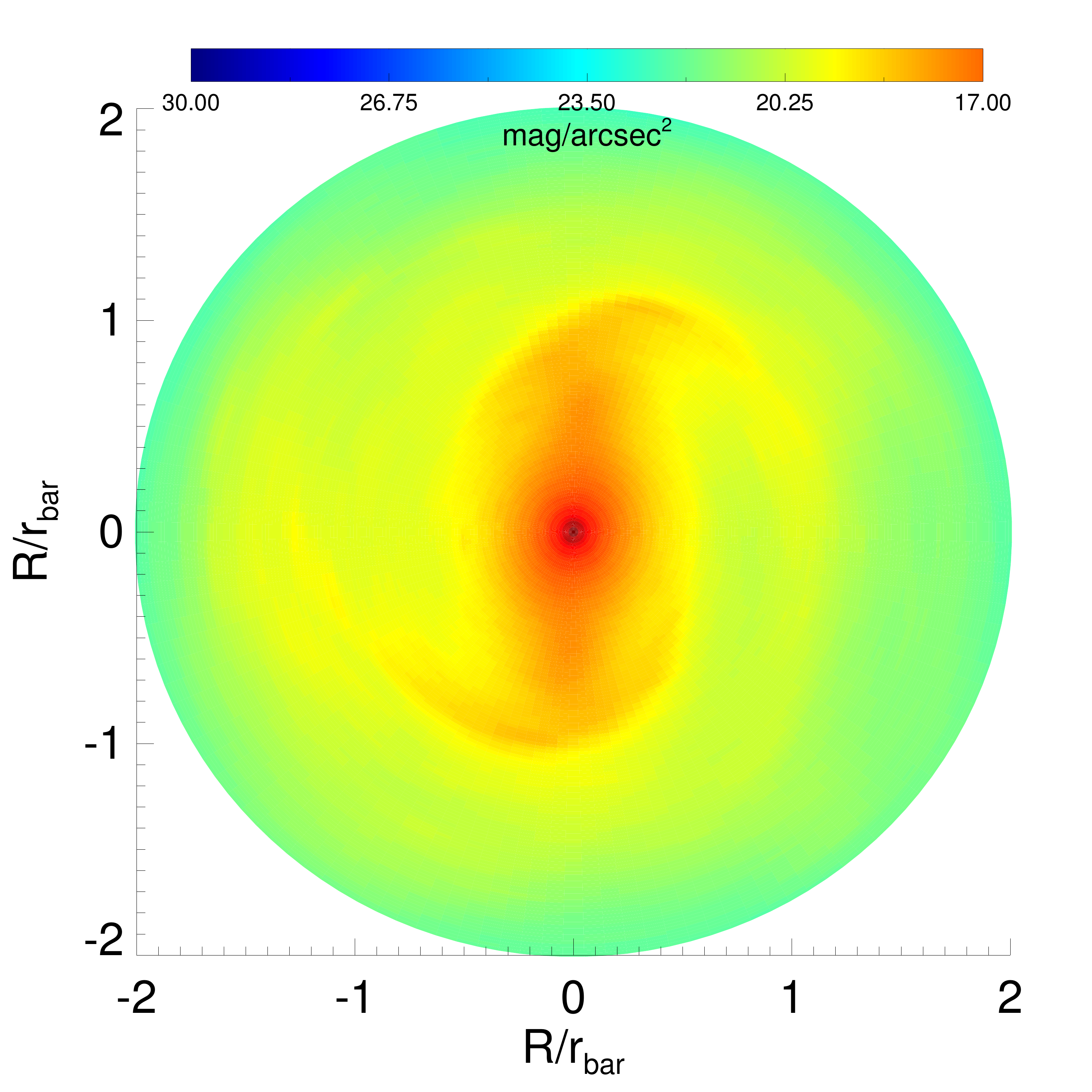}
\end{tabular}
\caption{
Rescaling of the images prior to obtaining bar stacks.
\emph{Upper panel:}
Image of NGC$\,$4548 in magnitude scale with range 17-25 $\mu_{3.6\mu \rm m}(AB)$. The frame is aligned such that the \emph{y-axis} points north and east is to the left. Axes are in units of arcsec.
The blue dashed line corresponds to the visual measurement of the bar length and position angle from HE2015.
The red dotted line corresponds to an ellipse of semi-major axis equal to $2\cdot r_{\rm bar}$, whose PA and ellipticity are determined from the disk orientation parameters from P4. 
\emph{Lower panel:}
De-projected image of NGC$\,$4548 reconstructed from the Fourier components in a polar grid on the disk plane.
The image has been reoriented with respect to the bar major axis ($PA_{\rm bar}=0$), geometrically reflected across the bar major axis (to make the spiral arms wind clockwise), 
and resized to a grid of radius 3 times the bar size.
}
\label{stack1}
\end{figure}
%
%
%
\section{Image scaling and stacking}\label{scaling-bars}
%
%
%
In this section we explain how the 3.6~$\mu$m frames are processed in order to obtain the stacks that
we  use to characterize the disks and bars of our sample.
%
%
\subsection{Fourier decomposition and scaling of the decomposed images}\label{bar_disk_method}
%
%
Under the assumption of an intrinsically circular and infinitesimally thin disk, the 3.6~$\mu$m images are deprojected to face-on. 
We use the disk orientation parameters from P4 (outer orientation angle and axial ratio), 
which were derived from the outermost isophotes of the galaxies after ellipse fitting was performed using the IRAF task \emph{ellipse} \citep{1987MNRAS.226..747J}.

The Fourier decompositions of the light distribution of the galaxy images (up to 40 modes) were performed in a polar grid with 128 bins in the azimuthal direction \citep[][]{1999AJ....117..792S}:
\begin{equation}\label{fourierdec}
I(r,\phi)=I_{0}(r)\left[1+\sum_{m=1}^{m=40} A_{m}(r)cos[m\big(\phi-\phi_{m}(r)\big)]\right],
\end{equation}
where $A_{m}=I_{m}/I_{0}$ are the normalized Fourier amplitudes and $I_{0}$ refers to the $m=0$ surface density component. 
The radial extent of the polar grid is the rough estimate of the galaxy outer radius used in P4 to define the image region for the GALFIT decomposition, called as $R_{\rm lim}$. 
Beyond this radius the galaxy flux at 3.6~$\mu$m is negligible. 
Unlike in 2-D multi-component decompositions, Fourier decompositions demand full azimuthal coverage of the galaxy images at all radii. 
For this reason, for a dozen of systems $R_{\rm lim}$ was reduced to avoid cropped sections of the galaxy images in the mosaics.

For the averaging of stellar disks, the only relevant Fourier mode is $m=0$, comprising the asymmetric component of the light distribution of the galaxies.
Using spline interpolation, the $I_{0}$ profiles are resized to a common frame determined by 
 the scalelength of the disk (up to $6\cdot h_{\rm R}$, using a 0.04~$h_{\rm R}$ wide radial bin) 
and  the extent of the disk in physical units (up to 25 kpc, using a 0.125~kpc wide radial bin).

For the stellar bars, the $m=2$ amplitude covers a large fraction of the bar flux, 
although higher order even modes ($m$ = 4, 6, and 8) also have  a significant contribution among barred lenticulars,
and early- and intermediate-type spirals \citep[][DG2016]{1996ASPC...91...37O,2002MNRAS.331..880L}. 
The images of the barred galaxies in our sample are reconstructed from Eq.~\ref{fourierdec}. 
By means of this decomposition+reconstruction sequence, we filter the 3.6~$\mu$m photometry by removing sharp density features, 
for instance those caused by star-forming regions and dust lanes that pollute the 3.6~$\mu$m emission \citep[e.g.][]{2015ApJS..219....5Q}. 
An image of the barred galaxy NGC$\,$4548 and that reconstructed from the Fourier modes are shown in Fig.~\ref{stack1}. 

Prior to doing the bar stacks, barred galaxies need to be scaled and oriented with respect to the stellar bars. 
The fact that we are using a polar grid facilitates this process, consisting of the following steps (see Fig.~\ref{stack1}): 
\begin{enumerate}
 \item Rotation of the image with respect to the bar major axis, so that the final bar position angle is zero;
 \item Geometric reflection across the bar major axis to make the spiral arms wind clockwise (in case they wind counterclockwise in the sky);
 \item Scaling of the reoriented image to a grid of radius $3 \cdot r_{\rm bar}$ (bins width of 0.02~$r_{\rm bar}$) using linear interpolation. 
\end{enumerate}
%
%
\begin{figure}
\centering
\includegraphics[width=0.5\textwidth]{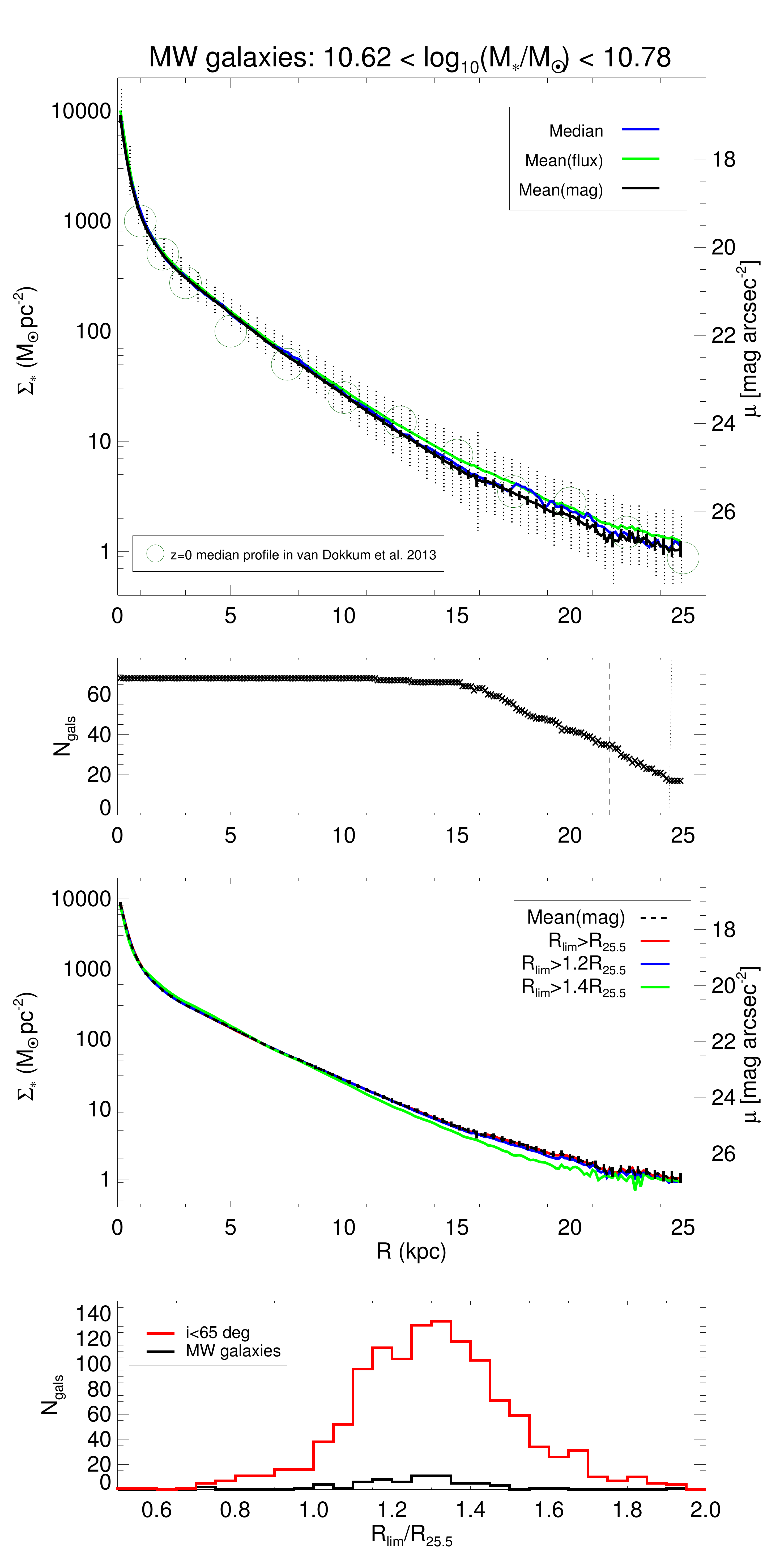}\\
\caption{
\emph{First panel:} 
Average surface mass stellar density profiles obtained from the stacking of 3.6~$\mu$m $I_{0}$ profiles, resized to a common physical scale in kpc, 
of 79 disk galaxies with total stellar masses comparable ($\pm 0.1$ dex) to that of the Milky Way ($M_{\rm MW}\approx5\cdot10^{10}M_{\odot}$). 
Different ways of averaging the stacks are compared, as indicated in the upper right legend. 
The sample dispersion ($\sigma$) and the standard deviation of the mean ($\sigma/\sqrt{N_{\rm gals}}$) are indicated with error bars (dashed and solid vertical lines, respectively), centred on the mean profile stacked in units of mag. 
The median profile for MW galaxies at $z\approx0$ from \citet[][]{2013ApJ...771L..35V} (traced from their Fig.~3) is also overplotted. 
\emph{Second panel:} 
For the same MW-type stellar mass bin, the fraction of galaxies whose radial profile extends to a given galactocentric radius. 
The $75 \%$, $50 \%,$ and $25 \%$ sample completeness levels are highlighted with vertical solid, dashed, and dotted lines, respectively.
\emph{Third panel:} 
Mean $\Sigma_{\ast}$ of MW galaxies sampled based on how far they extend (estimated from $R_{\rm lim}$) relative to $R_{25.5}$.
\emph{Fourth panel:} 
For the systems in our sample, histogram of the distributions of the image outer radius relative to $R_{25.5}$.
}
\label{stack_MW_dokum_methods}
\end{figure}
%
%
\subsubsection{Gravitational potential and scaling of $V_{\rm 3.6\mu m}$ and $A_{\rm 2}$}\label{bar_disk_method_vrot}
%
%
We use gravitational potentials ($\Phi$) obtained by DG2016 from the Fourier decomposed 3.6~$\mu$m images, applying the NIR-QB code \citep[][]{1999AJ....117..792S,2002MNRAS.337.1118L}. 
A constant mass-to-light ratio and an exponential vertical profile were assumed, and the scaleheight was estimated by applying the scaling relation between $h_{\rm z}$ and $h_{\rm R}$ given in \citet[][]{1998MNRAS.299..595D}. 
The stellar contribution to the circular velocity curve ($V_{\rm 3.6\mu m}$) of all the galaxies in our sample was estimated as 
\begin{equation}\label{eq-rot-cur}
V_{\rm 3.6\mu m}(r)=\sqrt{\Upsilon_{3.6 \rm \mu m}\left<F_{\rm R}(r)\right> r}, 
\end{equation}
where $r$ is the galactocentric radius and $F_{\rm R}=\frac{\partial \Phi}{\partial r}$ is the radial force obtained for a mass-to-light ratio $M/L=1$. 
A detailed analysis of the uncertainties in $V_{\rm 3.6\mu m}$ associated with the constant $M/L$ and scaleheight was presented in DG2016; 
these uncertainties are not larger than the nominal uncertainty on the value of $M/L$ \citep[see][]{2012AJ....143..139E,2014ApJ...788..144M}.

In DG2016 the tangential-to-radial force ratio \citep[$Q_{\rm T}$;][]{1981A&A....96..164C,2001A&A...375..761B,2001ApJ...550..243B} was also obtained. 
For most of the barred galaxies in our sample, the maxima of $Q_{\rm T}$ and $A_{2}$ in the bar region were calculated (denoted  $Q_{\rm b}$ and $A_{2}^{\rm max}$, respectively) and  used as proxies of bar strength. 
For all the galaxies in our sample, $A_{2}$ and $V_{\rm 3.6\mu m}$ radial profiles are also rescaled to a common frame in an identical fashion as $I_{0}$.
%
%
%
\subsection{Coadding galaxy disks and bars}\label{coaddbars}
%
%
Having rescaled and reoriented the disks, we are in the position to co-add them and obtain the average light profiles in bins of $M_{\ast}$, $T$, and galaxy family.

In the upper panel of Fig.~\ref{stack_MW_dokum_methods} we show the 1-D surface brightness profile resulting from stacking 79 galaxy images, resampled to a common physical scale in kpc,
with total stellar masses similar ($\pm 0.1$ dex) to that of the Milky Way ($M_{\rm MW}\approx5\cdot10^{10}M_{\odot}$). 
We also show the agreement of our results with the median profile at $z=0$ obtained by \citet[][]{2013ApJ...771L..35V} for MW-type galaxies using SDSS $g$-band photometry 
(we have corrected the factor of 2 error in the normalization for their Fig.~3; private communication with Pieter G. van Dokkum).

We also compare different methodological aspects for the stacks, namely the units when co-adding the light of each individual galaxy (flux or magnitudes), the  measure of central tendency used (mean or median), 
and the treatment of the outskirts when some of the sampled galaxies do not extend as far as the rest. 
Whether we use units of flux (MJy/sr) or magnitudes, the ranking of the data points is the same, hence median profiles always give the same results. 
Because of the greater sensitivity of the mean $\Sigma_{\ast}$ to the brightest systems, mean profiles computed in units of flux are slightly denser than when using median or mean magnitude. 
Since many of the images do not extend as far as 20-25 kpc, in the second panel of Fig.~\ref{stack_MW_dokum_methods} we show the sample completeness level (calculated from $R_{\rm lim}$) as a function of radius. 
In the outer regions of the common frame, the calculation of mean/median $\Sigma_{\ast}$ in the outskirts is only done using the systems whose $R_{\rm lim}$ extends that far (and when using magnitude units, with positive $I_{0}(r)$ values). 
As a result, up-bending sections may appear (for MW galaxies a break appears at $\sim 25.5$ mag arcsec$^{-2}$) which cannot be considered statistically representative. 
In addition, as shown in the bottom panel of Fig.~\ref{stack_MW_dokum_methods}, there are systems which do not reach as far as $R_{25.5}$ (see the S$^4$G pipelines for details). 
Sampling the galaxies based on the value of $R_{\rm lim}$ relative to $R_{25.5}$ (third panel in the same figure) can result in subtly fainter mean $\Sigma_{\ast}$, 
although this can introduce a bias because  less centrally concentrated galaxies extend further (relative to $R_{25.5}$, for a fixed $M_{\ast}$).

Unless stated otherwise, hereafter we use mean radial $\Sigma_{\ast}$ profiles resulting from the stack of $I_{0}$ amplitudes in units of mag resized to a common physical scale in kpc or normalized to $h_{\rm R}$. 
To avoid uncertainties related to the surface brightness depth and radial extent of the individual galaxies discussed above, 
we limit the analysis of $\Sigma_{\ast}$ up to the radius enclosing $75~\%$ of the (sub)sample coverage. 
In addition, the sample selection was limited to galaxies having $R_{\rm lim} > R_{25.5}$. 
Uncertainties on the stacks are estimated from the standard deviation of the mean, $\sigma/\sqrt{N_{\rm gals}}$, where $N_{\rm gals}$ refers to the number of galaxies inside a certain bin.

\begin{figure}
\centering   
\includegraphics[width=0.475\textwidth]{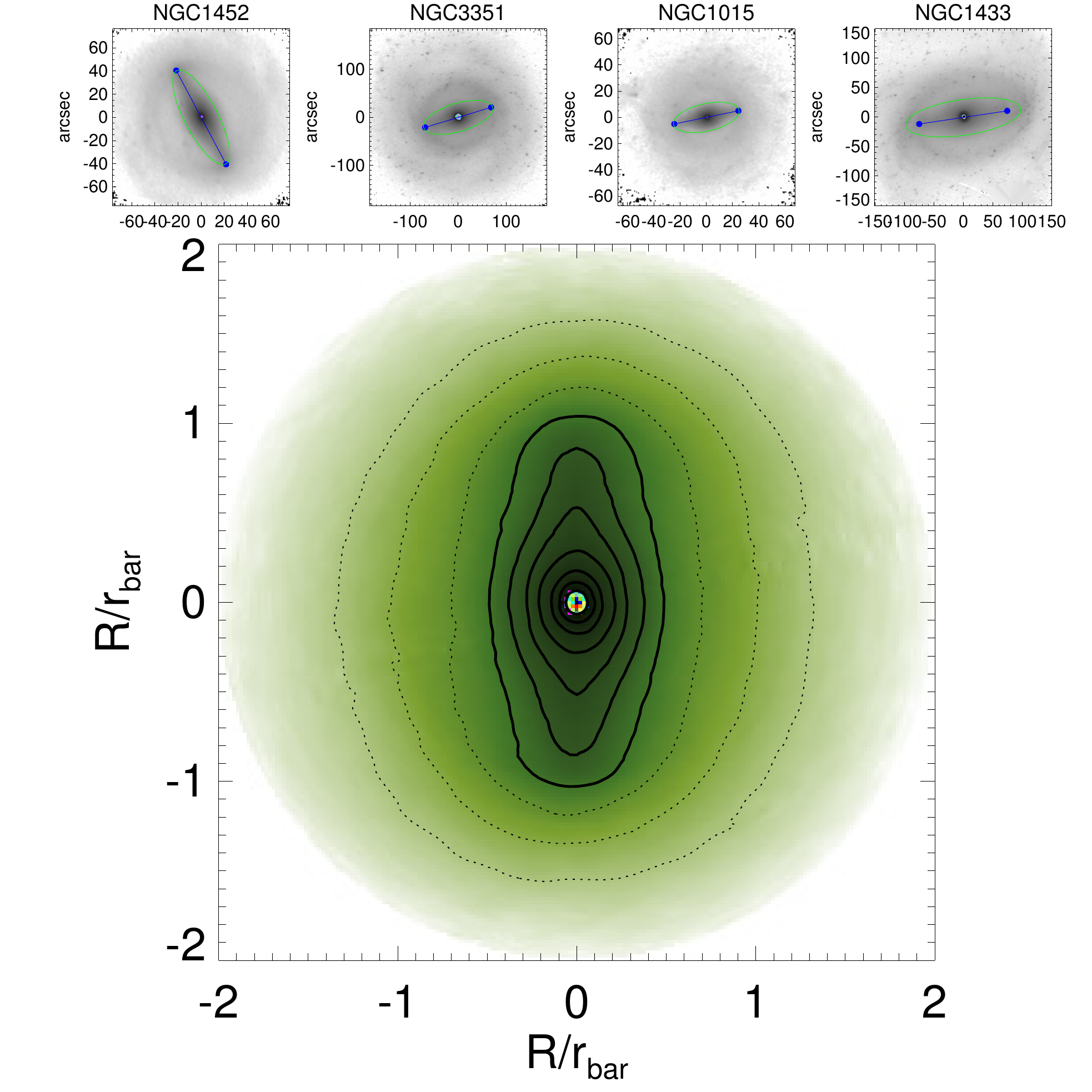}
\caption{
Co-added 2-D bar in magnitude scale with range 17-25 $\mu_{3.6\mu \rm m}(AB)$ and cropped to a radius $\sim2r_{\rm bar}$, 
resulting from the stacking of galaxies classified as SB in B2015 with Hubble types -1~$\le$~T~$\le$~1 (38 systems).
The black lines show the isocontours corresponding to the surface brightness levels 18-21 (solid line) and 21.5-22.5 (dotted line) mag arcsec$^{-2}$ with a step of 0.5 mag arcsec$^{-2}$.
Some of the galaxies used in the stack are shown in the uppermost panels with the bar elliptical isophote and visual bar measurement from HE2015 highlighted in green and blue, respectively.
}
\label{stack1-bars}
\end{figure}

In Fig.~\ref{stack1-bars} we present the result from co-adding all the bars hosted by SB galaxies with Hubble types $-1 \le T \le 1$ (38 systems). 
From the resulting 2-D image we can distinguish a well-defined bar, bulge, barlens, and a hint of an ansae shape. 
The spiral arms arising from the bars were systematically reoriented to wind clockwise in each of the binned galaxies. 
Consequently, the outermost bar isophotes of the stack may show a small bending with respect to the bar major axis towards the spiral arms. 
The effect is subtle in this example, but it is more noticeable for later-type spiral galaxies because of their larger pitch angles (see Sect.~\ref{bars-shape}). 
In addition to the spiral arms, other stellar structures such as lenses \citep[e.g.][]{1979ApJ...227..714K,2009ApJ...692L..34L} and rings \citep[e.g.][]{2014A&A...562A.121C} 
are intrinsically elliptical and appear typically aligned with the bar among the early-type galaxies, which explains why the isocontours beyond the bar in Fig.~\ref{stack1-bars} are not circular. 
However, in this work we focus the analysis of bar stacks exclusively on the bar region ($R\lesssim 1.5r_{\rm bar}$) and not in the outer parts. 
For this reason, the sample completeness in the radial direction is not an issue here.

\begin{figure}
\centering
\begin{tabular}{c c}
   \includegraphics[width=0.5\textwidth]{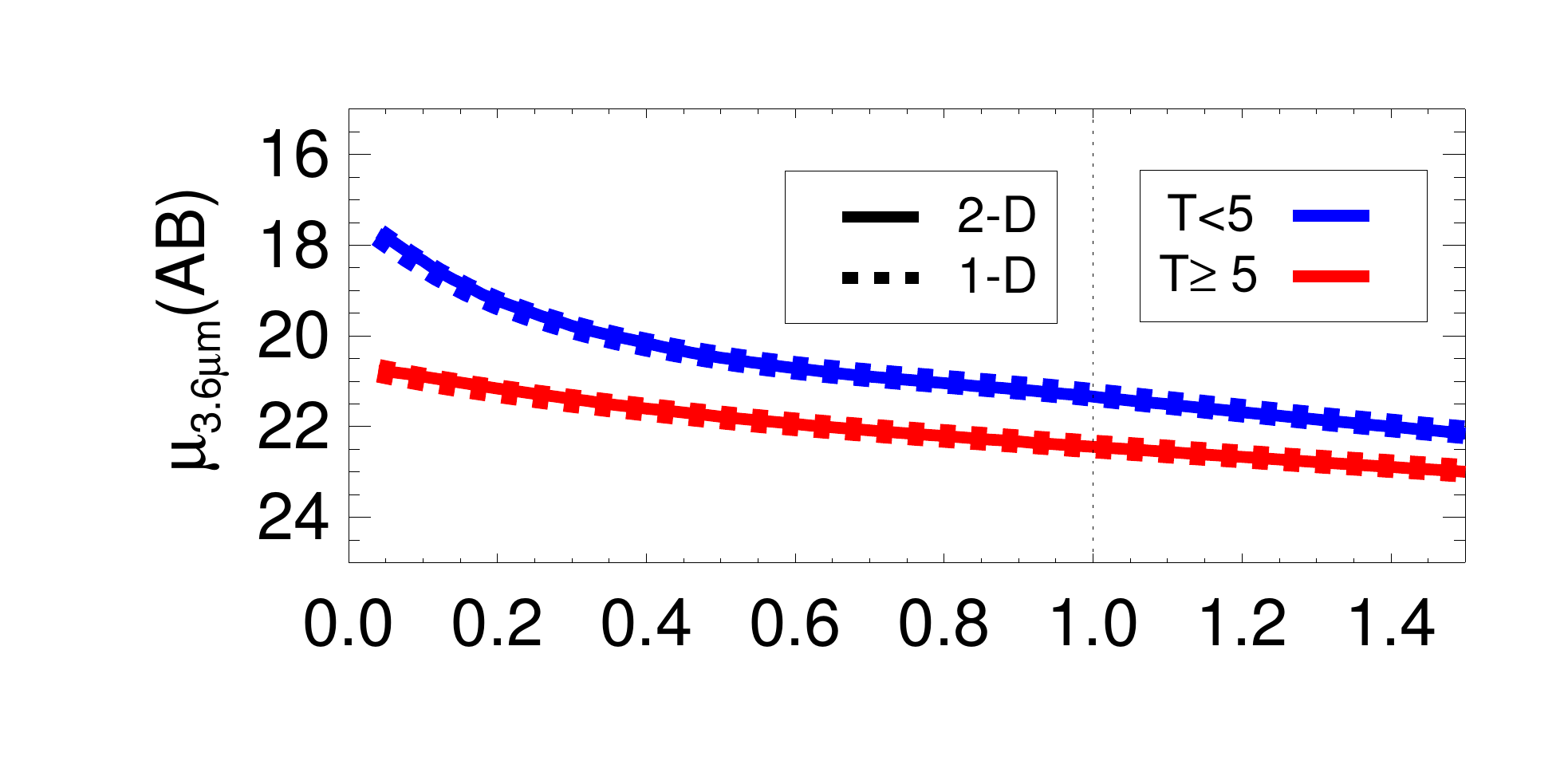}\\[-8ex]
   \includegraphics[width=0.5\textwidth]{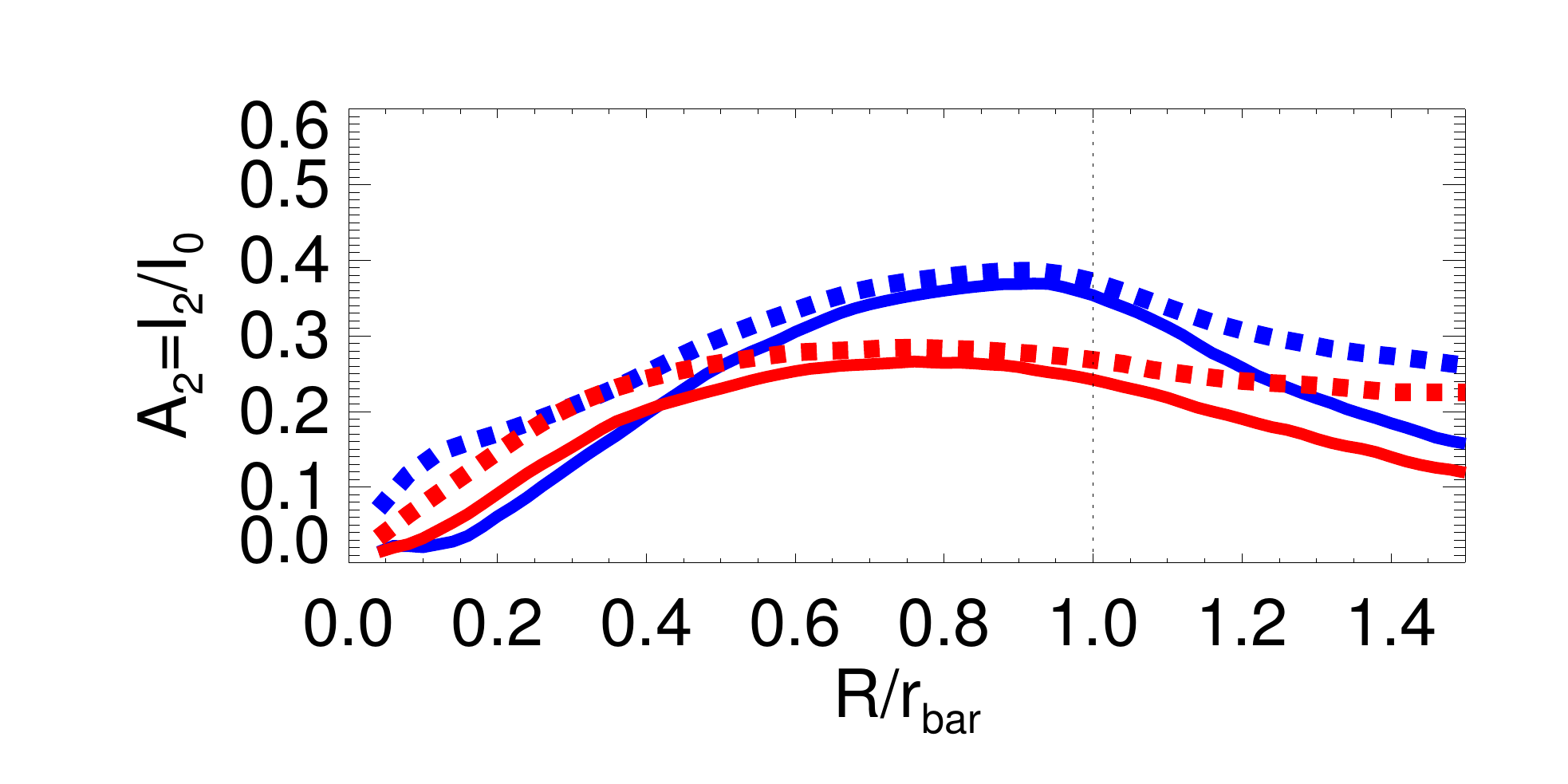}
\end{tabular}
\caption{
Mean 3.6~$\mu$m surface brightness and $A_{2}$ profiles obtained via 2-D (solid line) and 1-D (dashed line) stacking of early- ($T<5$, blue) and late-type ($T\ge5$, red) S$^4$G barred galaxies. 
From the 2-D bar stacks we show the azimuthally averaged $\mu_{3.6\mu \rm m}(AB)$ profile. 
Bars in early-type systems are more centrally concentrated and present larger density amplitudes than their late-type counterparts.
}
\label{1d2dcomp}
\end{figure}

In Fig.~\ref{1d2dcomp} we test the robustness of the 2-D and 1-D stacking methods by comparing their output, for the early- ($T<5$) and late-type ($T\ge5$) barred systems in our sample. 
As expected, whether we stack the bars in 2-D and obtain the azimuthally averaged surface brightness or  co-add the 1-D $I_{0}$ profiles of individual galaxies rescaled to $r_{\rm bar}$, the result is the same (upper panel). 
However, each of the methods offers different possibilities. Two-dimensional stacking allows the  mean bars to be calculated,  thus providing   a non-parametric characterization of the bar shape,  which may be useful for comparison with $N-$body models. 
By averaging 1-D stellar density profiles, we can directly obtain the dispersion of the radial profiles for a given subsample. In addition, 1-D stacking allows the mean $V_{\rm 3.6\mu \rm m}$ to be calculated directly. 
We also compare the mean $A_{2}$ profiles derived from the 2-D bar stacks with those obtained from 1-D stacking of the individual $A_{2}$ profiles (lower panel of Fig.~\ref{1d2dcomp}). 
The differences in the mean $A_2$ are fairly small at their maximum value ($\sim 5\%$). The 1-D stacking produces slightly larger values in the central parts  owing to non-axisymmetric inner structures (e.g. double barred galaxies). 
Beyond the bar region, larger $A_{2}$ amplitudes are mostly the result  of spiral modes. Inner and outer non-axisymmetric structures fade away during the 2-D stacking.
%
%
%
%
%
%
%
\section{One-dimensional characterization of stellar disks}\label{disk-char}
%
%
In this section we describe the 1-D stellar density profiles of the disk stacks after binning the sample based on the total stellar mass of the host galaxies and their morphological type. 
To this end, we separate S0s ($-3 \le T < 0$), early-type spirals ($0 \le T < 3$), intermediate-type spirals ($3 \le T < 5$), late-type spirals ($5 \le T \le 7$), and irregulars and Magellanics ($7 < T \le 10$). 
Occasionally, we also study early- ($T<5$) and late-type ($T\ge5$) systems separately. 
In addition, we characterize the average stellar contribution to the circular velocity of these synthetic disks. 
%
%
\begin{figure}
\centering
\begin{tabular}{c c c c}
   \includegraphics[width=0.416\textwidth]{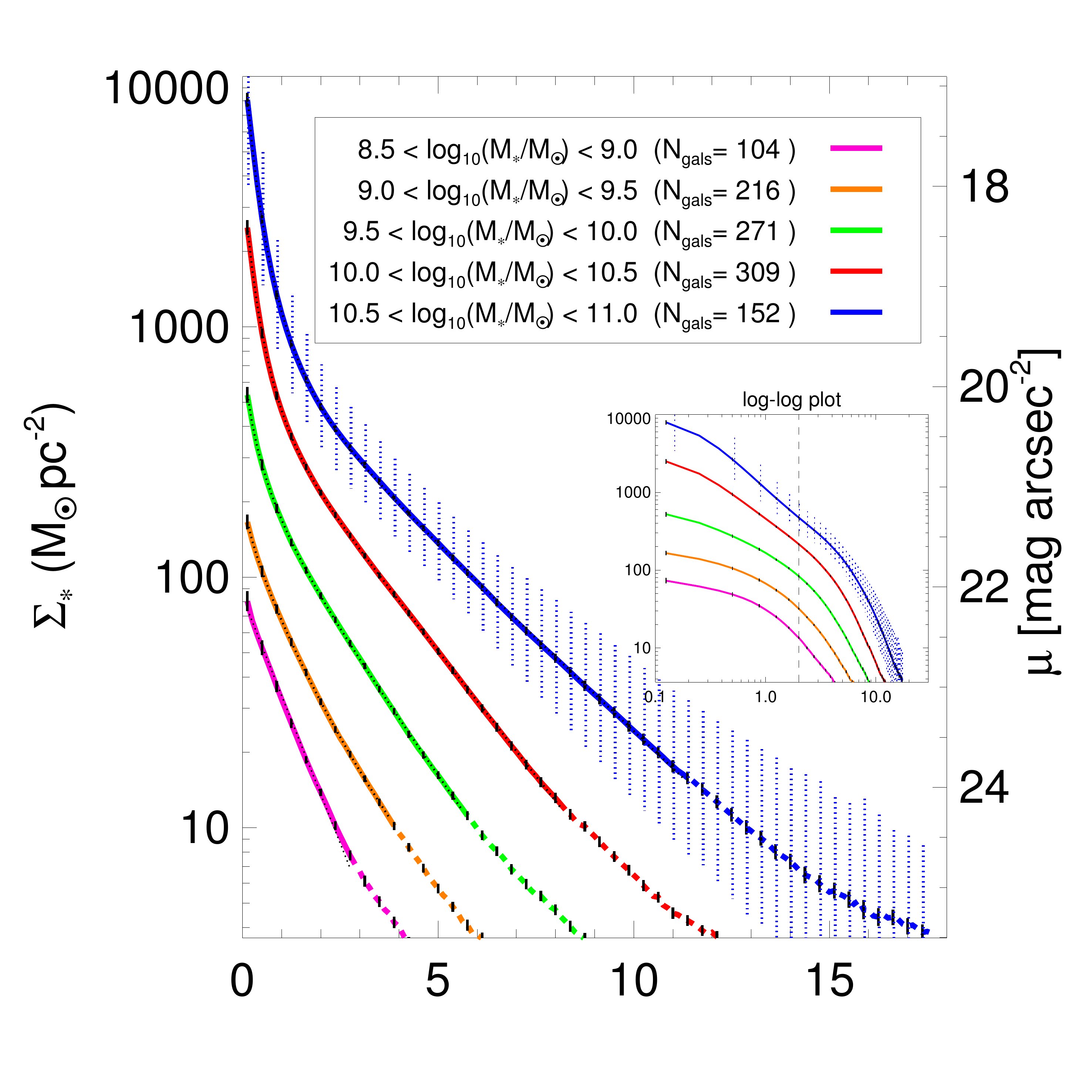}\\[-7.5ex]
   \includegraphics[width=0.416\textwidth]{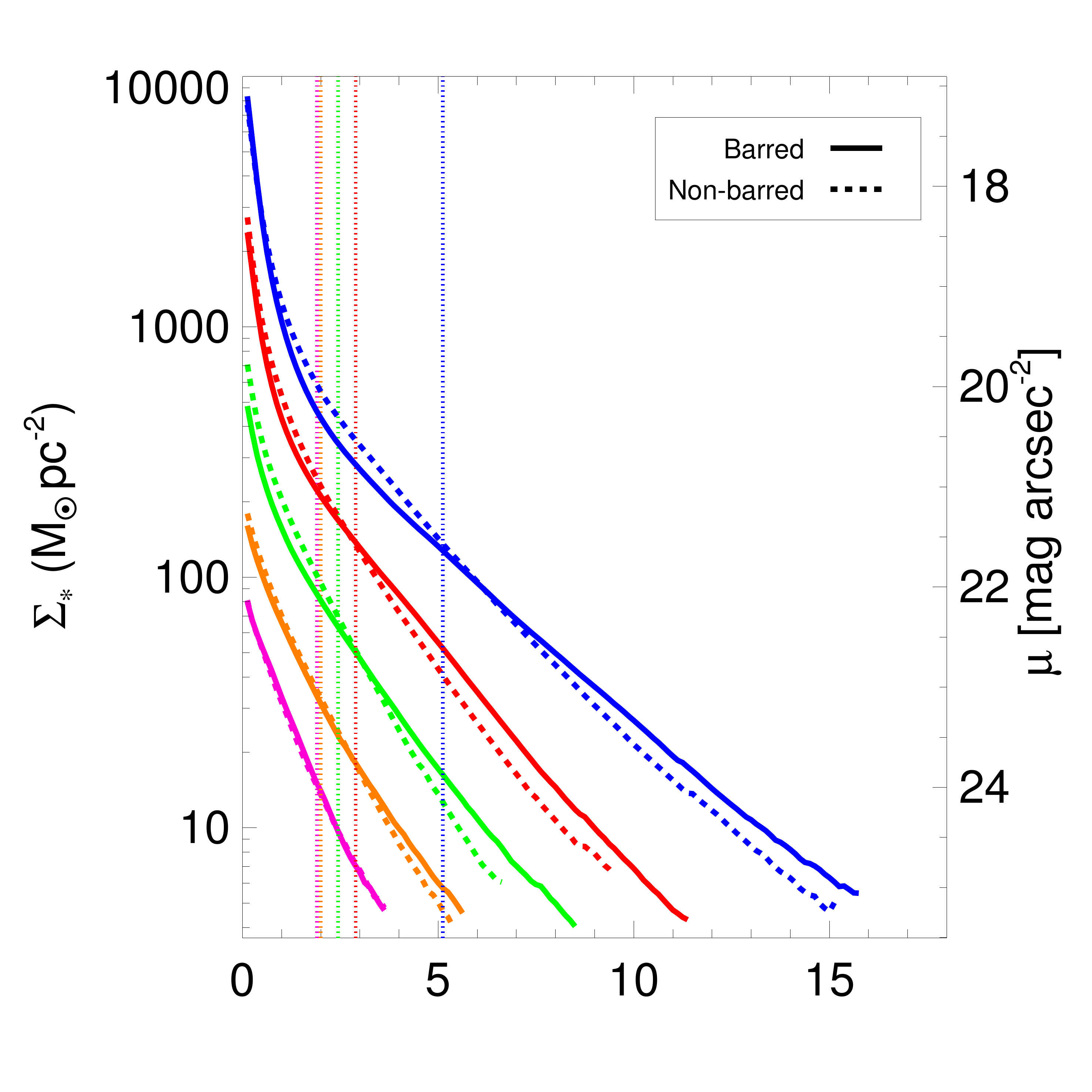}\\[-7.5ex]
   \includegraphics[width=0.416\textwidth]{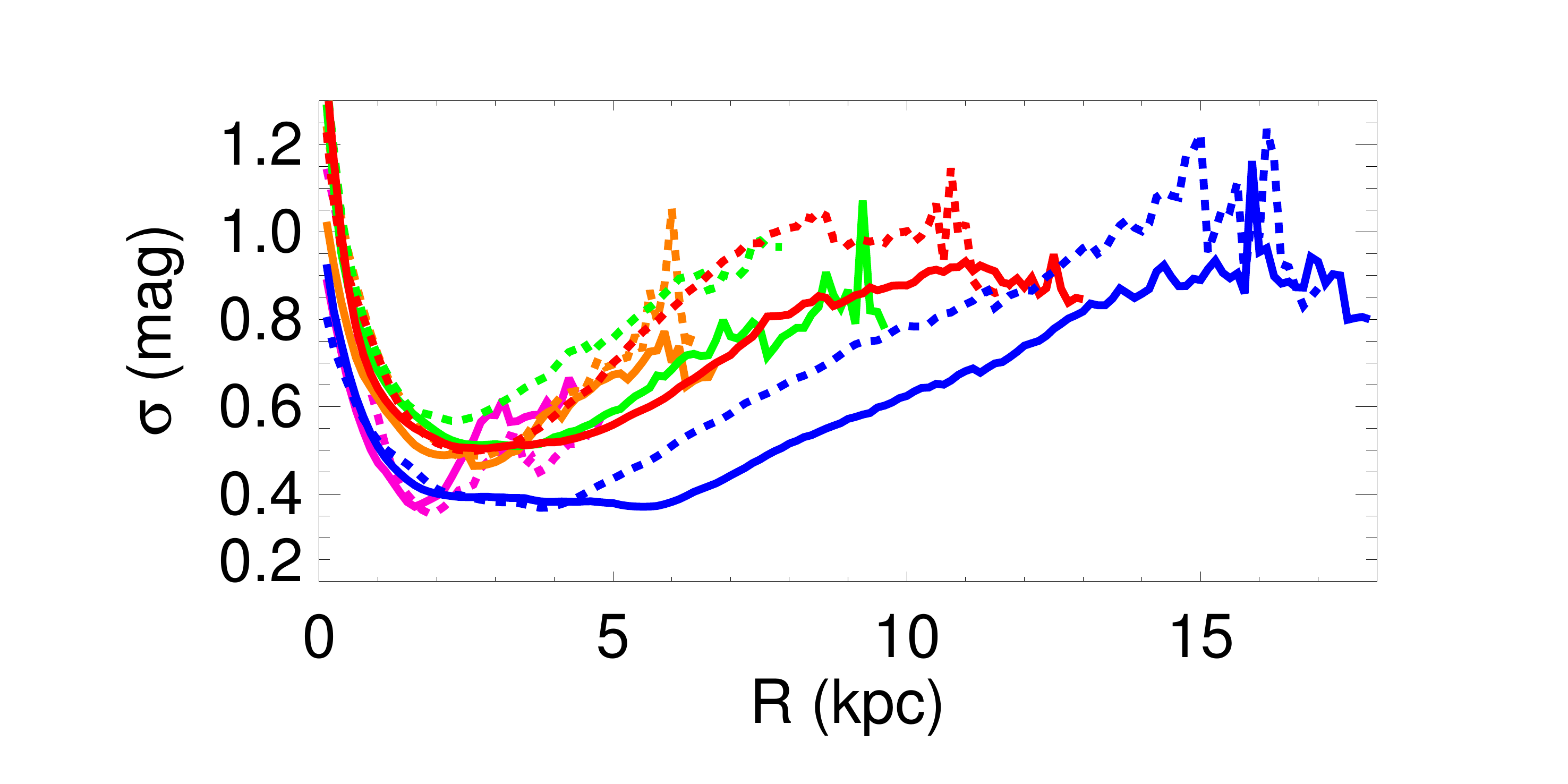}\\[-1.0ex]
\end{tabular}
\caption{
\emph{Upper panel:} Average stellar mass density profiles obtained by binning the sample in total stellar mass, rescaling the 3.6~$\mu$m $I_{0}$ profiles in physical units and stacking them 
(the inset shows the same in a log-log scale, the $R=2$ kpc distance highlighted with a dashed line). The solid lines trace $\Sigma_{\ast}$ within the radial range with a $100\%$ sample coverage. 
The dashed lines show $\Sigma_{\ast}$ when the sample coverage is let to be greater than $75\%$. 
The dotted black lines correspond to the two-component Sérsic+exponential fit to the mean disks in the range with a $100\%$ sample completeness level (see the text). 
The resulting disk parameters are listed in Table~\ref{disk_fits_params}. 
The small vertical lines along the profiles indicate the standard deviation of the mean ($\sigma/\sqrt{N_{\rm gals}}$). For the bin with the largest $M_{\ast}$, 
the vertical dotted lines indicate the sample dispersion ($\sigma$). 
\emph{Central panel:} As in the upper panel, but separating barred (solid line) and non-barred (dashed line) systems ($90\%$ coverage). 
The vertical dotted lines indicate the mean bar size of the barred galaxies in each of the $M_{\ast}$-bins. 
\emph{Lower panel}: Statistical dispersion of the rescaled luminosity profiles among the galaxies in each of the $M_{\ast}$-bins used in the upper panel. Barred (solid lines) and non-barred (dashed lines) galaxies are studied separately. 
}
\label{stack_kpc_disk_mass}
\end{figure}
%
%
\begin{figure}
\centering
\includegraphics[width=0.416\textwidth]{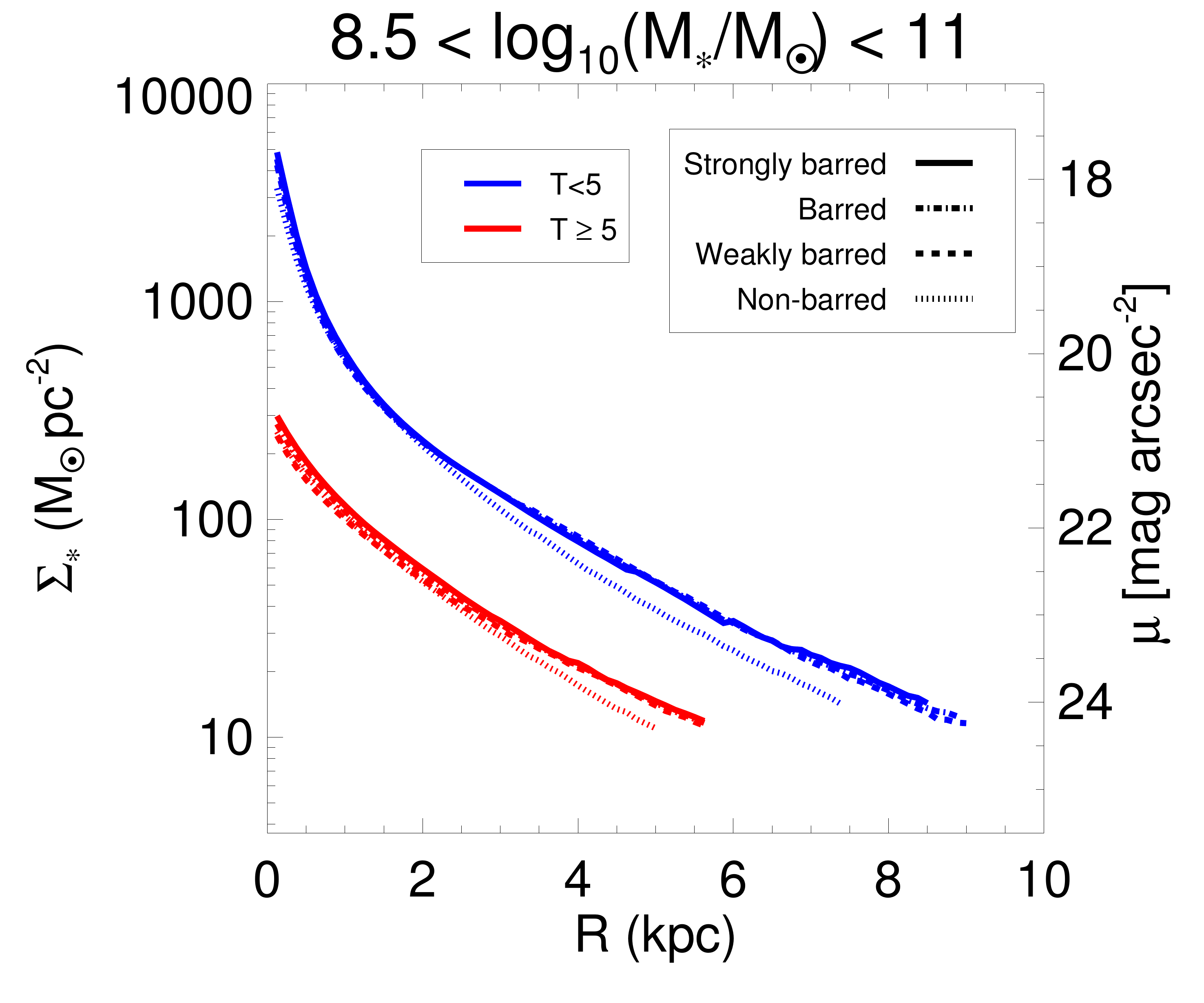}
\caption{Mean stellar mass density profiles for barred and non-barred disk galaxies (selected according to B2015 classification) 
and for weakly ($\rm S\underline{A}$B+SAB) and strongly ($\rm SA\underline{B}$+SB) barred galaxies with $10^{8.5}\lesssim M_{\ast}/M_{\odot}\lesssim10^{11}$ ($90\%$ sample coverage). 
Early-type ($T<5$) and late-type ($T\ge5$) systems are studied separately. 
}
\label{stack_kpc_disk_ttype}
\end{figure}
%
%
\subsection{Mean density profiles in physical units}\label{disk_kpc}
%
%
In the top panel of Fig.~\ref{stack_kpc_disk_mass} we show the mean $\Sigma_{\ast}$ obtained by binning the sample in $M_{\ast}$ and rescaling the density profiles to common frame in physical units.
It is clear that the central density and scalelength (slope) of the mean $\Sigma_{\ast}$ increase with increasing $M_{\ast}$. 
In the inner parts the contribution of the bulge makes the profiles steeper for the more massive galaxies. 
Among the massive systems, we checked that all the trends in Fig.~\ref{stack_kpc_disk_mass} are practically the same even when S0 galaxies are excluded.

In order to parameterize the mean profiles in physical units, we performed 1-D chi-square minimization fitting to $\mu_{3.6\mu \rm m}$ in units of magnitudes (dotted lines in the upper panel of Fig.~\ref{stack_kpc_disk_mass}). 
We assigned the same weight to all the data points and used two components which were modelled with exponential and Sérsic profiles:
\begin{equation}
\Sigma_{\ast}(r)=\Sigma_{\rm b\circ}{\rm exp}(-(r/R_{\rm b})^{1/n})+\Sigma_{\circ}{\rm exp}(-r/h_{\rm R}),
\end{equation}
where $\Sigma_{\circ}$ and $h_{\rm R}$ are the disk extrapolated central stellar density and scalelength, respectively, while $\Sigma_{\rm b\circ}$ and $R_{\rm b}$ refer to the central density and scale parameter of the Sérsic component, 
and $n$ is the Sérsic index (shape parameter). For the bin with $M_{\ast}< 10^{9}M_{\odot}$ we use a one-component exponential fit. 
The resulting parameters from the decomposition of all the $\Sigma_{\ast}$ profiles are listed in Table~\ref{disk_fits_params}. 
By giving these decomposition quantities, we only aim at parameterizing the mean profiles.
We also calculated the dispersion ($\sigma$) of the luminosity profiles at every radius (bottom panel of Fig.~\ref{stack_kpc_disk_mass}), which is $\lesssim 1$~mag on average. 
Interestingly, the scatter is larger for non-barred galaxies than for  the barred systems when $M_{\ast}\ge10^{9.5}M_{\odot}$.

%
%
As shown in the central panel of Fig.~\ref{stack_kpc_disk_mass}, there is a significant difference between the mean surface brightness of barred and non-barred galaxies for systems with total stellar masses $M_{\ast}\ge10^9M_{\odot}$. 
For the lowest mass galaxies the difference is not found. In the outer disk, barred galaxies show typically brighter $\Sigma_{\ast}$ profiles. 
On the contrary, in the inner parts non-barred galaxies are on average denser at a given total stellar mass. 
In addition, for a given stellar mass interval, we find that the intersecting point of the mean $\Sigma_{\ast}$ profiles of barred and non-barred galaxies is located very close to the mean bar size. 
In Fig.~\ref{stack_kpc_disk_ttype} we checked that this trend is maintained also when the samples are split into early-type ($T<5$) and late-type ($T\ge5$) systems. 
Furthermore, we do not find any significant difference in the mean $\Sigma_{\ast}$ between strongly and weakly barred galaxies.
%
%
\subsection{Mean density profiles normalized by $h_{\rm R}$}\label{ttypesprofs}
%
%
\begin{figure*}
\centering   
\begin{tabular}{c c c}
      \includegraphics[width=0.49\textwidth]{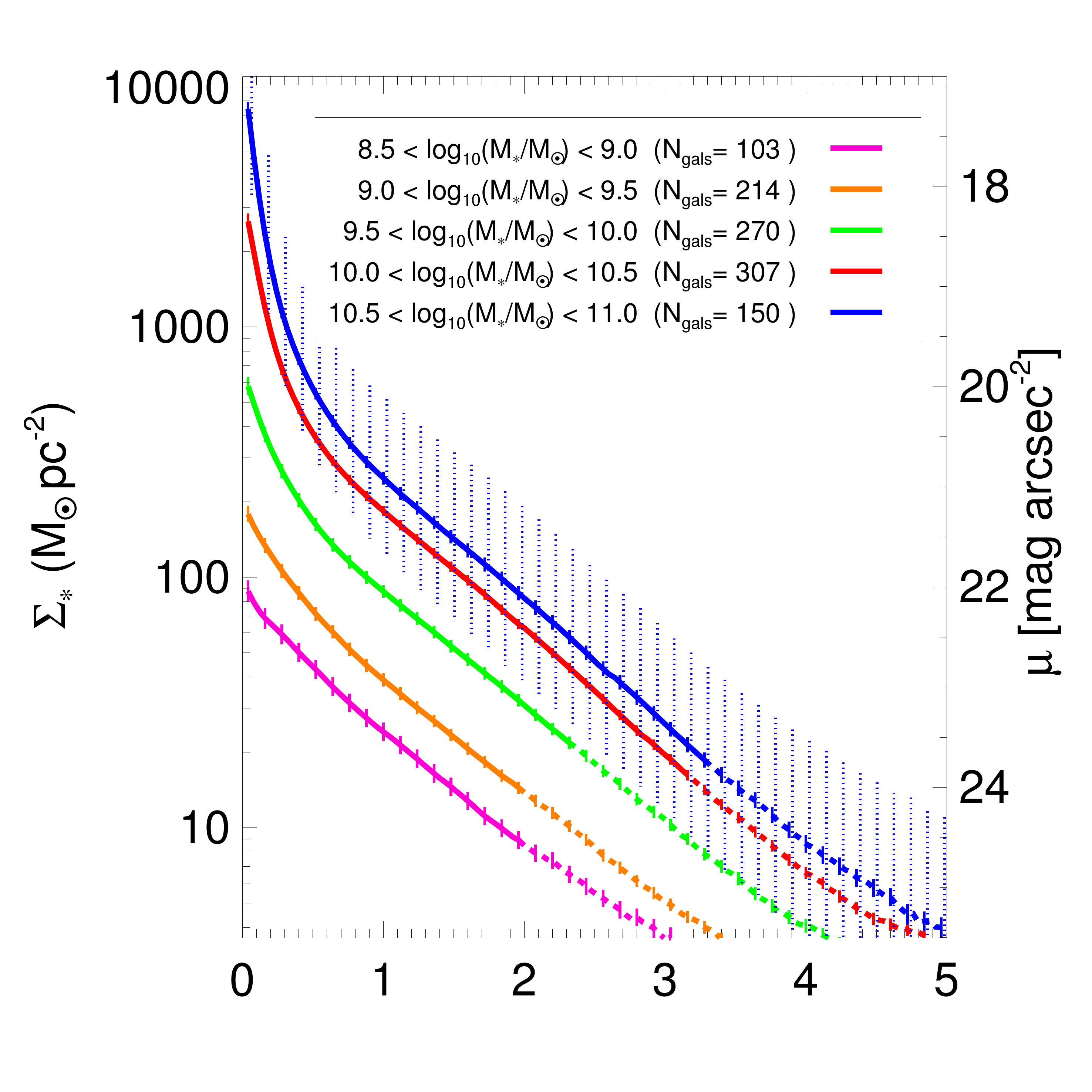}
      \includegraphics[width=0.49\textwidth]{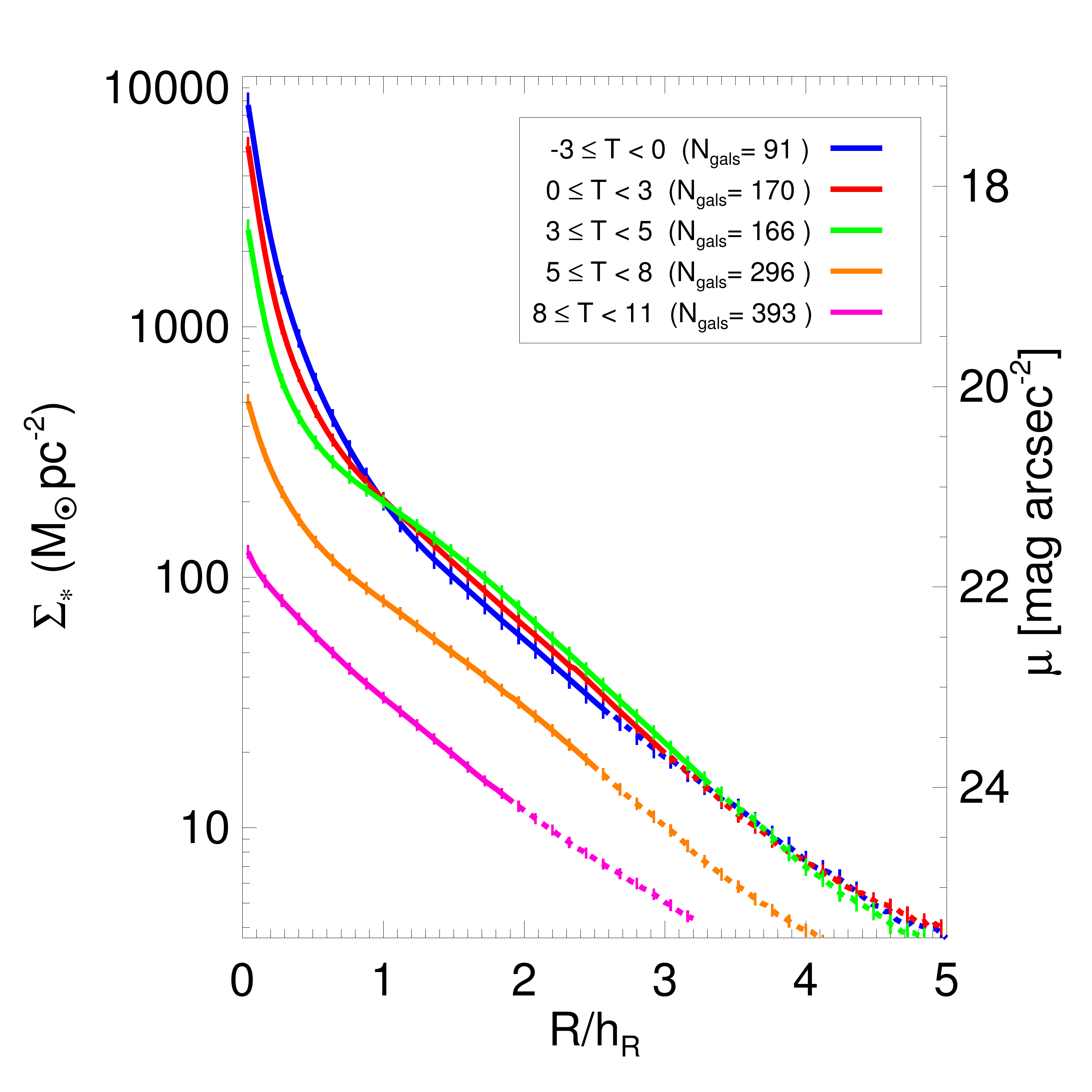}\\[-5ex]
      \includegraphics[width=0.49\textwidth]{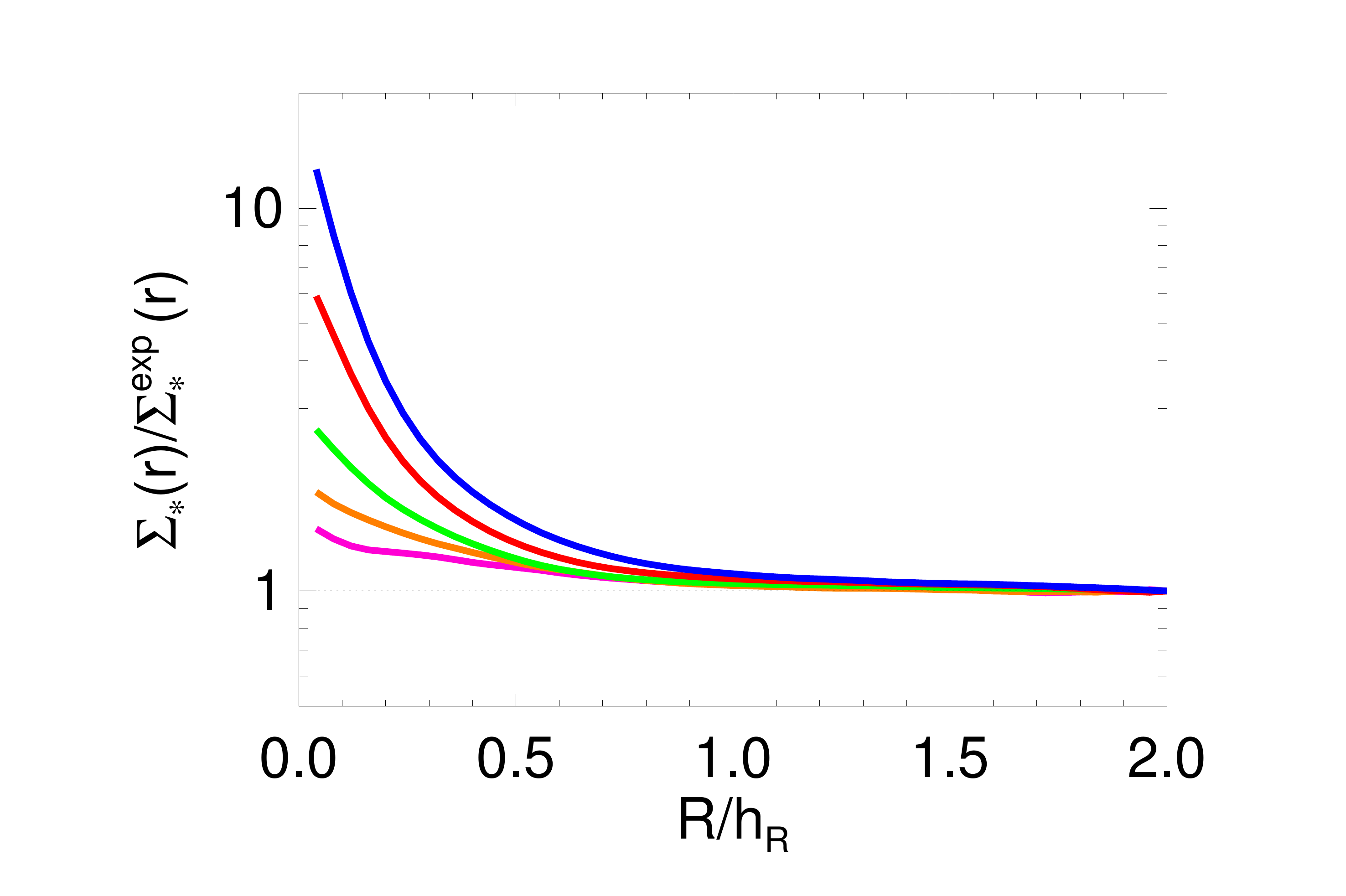}
      \includegraphics[width=0.49\textwidth]{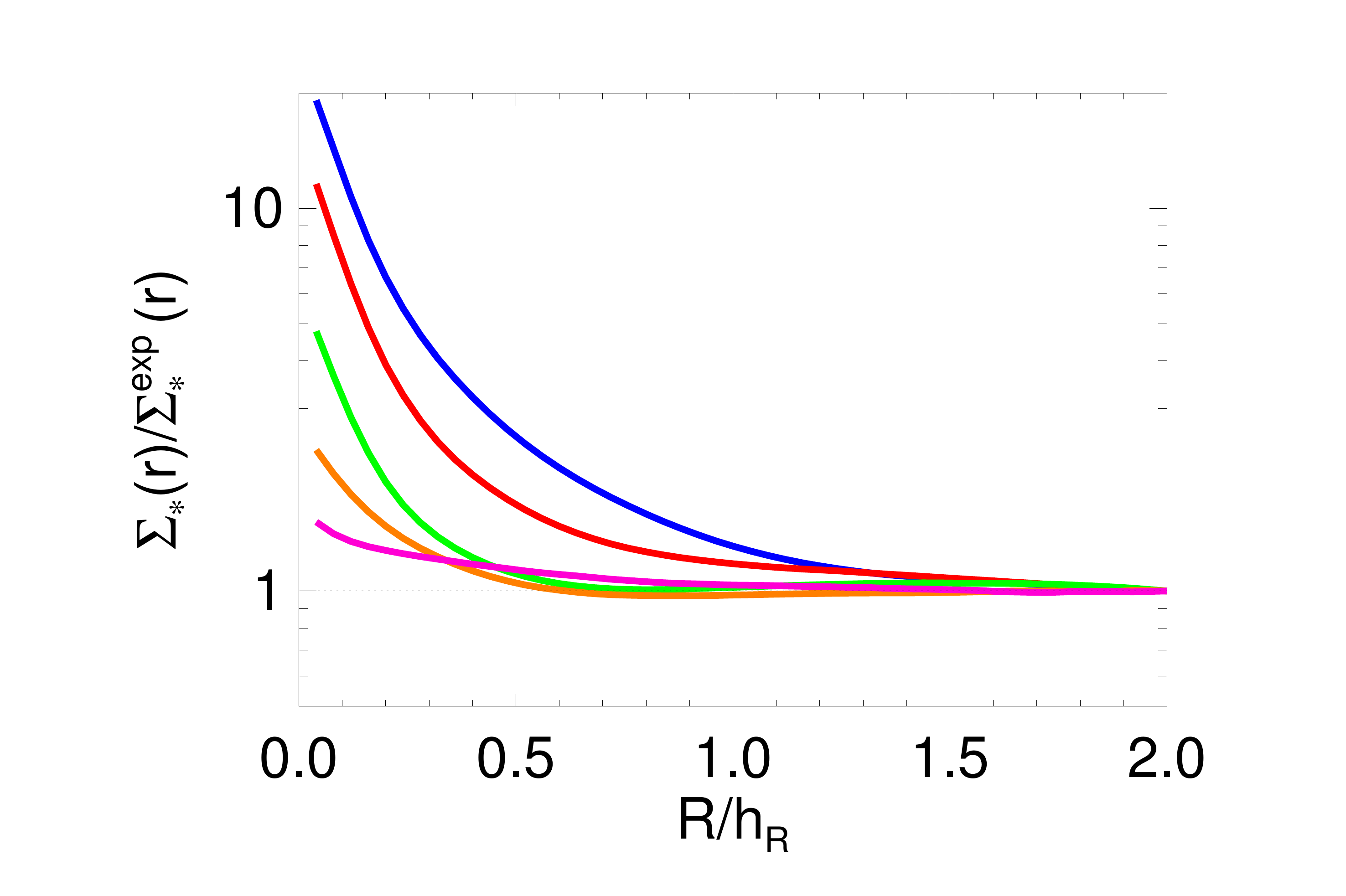}
\end{tabular}
\caption{
\emph{Left column:} 
Mean $\Sigma_{\ast}$ derived by stacking 3.6~$\mu$m 1-D density profiles (S$^{4}$G galaxies binned in total stellar mass) resized to the disk scalelength (\emph{upper panel}), 
and deviation from an exponential disk (normalized to fit $\Sigma_{\ast}$ at $2 \cdot h_{\rm R}$) of the mean $\Sigma_{\ast}$ within the central regions ($r \le 2h_{\rm R}$) (\emph{lower panel}). 
\emph{Right column:} 
As in the left column, but taking subsamples in terms of the revised Hubble type of the galaxies. 
For the upper panels, the different line styles indicate the $75\%$ and $100\%$ sample completeness level in the same fashion as in Fig.~\ref{stack_kpc_disk_mass}, 
and the vertical lines correspond to the standard error of the mean. In the upper left panel, for the bin with largest $M_{\ast}$, we indicate the sample dispersion with vertical dotted lines.}
\label{stack_1}
\end{figure*}
%
%
\begin{table*}
\centering
\tiny
\begin{tabular}{|| c|| c| c| c| c| c||}
\hline
\multicolumn{1}{||c}{\centering Disk parameters} & \multicolumn{1}{|c}{\centering $8.5<\rm log_{10}M_{\ast}/M_{\odot}<9$} & \multicolumn{1}{|c}{\centering $9<\rm log_{10}M_{\ast}/M_{\odot}<9.5$} &
\multicolumn{1}{|c}{\centering $9.5<\rm log_{10}M_{\ast}/M_{\odot}<10$} & \multicolumn{1}{|c||}{\centering $10<\rm log_{10}M_{\ast}/M_{\odot}<10.5$} & \multicolumn{1}{|c||}{\centering $10.5<\rm log_{10}M_{\ast}/M_{\odot}<11$} \tabularnewline
\hline
\hline
$\Sigma_{\circ}$ ($M_{\odot}$ pc$^{-2}$)&  83.5&  51.1& 192.6& 500.3& 733.9 \\
$\mu_{\circ}$ (mag arcsec$^{-2}$)&22.10 (0.94)&22.63 (0.91)&21.19 (1.04)&20.15 (1.01)&19.74 (1.09) \\
$h_{\rm R}$ (kpc)& 1.102 (0.689)& 1.146 (0.870)& 1.975 (1.073)& 2.182 (1.184)& 2.934 (1.496) \\
\hline
$\Sigma_{\rm b\circ}$ ($M_{\odot}$ pc$^{-2}$)&...& 178.6& 827.0&  4674.2& 24011.7 \\
$\mu_{\rm b\circ}$ (mag arcsec$^{-2}$)&...&21.27&19.61&17.73&15.95 \\
$R_{\rm b}$ (kpc)&...&0.597&0.171&0.167&0.101 \\
$n$&...&1.681&1.752&1.444&1.773 \\
\hline
\end{tabular}
\caption{
The 1D Sérsic+exponential fit (bulge+disk) parameterization is listed for the different mean profiles binned in terms of $M_{\ast}$ (Fig.~\ref{stack_kpc_disk_mass}). 
For both components we present the scalelength and central surface brightness and stellar mass density. For the bulge, the Sérsic index is also given. 
In parentheses, we show the dispersion of outer disk parameters based on the multicomponent decompositions of the corresponding individual galaxies from \citet[][]{2015ApJS..219....4S}.
}
\label{disk_fits_params}
\end{table*}

%
%
In Fig.~\ref{stack_1} we show the mean radial $\Sigma_{\ast}$ derived by stacking the 1-D density profiles after resizing them with respect to the disk scalelength, grouping the galaxies based on their total stellar mass and Hubble type. 
Unlike in Sect.~\ref{disk_kpc}, the resulting disks show a common slope, which is expected because of the used scaling and by definition of $h_{\rm R}$. 
This means that a single exponential scalelength (as fitted in P4) describes fairly well the mean stellar disk surface brightness outside the central parts of the galaxies down to at least $\sim24.5$ mag arcsec$^{-2}$.

Furthermore, galaxies with total stellar masses $\ge10^{10}M_{\odot}$ show prominent central components in their profiles ($R \lessapprox h_{\rm R}$): 
the central density is $\sim 6-10$ times larger than the extrapolated disk central density. 
To make this clearer, in the lower panels of Fig.~\ref{stack_1} we show the deviation from an exponential slope (fitted at $2 \cdot h_{\rm R}$) of the mean profile (normalized to $h_{\rm R}$), focusing on the central regions. 
Galaxies of intermediate stellar masses ($10^{9}M_{\odot}<M_{\ast}<10^{10}M_{\odot}$) show less pronounced central mass concentrations: 
the central density is $\sim 2-3$ times larger than that extrapolated from the exponential disk. 
For the bin with the faintest systems ($M_{\ast}<10^{9}M_{\odot}$), the mean $\Sigma_{\ast}$ shows hardly any deviation from the exponential disk at all radii. 

When binning the sample as a function of $T$ (see the right column in Fig.~\ref{stack_1}), we show that the disks in galaxies with morphological types $T<5$ show similar density profiles 
(the average total stellar mass of these galaxies in the S$^4$G is roughly the same for all morphological type bins earlier than Sc: $\sim 2\cdot 10^{10}M_{\odot}$). 
They differ in the shape and prominence of their central mass, which increases with decreasing T. 
This is expected because the bulge-to-total ratio is one criterion to determine the Hubble type of a galaxy \citep[][]{1994cag..book.....S}. 
A more thorough analysis of the disks reveal that intermediate-type spirals ($3 \le T < 5 $) show flatter profiles in the inner regions, 
probably owing to the effect of bars (typically $\sim 1 h_{\rm R}$), and they decline exponentially afterwards.

\begin{figure}
\centering
   \includegraphics[width=0.5\textwidth]{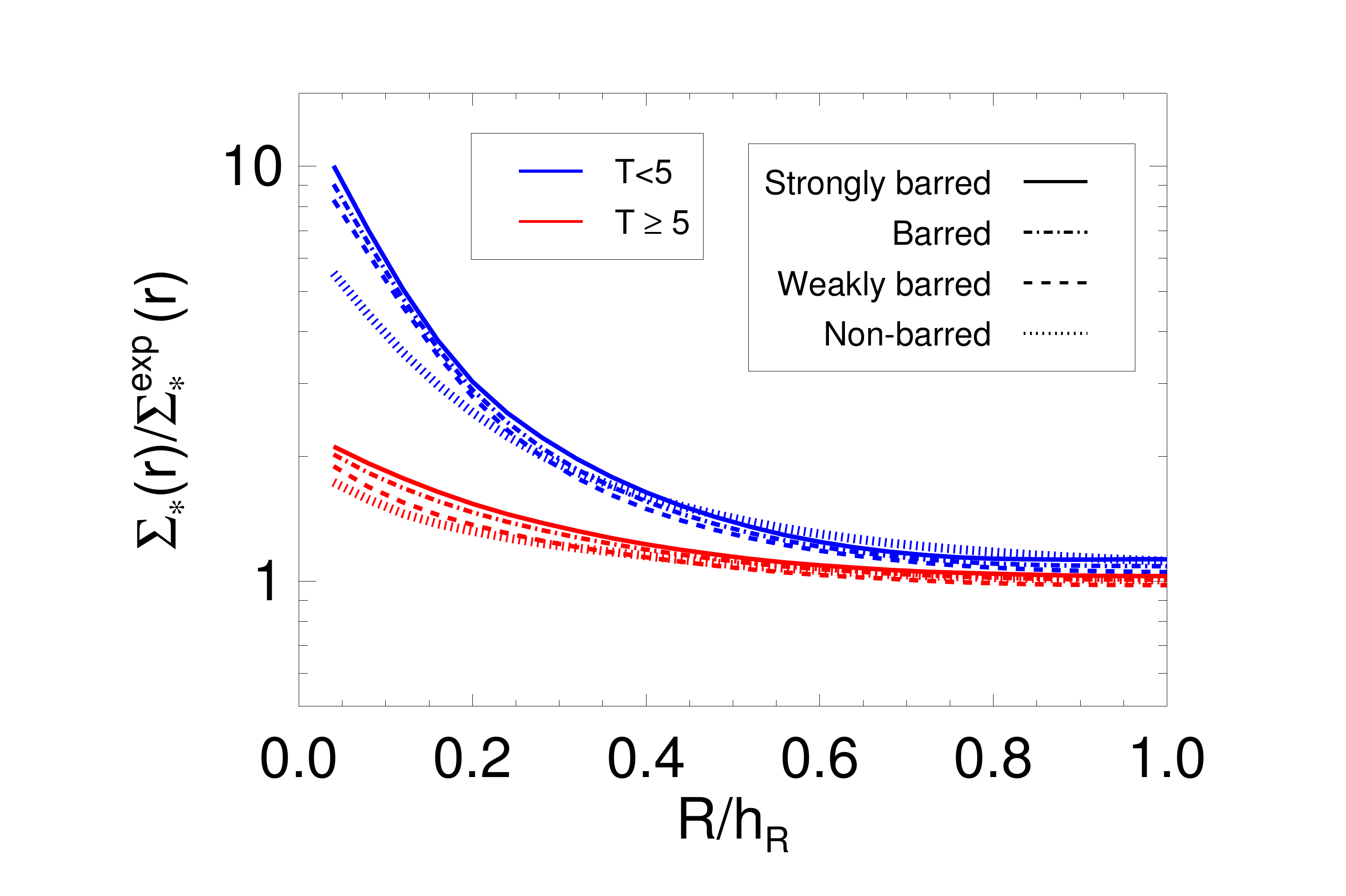}
\caption{
As in the lower panels of Fig.~\ref{stack_1}, but binning the sample into early- ($T<5$) and late-type ($T\ge5$) systems and separating barred and non-barred galaxies. 
Weakly (S$\underline{A}$B+SAB) and strongly (S$A\underline{B}$+SB) barred galaxies with $10^{8.5}\lesssim M_{\ast}/M_{\odot}\lesssim10^{11}$ are also studied independently.
}
\label{stack_families_inner}
\end{figure}

In Fig.~\ref{stack_families_inner} we study the central mass concentration of the disk stacks corresponding to barred and non-barred galaxies, splitting the sample into early- ($T<5$) and late-type ($T\ge5$) systems. 
We find that the central mass concentration of barred galaxies is on average larger than in their non-barred counterparts. 
Also, strongly barred galaxies are somewhat more centrally concentrated than weakly barred galaxies. 
This contrast is particularly clear (almost a factor~2 difference between barred and non-barred galaxies) for morphological types earlier than Sc. 
We checked that these trends were maintained when restricting the samples to non-active systems \citep[$\sim 92\%$ in our sample, based on the catalogue by][]{2010A&A...518A..10V}. 
This confirmed the lack of any bias (for $T<5$) due to the non-stellar emission at 3.6~$\mu$m associated with hot dust heated by active galactic nuclei (AGN) \citep[e.g.][]{2012ApJ...744...17M} 
since several studies have found AGNs to be more frequent among (early-type) barred galaxies \citep[e.g.][]{2000ApJ...529...93K,2002ApJ...567...97L,2004ApJ...607..103L,2009ASPC..419..402H} \citep[but see][]{2012ApJ...750..141L}. 
%
%
%
%
\begin{figure}
\centering   
\begin{tabular}{c c}
       \includegraphics[width=0.229\textwidth]{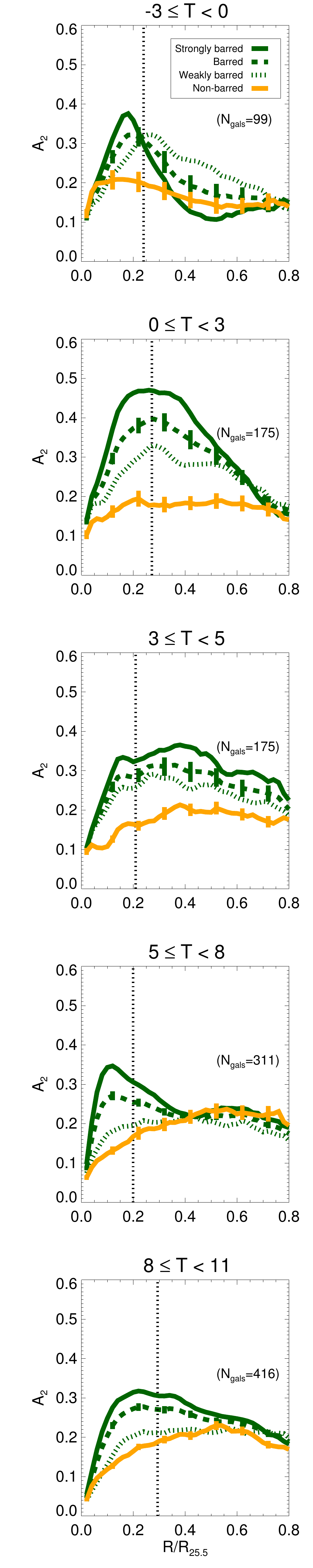}     
       \includegraphics[width=0.229\textwidth]{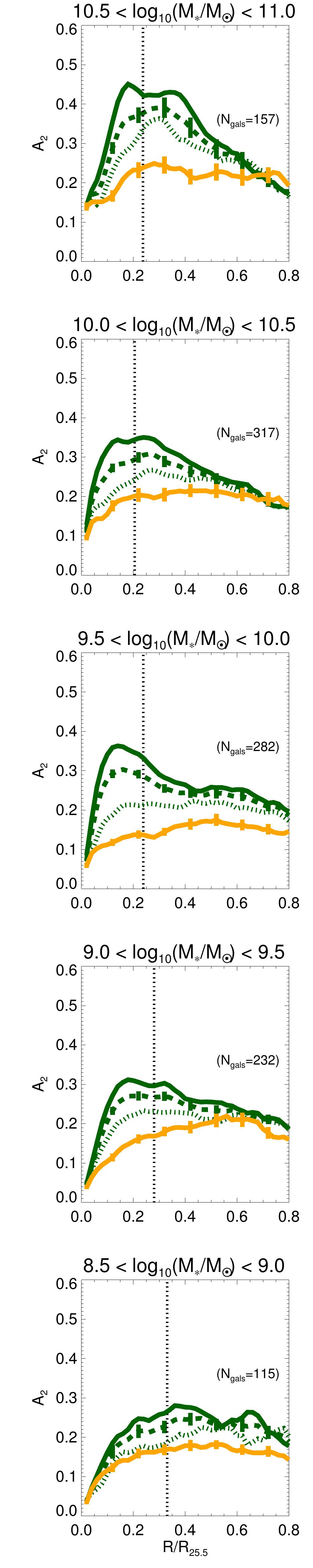}
\end{tabular}
\caption{
Mean $A_2$ radial profiles as a function the morphological type (left column) and the stellar mass (right column), computed from the Fourier amplitudes of individual galaxies rescaled to a common frame determined by $R_{25.5}$. 
Barred (green dashed line) and non-barred (orange solid line) systems are studied separately, based on the morphological classification by B2015. 
We also divided the bin of barred galaxies into weakly ($\rm S\underline{A}$B+SAB, dotted line) and strongly ($\rm SA\underline{B}$+SB, solid line) barred galaxies. 
The vertical dotted lines indicate the mean bar size, relative to $R_{25.5}$, of the galaxies in the bin. Ten galaxies whose disks did not extend as far as 0.8$R_{25.5}$ were excluded from this analysis. 
The standard deviation of the mean for barred and non-barred systems is indicated with vertical lines.
}
\label{a2_prof_ind}
\end{figure}
%
%
\subsection{Mean $A_{2}$ profiles of barred and non-barred systems}\label{a2_medi}
%
%
In Fig.~\ref{a2_prof_ind} we show the mean $A_{2}$ radial profiles, in subsamples defined by $T$ and $M_{\ast}$, computed from the $A_{2}$ in individual galaxies (from DG2016) scaled to $R_{25.5}$. 
Barred and non-barred galaxies are studied in separate bins. The subsamples are also split into weak ($\rm S\underline{A}$B+SAB) and strong ($\rm SA\underline{B}$+SB) bars.

All barred galaxies show a similar pattern regardless of the family: $A_{2}$ monotonically grows until a maximum due to the bar is reached.
In some cases such maxima are reached slightly beyond the typical bar length as a consequence of the superposition of the bar and spiral modes.
After this peak, other maxima in the outer disk region are most likely due to the spiral arms. 

The amplitude of the $A_{2}$ profiles are greater (lower) for strong (weak) bars within and also outside the bar region of spiral galaxies and for all stellar masses. 
In non-barred galaxies, low-amplitude ($\lessapprox 0.2$) local maxima appear within the inner regions ($R\lessapprox0.2-0.3R_{25.5}$) in the $A_{2}$ profiles when $M_{\ast}>10^{10}M_{\odot}$ and $T<5$, 
which may be due to inner spiral modes or to visually undetected weak bars (bars in DG2016 had $\left< r_{\rm bar}\right>\approx (0.2-0.3) R_{25.5}$). 
When $0 \le T < 5$ or $M_{\ast}>10^{9.5}M_{\odot}$, the mean $A_{2}$ of non-barred galaxies is systematically lower than  the barred systems at all radii. 
%
%
\begin{figure*}
\centering
\begin{tabular}{c c c c}
     \includegraphics[width=0.5\textwidth]{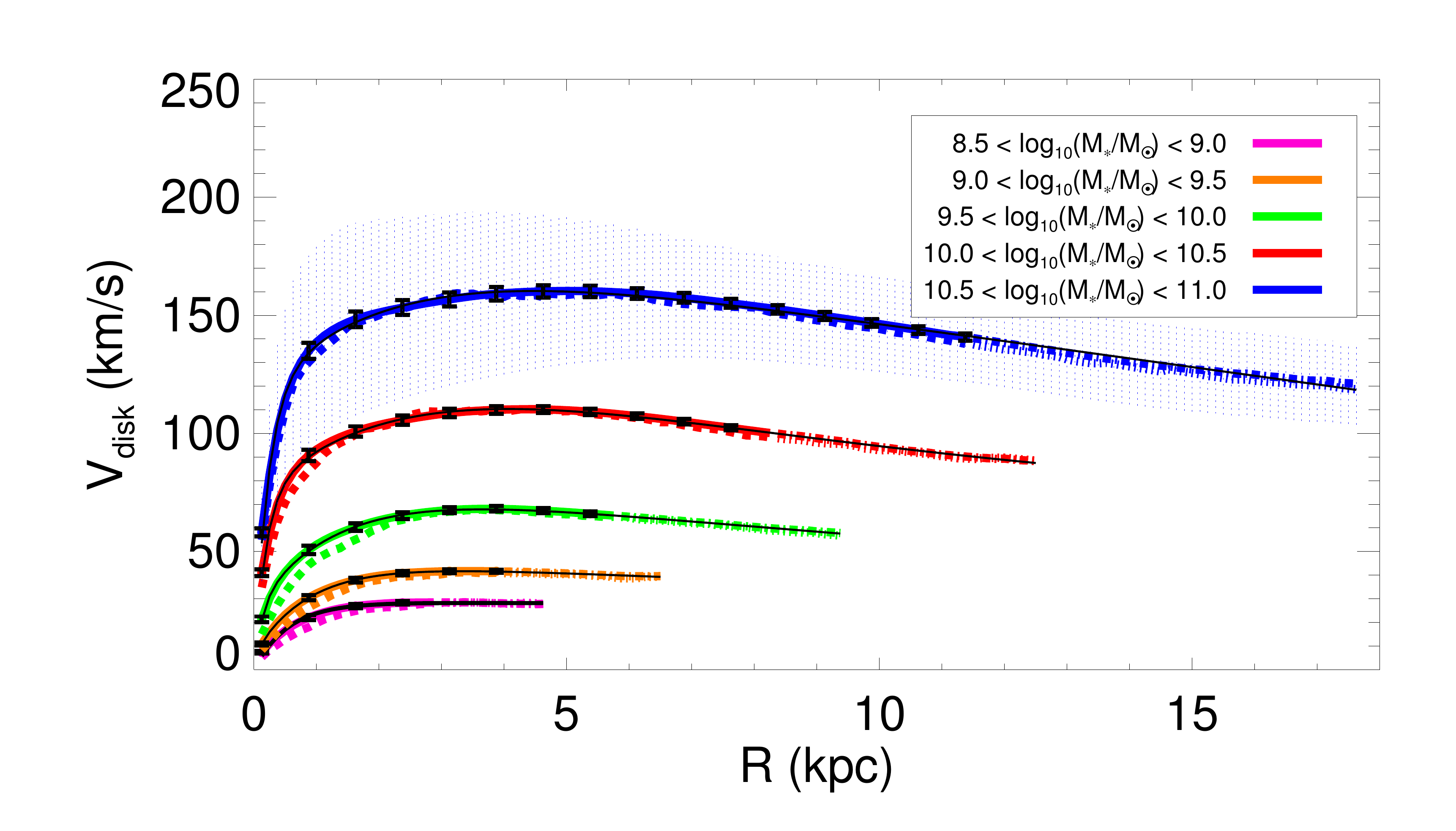}
     \includegraphics[width=0.5\textwidth]{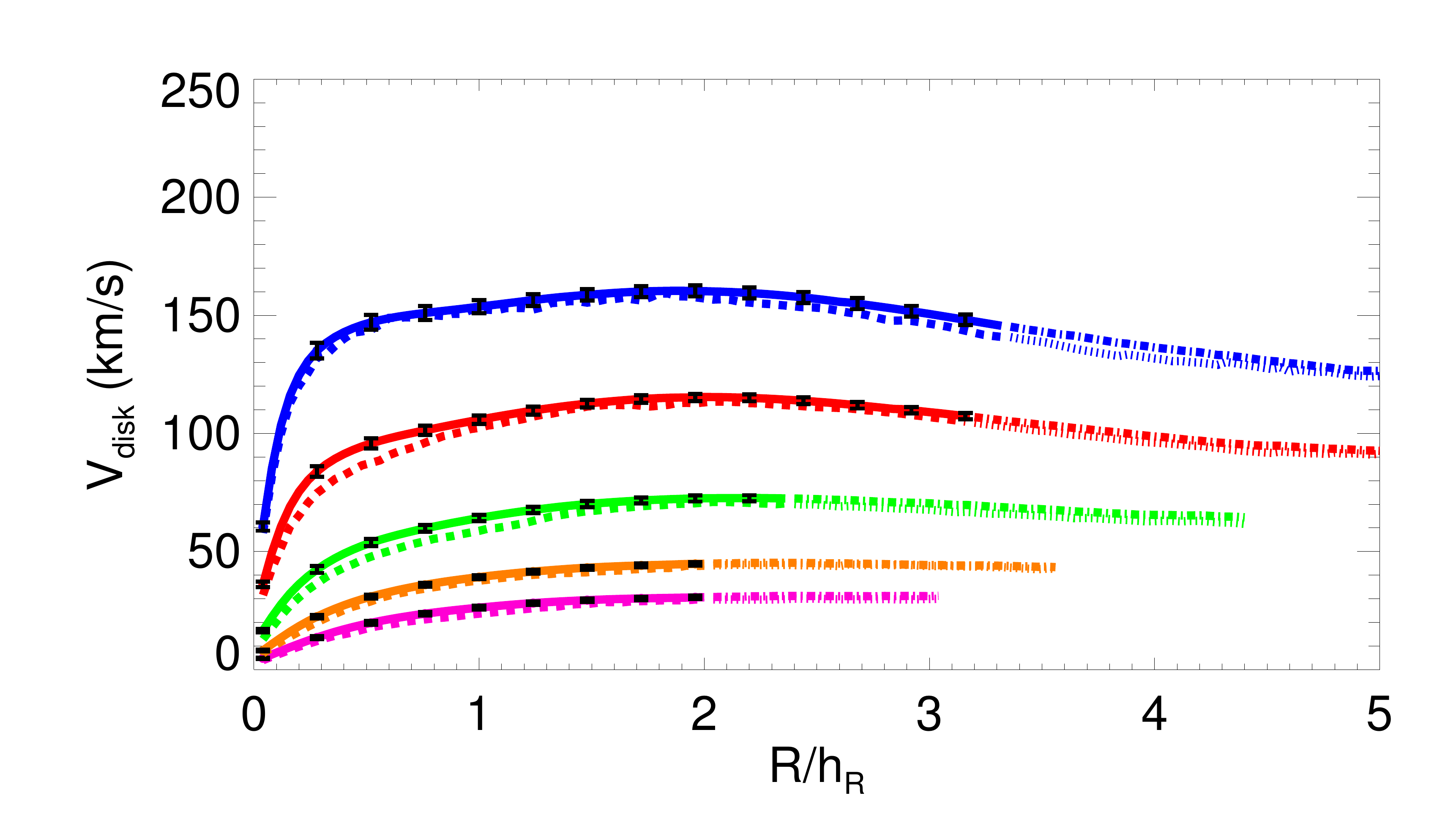}\\
     \includegraphics[width=0.5\textwidth]{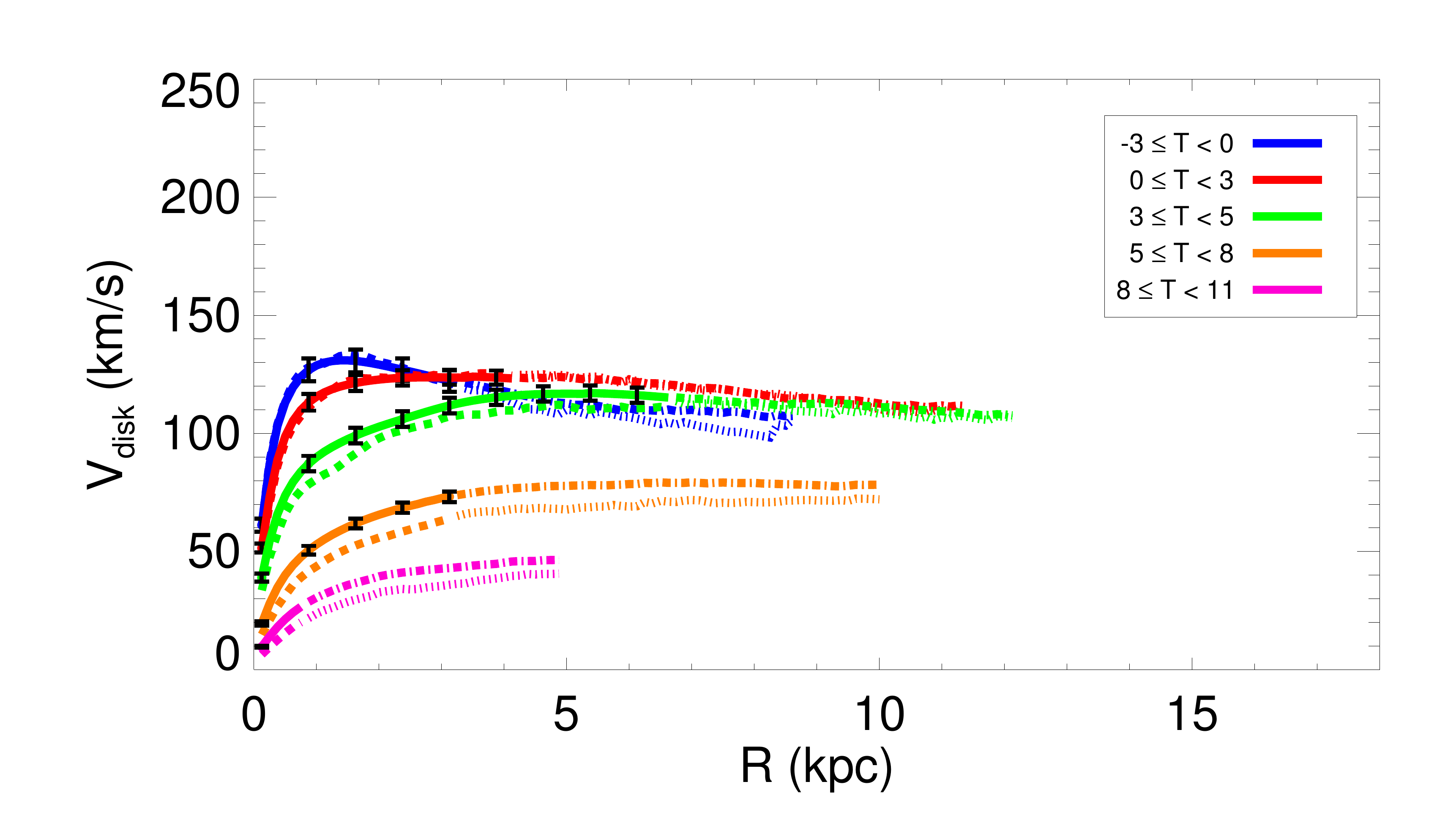}
     \includegraphics[width=0.5\textwidth]{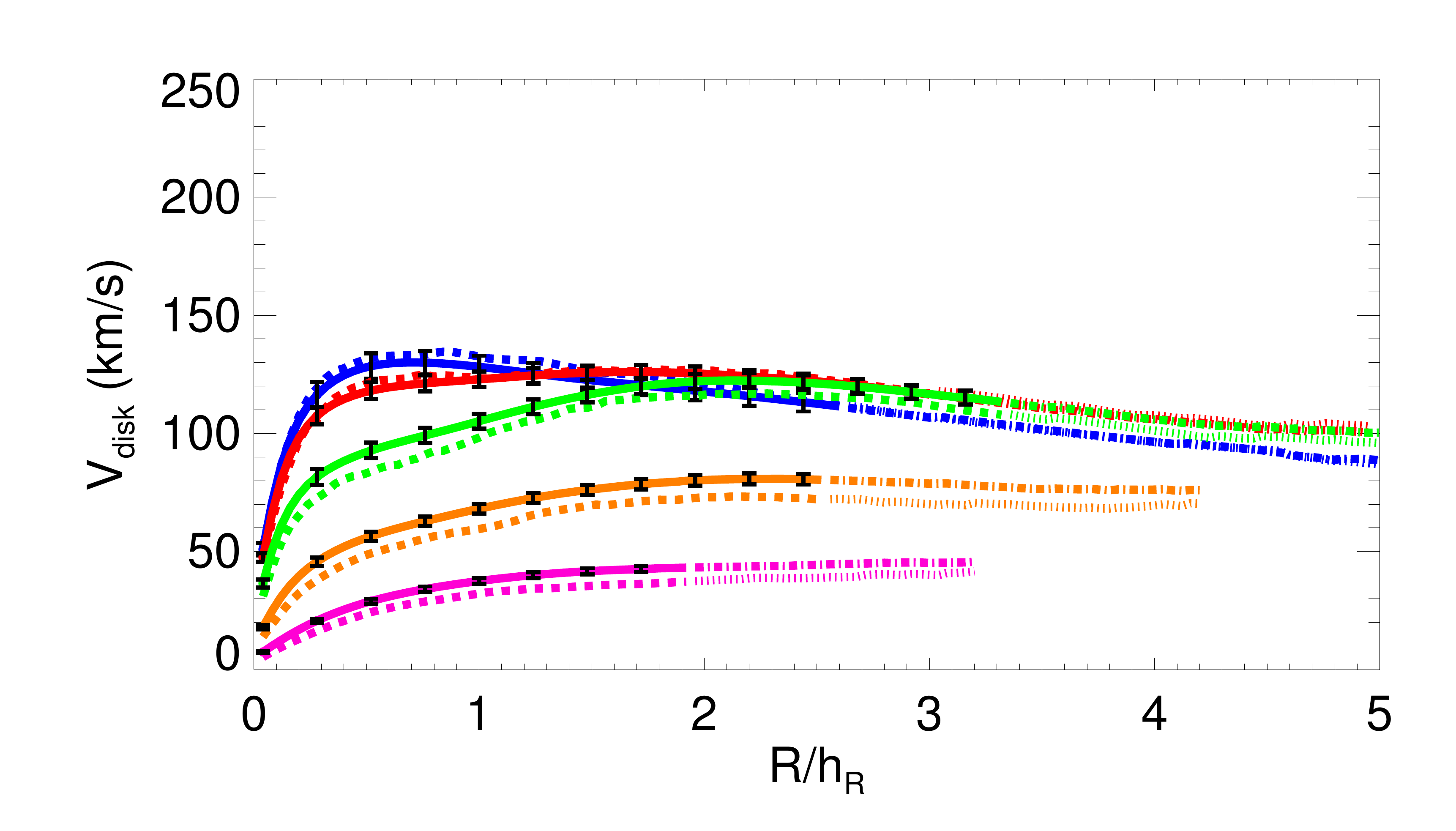}     
\end{tabular}
\caption{
Mean and median stellar contribution to the circular velocities curves (solid and dashed lines, respectively) inferred from the individual radial force profiles resized in physical units (left) and normalized to the disk scalelength (right), 
with the subsamples defined based on the total stellar mass (upper panels) and revised Hubble type (lower panels). 
The $100\%$ and $75\%$ sample coverage in the radial direction are indicated with different line styles. 
For the upper left panel, the black solid lines correspond to the chi-square minimization fit with a double-component {\it Polyex} function (see  text), with the resulting parameters listed in Table~\ref{rcur_fits_params}. 
The vertical error bars, centred in the mean profile, indicate the standard deviation of the mean. In the upper left panel, for the bin with largest $M_{\ast}$ we also show the sample dispersion with dotted lines.
}
\label{stack_vdisk_comp}
\end{figure*}
%
%
\begin{figure*}
\centering   
\begin{tabular}{c c c c}
     \includegraphics[width=0.5\textwidth]{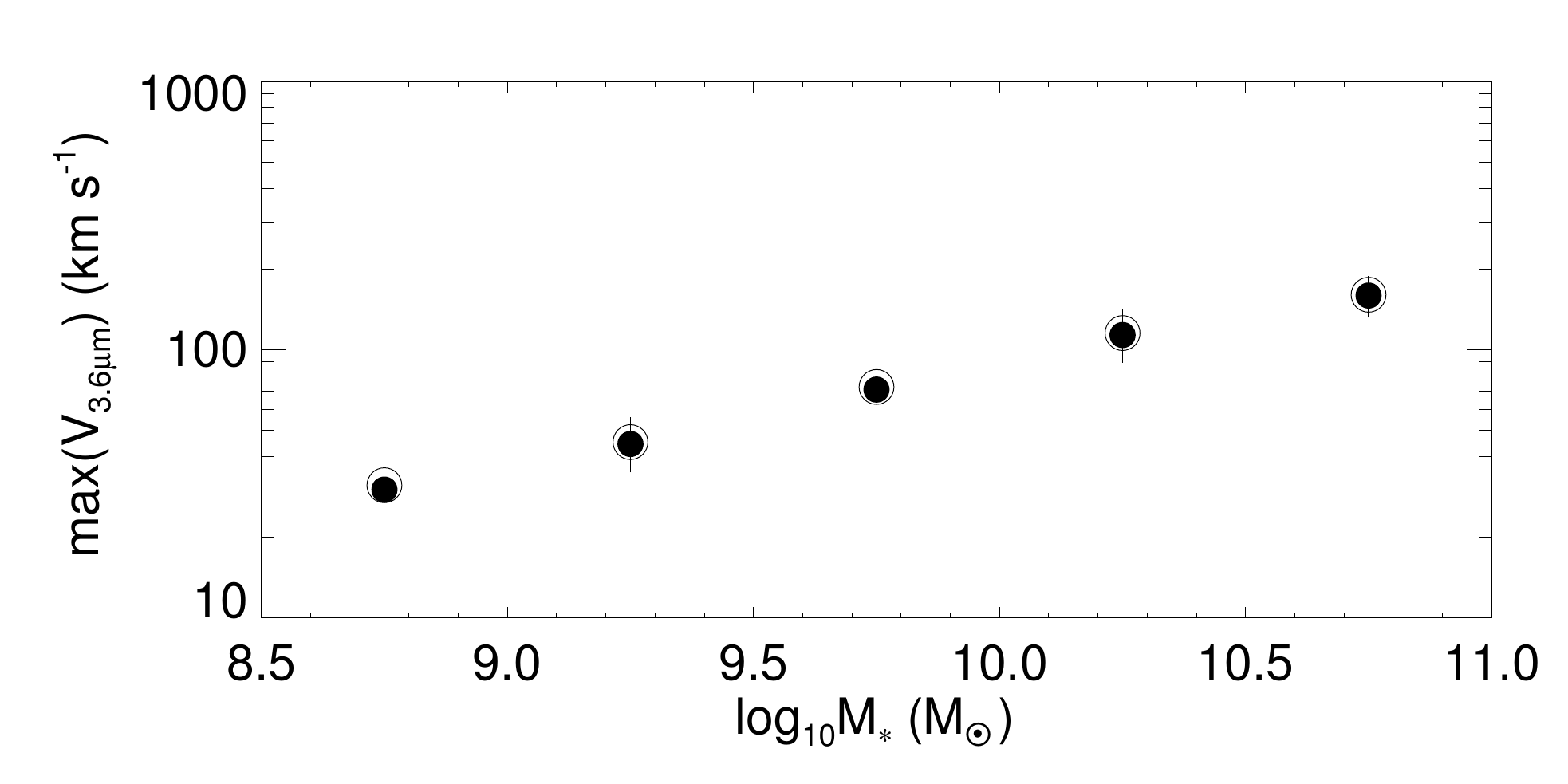}
     \includegraphics[width=0.5\textwidth]{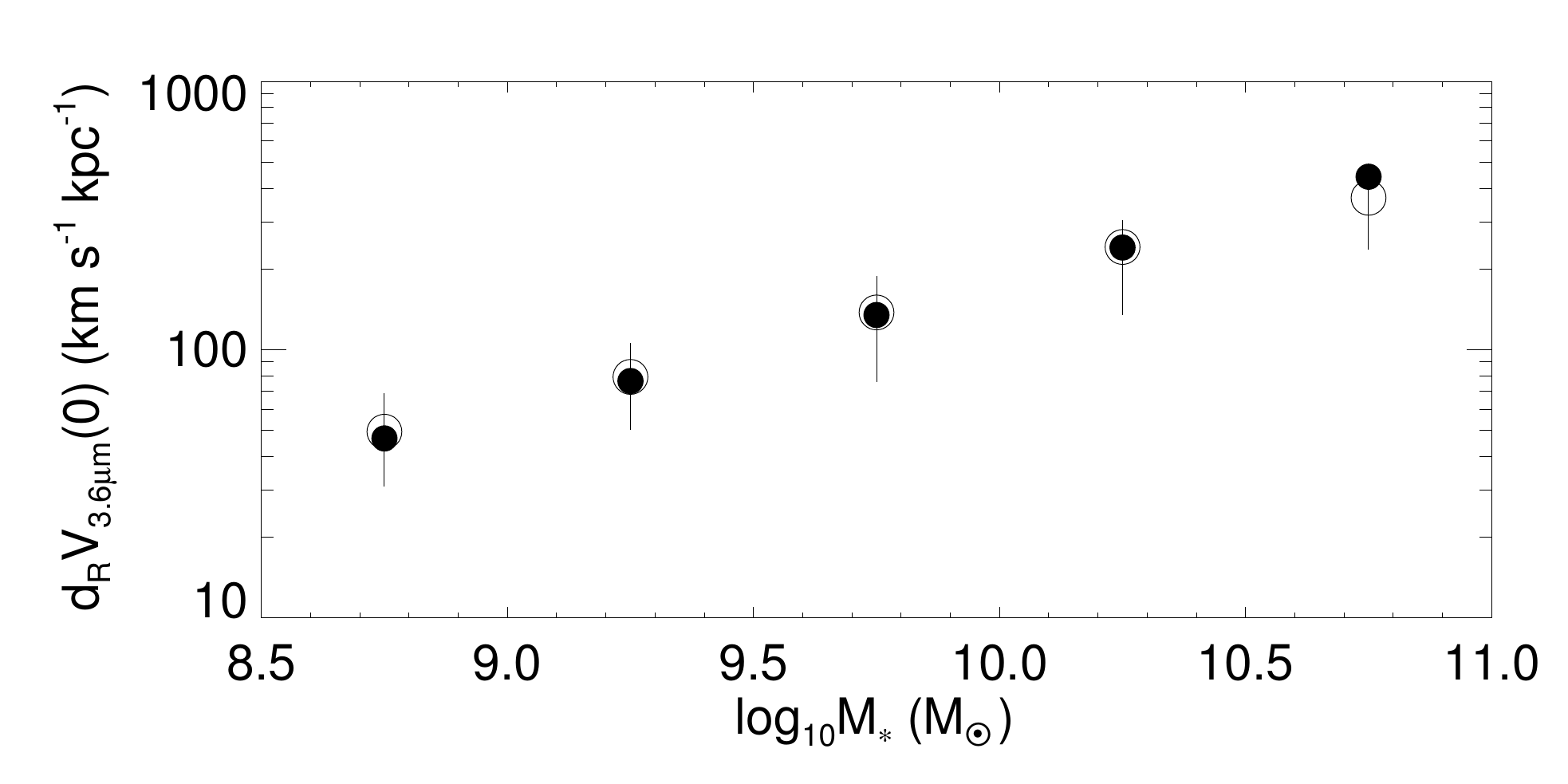}\\
\end{tabular}
\caption{
 Maximum of the disk component velocity (\emph{left panel}) and the inner slope (\emph{right panel}) as a function of the central value of the $M_{\ast}$-bins where the mean surface brightness profiles and 
the stellar contribution to the circular velocity were inferred. The values resulting from the mean and median $V_{\rm 3.6\mu m}$ are indicated with open and filled circles, respectively. 
The vertical lines indicate the change in the maximum rotation velocity and the inner velocity gradient when using $<V_{\rm 3.6\mu m}>\,\pm\,\sigma$ instead of $<V_{\rm 3.6\mu m}>$.
}
\label{stellar-halo-ratio}
\end{figure*}
%
%
\begin{table*}
\centering
\scalebox{0.95}{
\tiny
\begin{tabular}{|| c|| c| c| c| c| c||}
\hline
\multicolumn{1}{||c}{\centering $V_{3.6\mu \rm m}$ parameters} & \multicolumn{1}{|c}{\centering $8.5<\rm log_{10}M_{\ast}/M_{\odot}<9$} & \multicolumn{1}{|c}{\centering $9<\rm log_{10}M_{\ast}/M_{\odot}<9.5$} &
\multicolumn{1}{|c}{\centering $9.5<\rm log_{10}M_{\ast}/M_{\odot}<10$} & \multicolumn{1}{|c||}{\centering $10<\rm log_{10}M_{\ast}/M_{\odot}<10.5$} & \multicolumn{1}{|c||}{\centering $10.5<\rm log_{10}M_{\ast}/M_{\odot}<11$} \tabularnewline
\hline
\hline
$V_{3.6\mu \rm m}^{\rm max}$ (km s$^{-1}$)& 31.2 (6.3)& 45.2 (10.6)& 72.7 (20.8)&115.4 (26.2)&160.4 (28.7) \\
d$V_{3.6\mu \rm m}(0)$ (km s$^{-1}$ kpc$^{-1}$)& 49.5 (19.1)& 79.2 (27.9)&137.8 (56.2)&241.7 (84.4)&368.8 (85.1) \\
d$V_{3.6\mu \rm m}(0)$ (km s$^{-1}$ $h_{\rm R}^{-1}$)& 77.6 (41.7)&135.7 (71.6)&285.7 (179.9)&651.4 (312.6)& 1134.1 (359.0) \\
\hline
\end{tabular}}
\caption{
Maximum and inner slope of the mean $V_{3.6\mu \rm m}$, derived from the stellar contribution to the rotation curves, rescaled with respect to $h_{\rm R}$ and the disk size in kpc.
Parameters are given for the different stellar mass bins. In parentheses, we indicate the mean change in the parameters when using $<V_{\rm 3.6\mu m}>\,\pm\,\sigma$ instead of $<V_{\rm 3.6\mu m}>$ (see error bars in Fig.~\ref{stellar-halo-ratio}).
}
\label{vcirc_mass}
\end{table*}

%
%
\begin{table*}
\centering
\tiny
\begin{tabular}{|| c|| c| c| c| c| c||}
\hline
\multicolumn{1}{||c}{\centering $V_{3.6\mu \rm m}$ parameters} & \multicolumn{1}{|c}{\centering $-3\le T<0$} & \multicolumn{1}{|c}{\centering $0\le T<3$} &
\multicolumn{1}{|c}{\centering $3\le T<5$} & \multicolumn{1}{|c||}{\centering $5\le T<8$} & \multicolumn{1}{|c||}{\centering $8\le T<11$} \tabularnewline
\hline
\hline
$V_{3.6\mu \rm m}^{\rm max}$ (km s$^{-1}$)&130.1 (48.5)&126.0 (38.7)&122.4 (40.8)& 80.8 (38.4)& 45.7 (23.2) \\
d$V_{3.6\mu \rm m}(0)$ (km s$^{-1}$ kpc$^{-1}$)&659.2 (443.6)&302.3 (156.8)&224.7 (85.7)&141.6 (71.1)& 73.0 (47.2) \\
d$V_{3.6\mu \rm m}(0)$ (km s$^{-1}$ $h_{\rm R}^{-1}$)&931.5 (432.5)&871.3 (411.1)&653.6 (349.4)&325.5 (228.6)&  124.6 (107.3) \\
\hline
\end{tabular}
\caption{
As in Table~\ref{vcirc_mass}, but for the five subsamples selected based on morphological type considered in this study.
}
\label{vcirc_ttype}
\end{table*}

%
%
\begin{table*}
\centering
\tiny
\begin{tabular}{|| c|| c| c| c| c| c||}
\hline
\multicolumn{1}{||c}{\centering Disk parameters} & \multicolumn{1}{|c}{\centering $8.5<\rm log_{10}M_{\ast}/M_{\odot}<9$} & \multicolumn{1}{|c}{\centering $9<\rm log_{10}M_{\ast}/M_{\odot}<9.5$} &
\multicolumn{1}{|c}{\centering $9.5<\rm log_{10}M_{\ast}/M_{\odot}<10$} & \multicolumn{1}{|c||}{\centering $10<\rm log_{10}M_{\ast}/M_{\odot}<10.5$} & \multicolumn{1}{|c||}{\centering $10.5<\rm log_{10}M_{\ast}/M_{\odot}<11$} \tabularnewline
\hline
\hline
$V_{\circ}^{\rm comp1}$ (km s$^{-1}$)& 28.50& 39.26& 50.39& 16.12&118.13 \\
$r_{\rm PE}^{\rm comp1}$ (kpc)& 0.55& 0.93& 1.73& 3.77& 0.23 \\
$\alpha^{\rm comp1}$&-0.001&-0.041&-0.378&-5.153&-0.038 \\
\hline
$V_{\circ}^{\rm comp2}$ (km s$^{-1}$)&...& 6.77&26.18&72.74&65.17 \\
$r_{\rm PE}^{\rm comp2}$ (kpc)&...& 0.06& 0.12& 0.18& 1.47 \\
$\alpha^{\rm comp2}$&...& 0.007& 0.042& 0.051& 0.357 \\
\hline
\end{tabular}
\caption{
Parameterization of the mean $V_{\rm 3.6\mu m}$ profiles binned in terms of $M_{\ast}$ (Fig.~\ref{stack_vdisk_comp}).
We use a double-component {\it Polyex} model (Eq.~\ref{polyexeq}) (only one component is used for the bin with $M_{\ast}<10^{9}M_{\odot}$).}
\label{rcur_fits_params}
\end{table*}

%
%
\subsection{Stellar contribution to the circular velocity}\label{rotcurs}
%
%
We obtain the mean and median $V_{\rm 3.6\mu m}$ profiles from the rescaled individual rotation curves as a function of $M_{\ast}$ and $T$-type, shown in Fig.~\ref{stack_vdisk_comp}. 
Given the scatter in total stellar mass for a fixed morphological type, the mean and median $V_{\rm 3.6\mu m}$ values are somewhat different ($<10\%$) when studied as a function of $T$. 
In order to characterize $V_{\rm 3.6\mu m}$, we derive the maximum rotation ($V_{\rm 3.6\mu m}^{\rm max}$), the radius of maximum rotation ($r_{\rm 3.6\mu m}^{\rm max}$),
and the inner velocity gradient of $V_{\rm 3.6\mu m}$, denoted as d$_{\rm R}V_{\rm 3.6\mu m}(0)$. 
This is done by fitting the inner rotation curve with a polynomial function of order $m=3$, and estimating the inner slope from the linear term \citep[following the method in][]{2013MNRAS.433L..30L}. 
Values of d$_{\rm R}V_{\rm 3.6\mu m}(0)$ and $V_{\rm 3.6\mu m}^{\rm max}$ for the different $T$- and $M_{\ast}$-bins are listed in Table~\ref{vcirc_mass} and Table~\ref{vcirc_ttype}.

As expected, the maximum velocities and the inner slopes of $V_{\rm 3.6\mu m}$ increase with stellar mass (see Fig.~\ref{stellar-halo-ratio} for the behaviour of the individual parameters). 
For the stacks of massive galaxies, the associated rotation curves show a maximum at $\sim2.2h_{\rm R}$, which is the radius of the maximum velocity of an exponential disk \citep[][]{1970ApJ...160..811F}. 
When the stellar contribution to the circular velocity is studied as a function of $T$ (see the lower panels of Fig.~\ref{stack_vdisk_comp}), 
the higher central concentration of lenticular galaxies manifests in a centrally peaked $V_{\rm 3.6\mu m}$ rotation curve, with a maximum at $\sim 0.75 h_{\rm R}$. 
The $T$-dependent segregation in the bulge prominence among the systems with morphological types earlier than Sc (shown in Sect.~\ref{ttypesprofs})
is also manifested in a increasing d$_{\rm R}V_{\rm 3.6\mu m}(0)$ with decreasing $T$ among the systems with similar maximum velocities ($V_{\rm 3.6\mu m}^{\rm max}\approx120-130$ km/s). 
Faint late-type galaxies do not show clear declining section in the average $V_{\rm 3.6\mu m}$ within the studied radial range ($75\%$ coverage). 

Finally, in order to parameterize $V_{\rm 3.6\mu m}$, we use the {\it Polyex} function \citep[][]{2002ApJ...571L.107G}:
\begin{equation}\label{polyexeq}
V_{\rm PE}(r)=V_{\circ}\big(1-{\rm exp}(-r/r_{\rm PE})\big)(1+\alpha r/r_{\rm PE}),
\end{equation}
where $V_{\circ}$ sets the amplitude of the rotation curve, $r_{\rm PE}$ corresponds to the exponential scale of the inner region, and $\alpha$ determines the slope of the outer part. 
This function is often used as a template for observed rotation curves \citep[e.g.][]{2006ApJ...640..751C}. 
For the disk component we use a double component chi-square minimization fit ($V_{\rm 3.6\mu m}=V_{\rm PE}^{\rm comp1}+V_{\rm PE}^{\rm comp2}$) that accounts for the inner hump produced by bulges in the stacks with the largest stellar masses. 
The resulting parameters, which have no physical meaning, are listed in Table~\ref{rcur_fits_params}. 
%
%
%
\section{Two-dimensional characterization of stellar bars}\label{bars-char}
%
%
%
DG2016 performed a statistical analysis of the lengths and strengths of stellar bars in the Hubble sequence and also in connection to the morphological family and total stellar mass of their host disks. 
In the  Sect.~\ref{a2_medi} we analysed the mean $A_2$ profiles for barred and non-barred galaxies based on the Fourier decomposition of individual galaxies from DG2016. 
Here, we characterize the morphologies and strengths of the average bars obtained by stacking 3.6 $\mu$m images in bins of stellar mass, Hubble type, and galaxy family. 
%
%
\subsection{Shape of average bars}\label{bars-shape}
%
%
\begin{figure*}
\centering   
\begin{tabular}{c c}
   \includegraphics[width=1.0\textwidth]{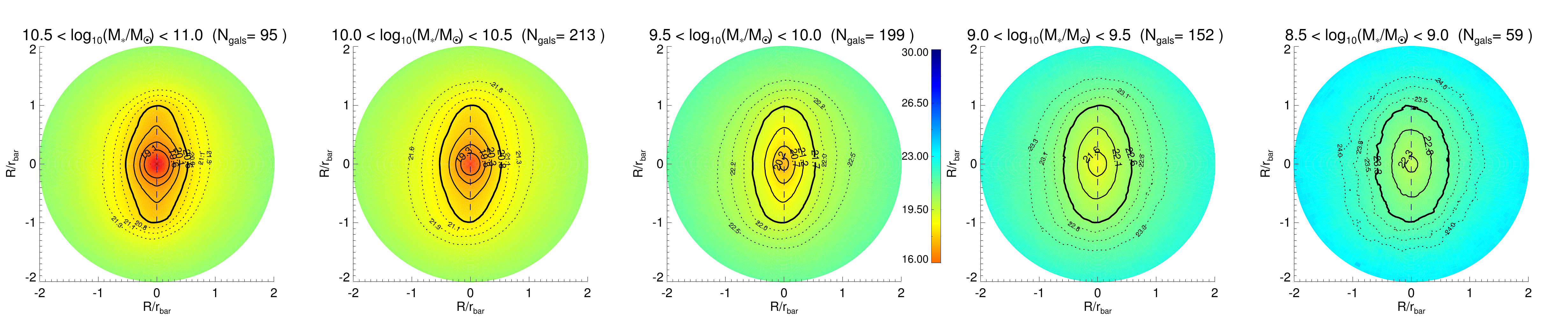}\\
   \includegraphics[width=1.0\textwidth]{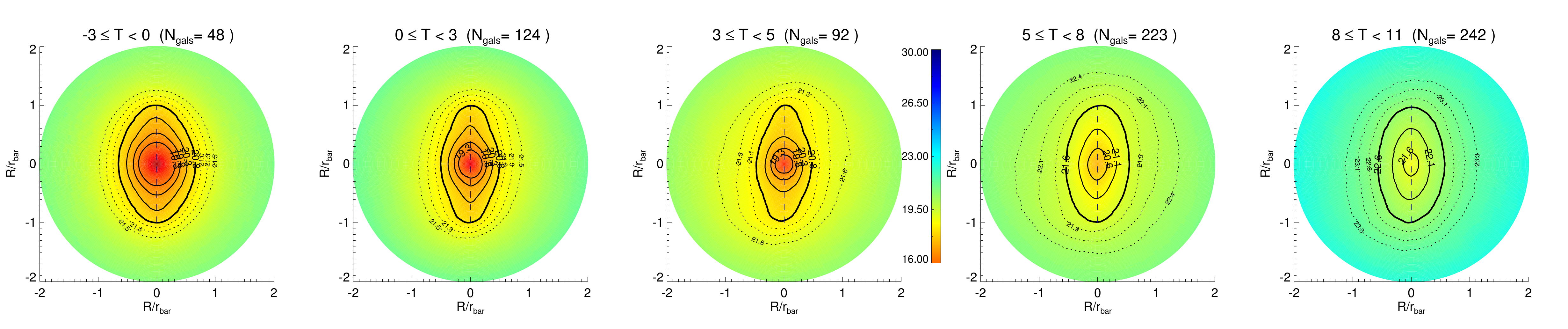}\\
\end{tabular}
\caption{
\emph{First row:}
2-D synthetic stellar bars constructed from co-added 3.6~$\mu$m images of barred galaxies oriented and rescaled with respect to the bars, grouped based on total stellar mass (increasing from right to left). 
Different surface brightness contours, starting at the bar end and going to brighter (solid lines) and fainter (dotted lines) levels, are displayed. The $\mu_{3.6\mu \rm m}$ levels are also indicated in units of mag arcsec$^{-2}$. 
The vertical dashed lines delimit the bar major axis.
\emph{Second row:}
As in the upper panels, but binning the sample based on revised Hubble type.
}
\label{stack3}
\end{figure*}
%
%
\begin{figure*}
\centering
   \includegraphics[width=1.0\textwidth]{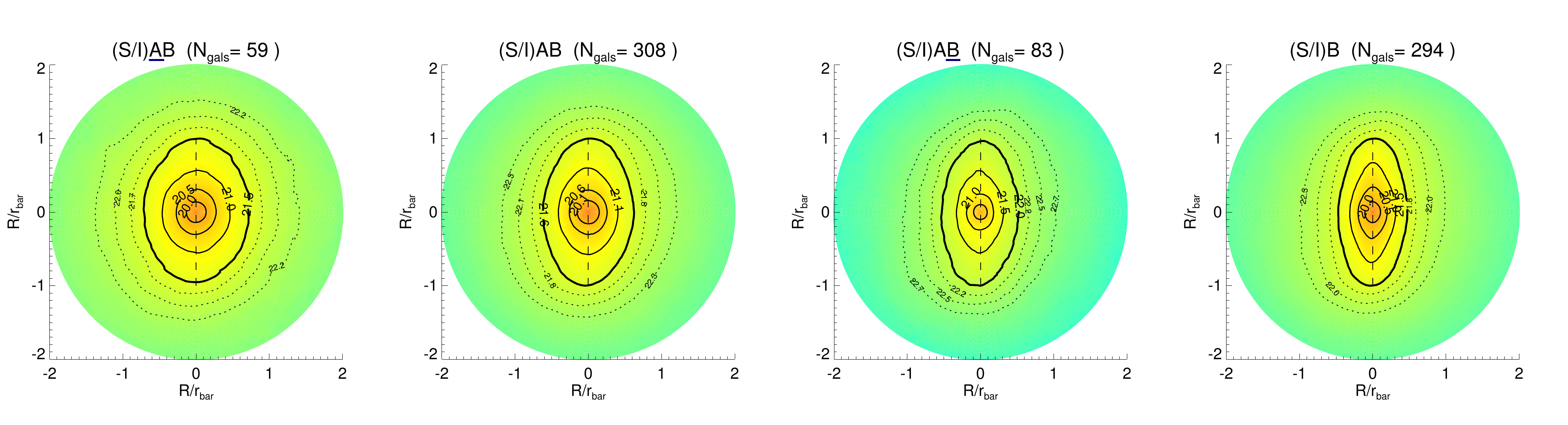}
\caption{
2-D synthetic images constructed from co-added 3.6~$\mu$m images of barred galaxies, displayed as in Fig.~\ref{stack3} with the galaxies grouped based on their morphological family.
}
\label{stack_families}
\end{figure*}
%
%
\begin{figure*}
\centering
   \includegraphics[width=1.0\textwidth]{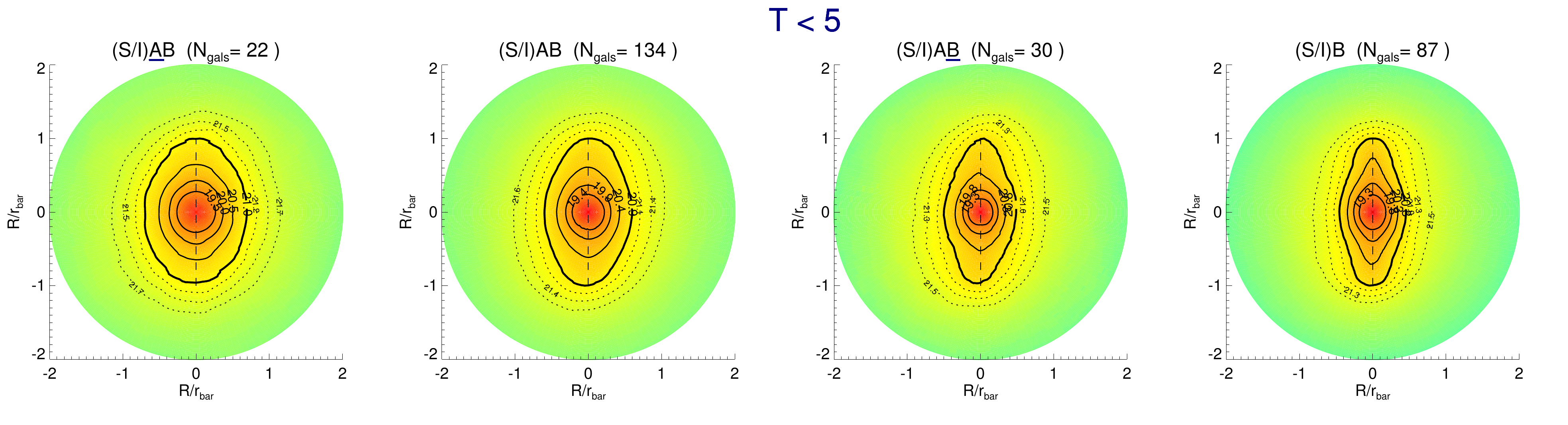}
   \includegraphics[width=1.0\textwidth]{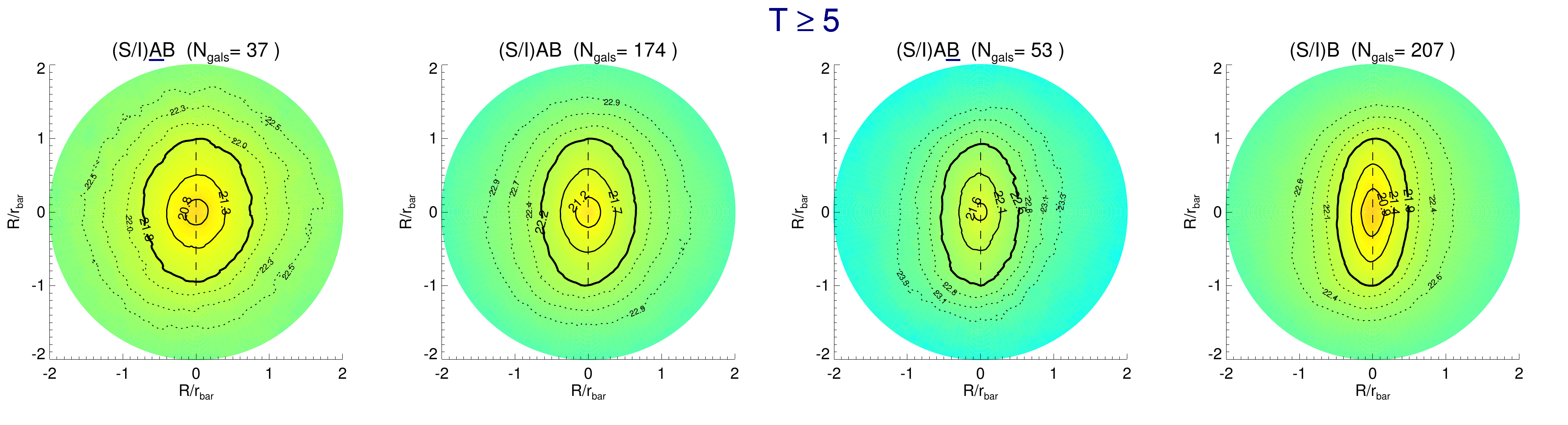}
\caption{
As in  Fig.~\ref{stack_families}, but separating systems with morphological types earlier than (top row) and equal to or later than (bottom row)  $T=5$ (Sc galaxies).
}
\label{stack_families_t5}
\end{figure*}
%
%
\begin{figure*}
\centering   
\begin{tabular}{c c c c c c}
     \includegraphics[width=0.33\textwidth]{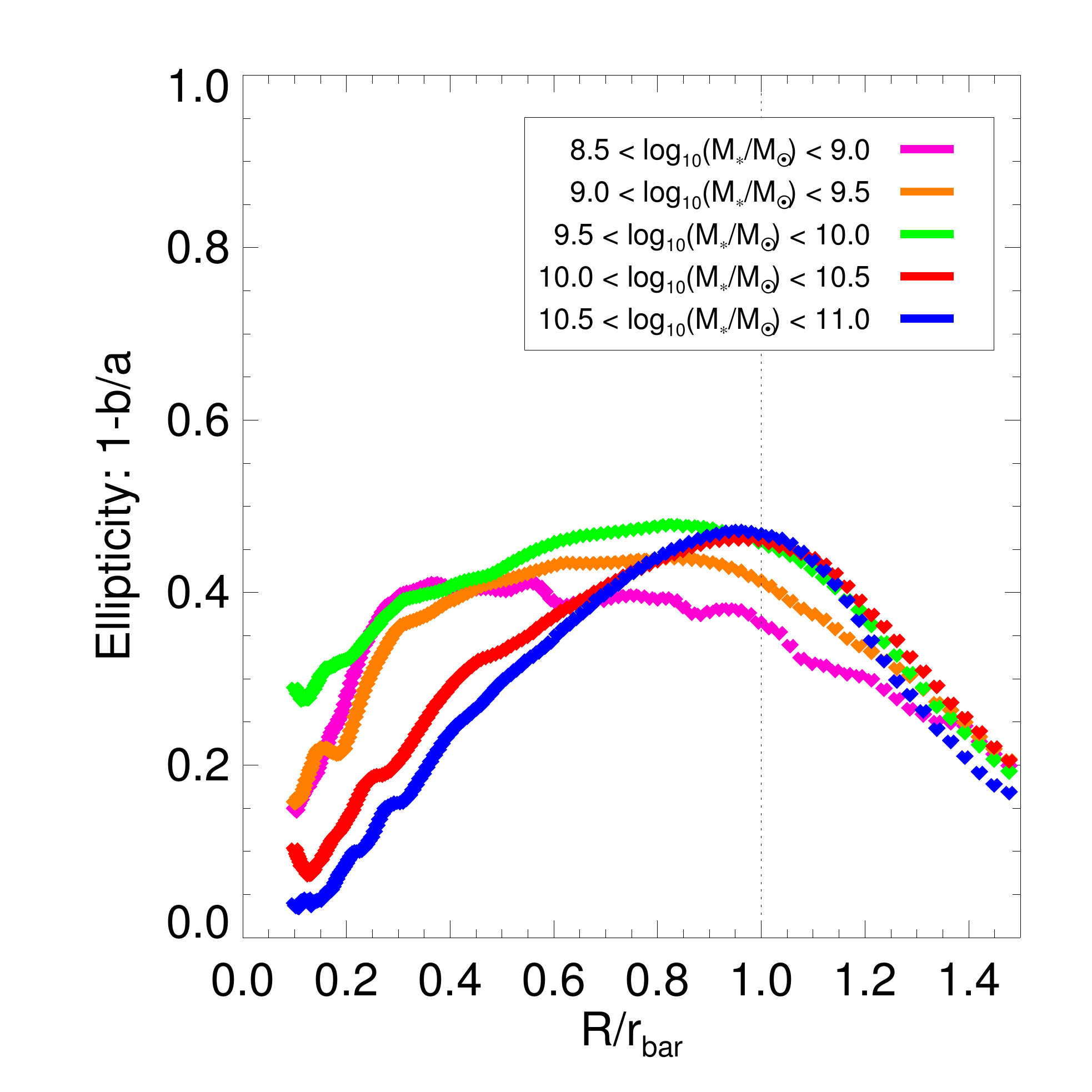}
     \includegraphics[width=0.33\textwidth]{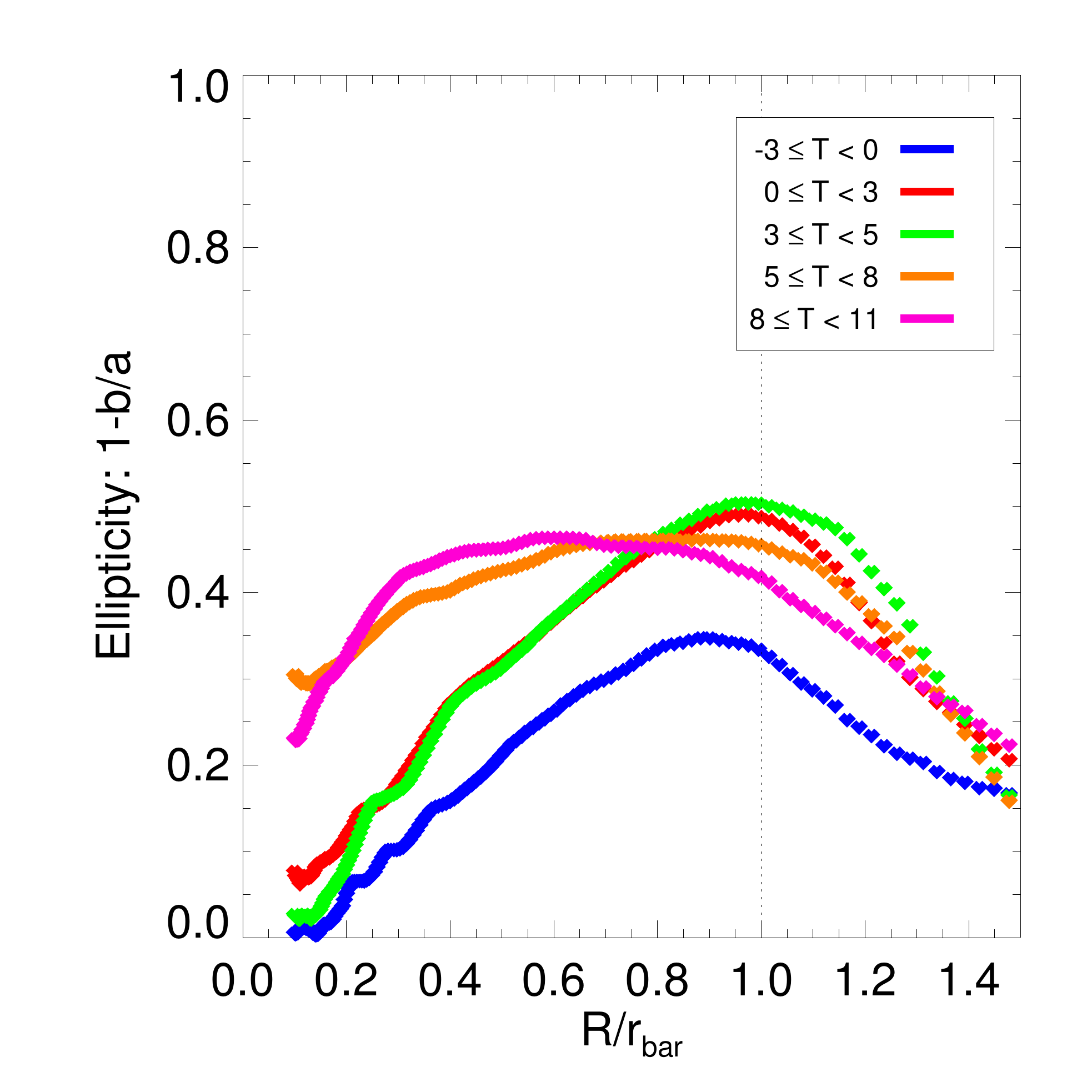}
     \includegraphics[width=0.33\textwidth]{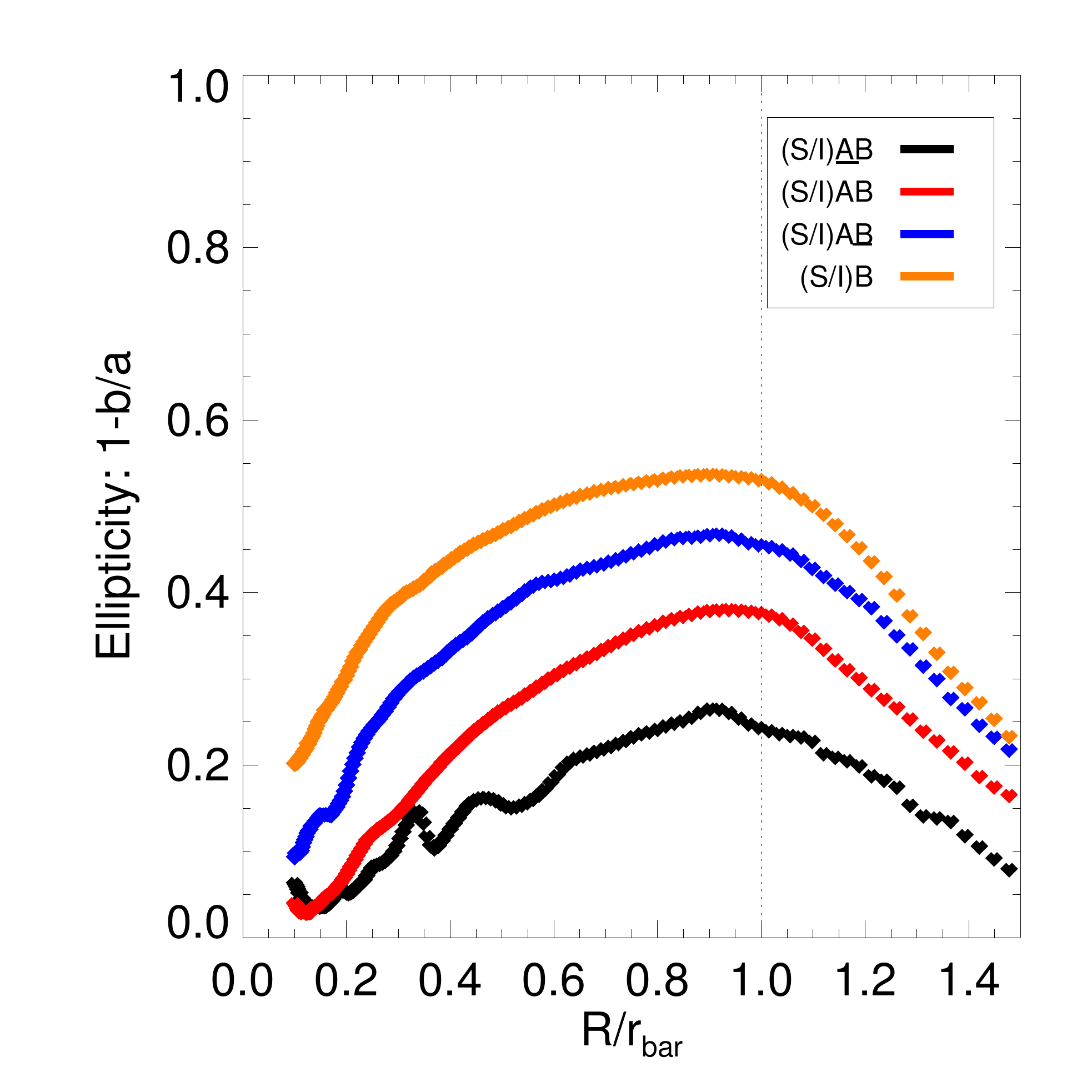}\\
     \includegraphics[width=0.33\textwidth]{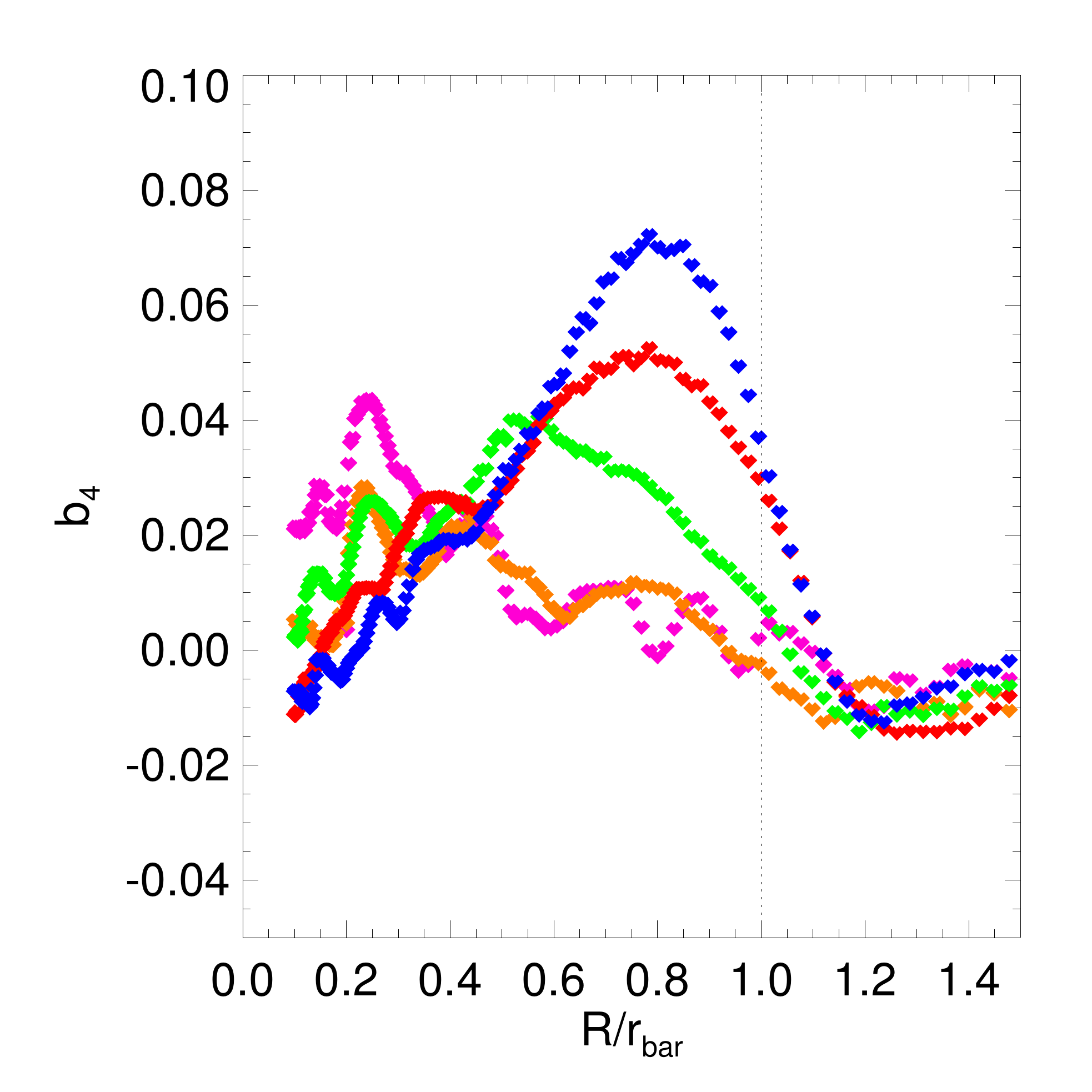}
     \includegraphics[width=0.33\textwidth]{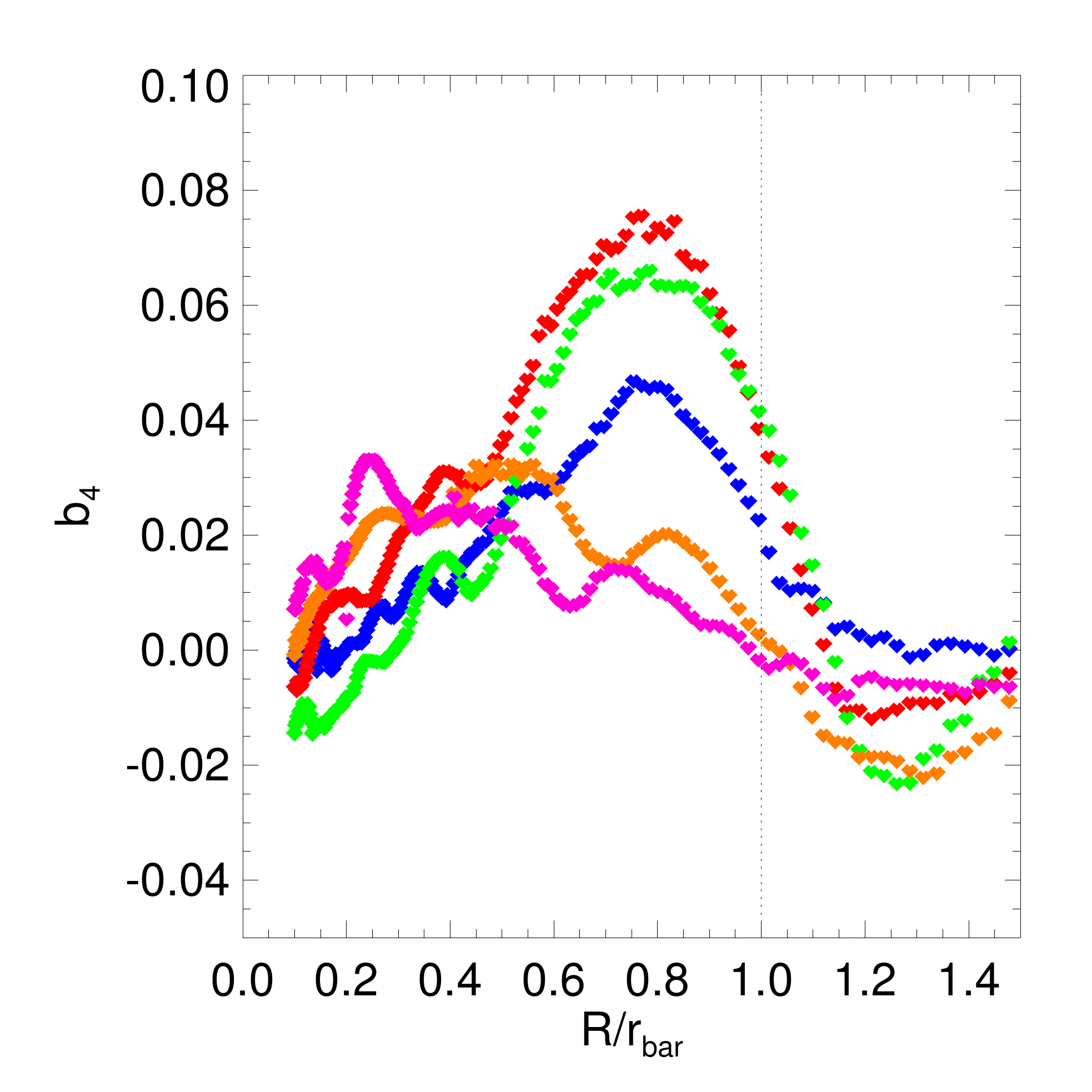}
     \includegraphics[width=0.33\textwidth]{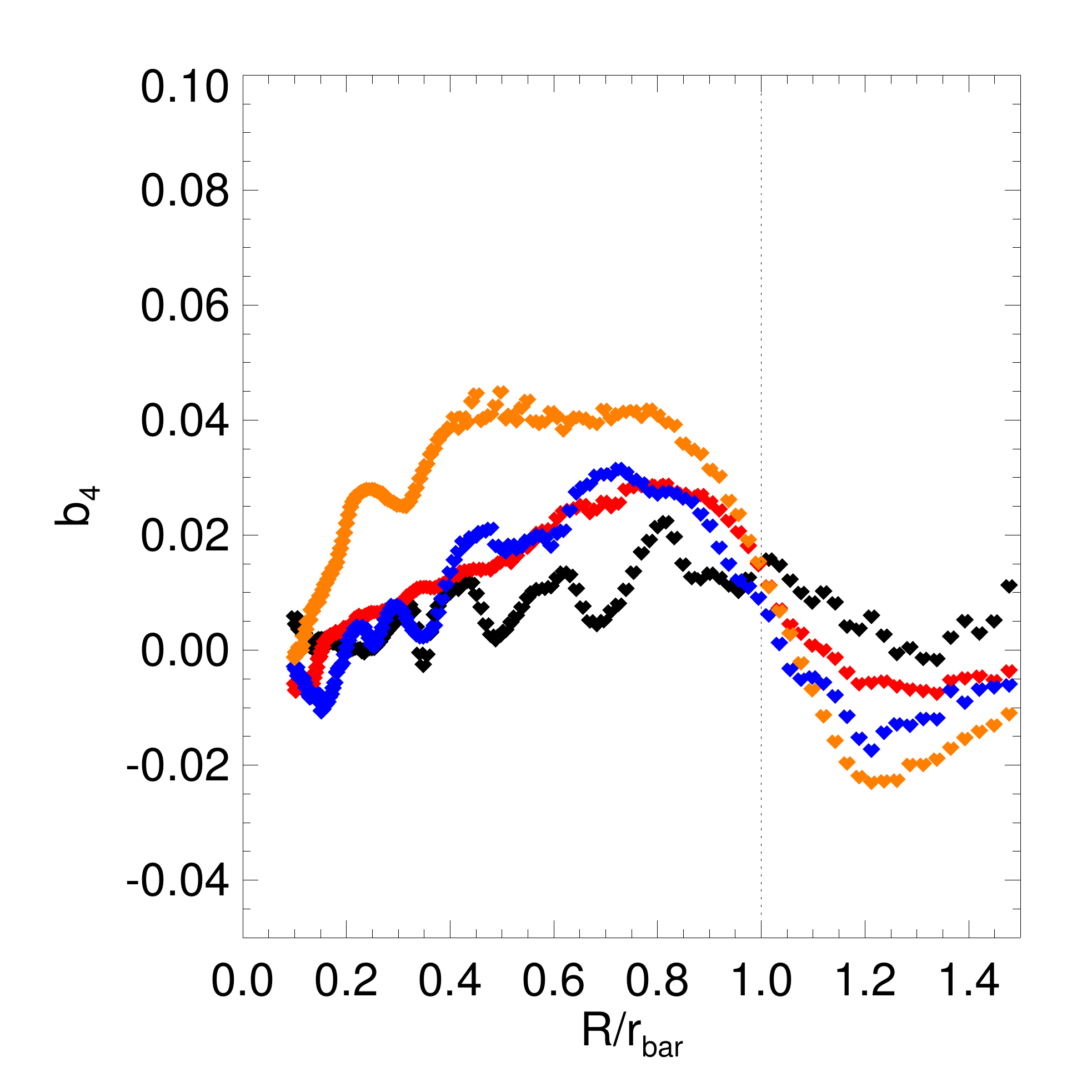}\\
\end{tabular}
\caption{
Radial profiles of ellipticity and $b_4$ parameter (4${\rm th}$ harmonic deviation from ellipse) of the elliptical isophotal fit to the bar stacks shown in Fig.~\ref{stack3} and Fig.~\ref{stack_families}.
}
\label{stack_bars_ellip_b4}
\end{figure*}
%
%
\begin{figure*}
\centering   
\begin{tabular}{c c c c}
     \includegraphics[width=0.25\textwidth]{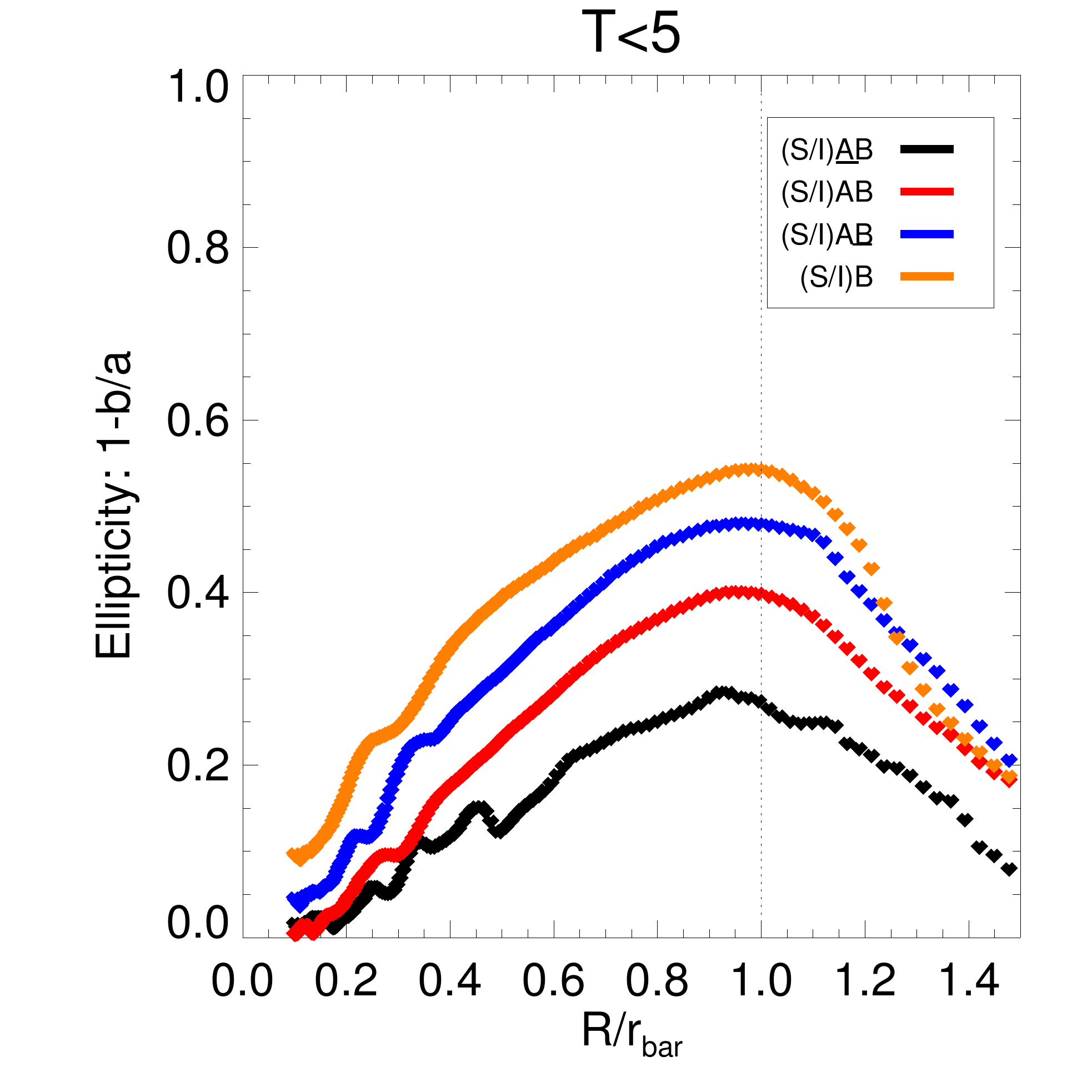}
     \includegraphics[width=0.25\textwidth]{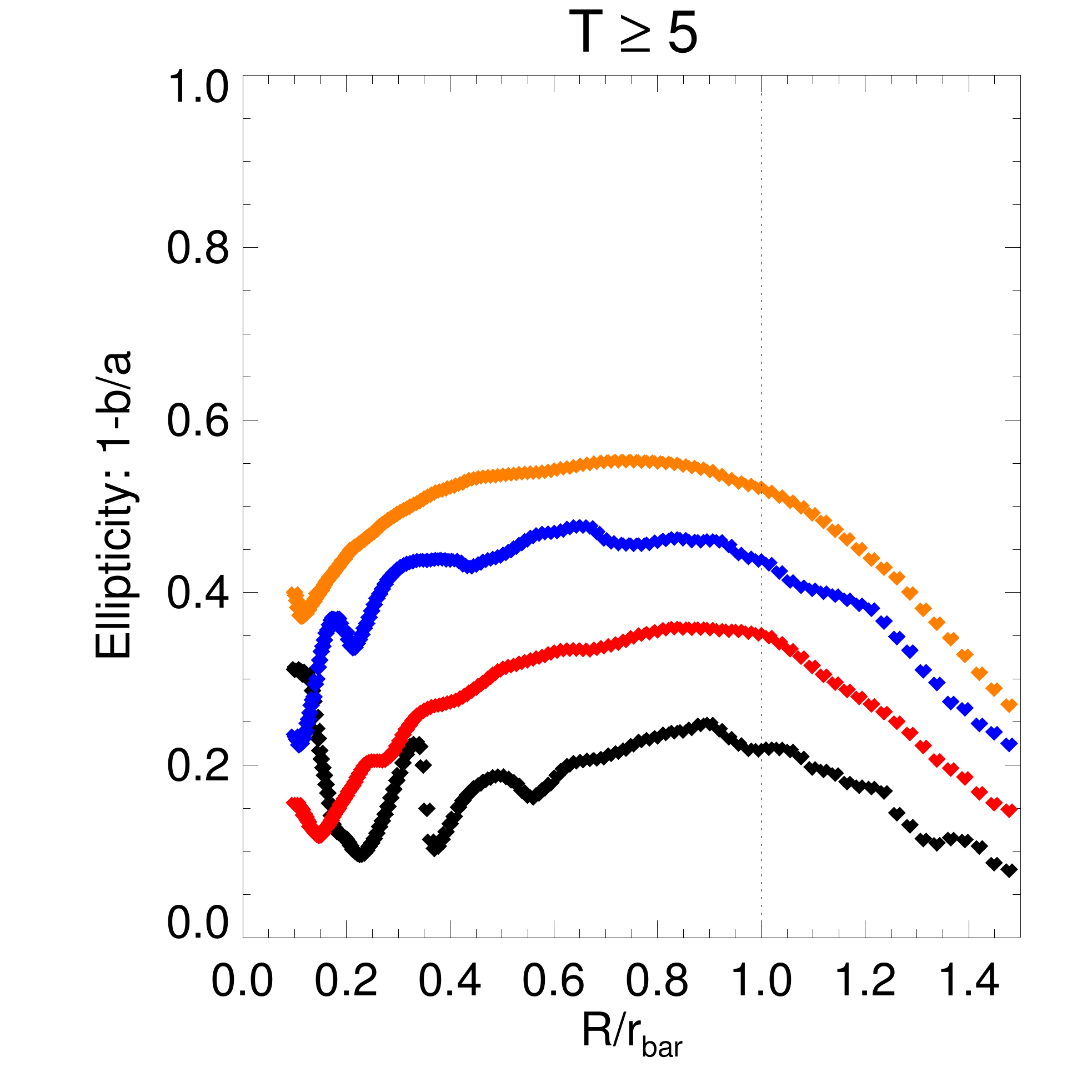}
     \includegraphics[width=0.25\textwidth]{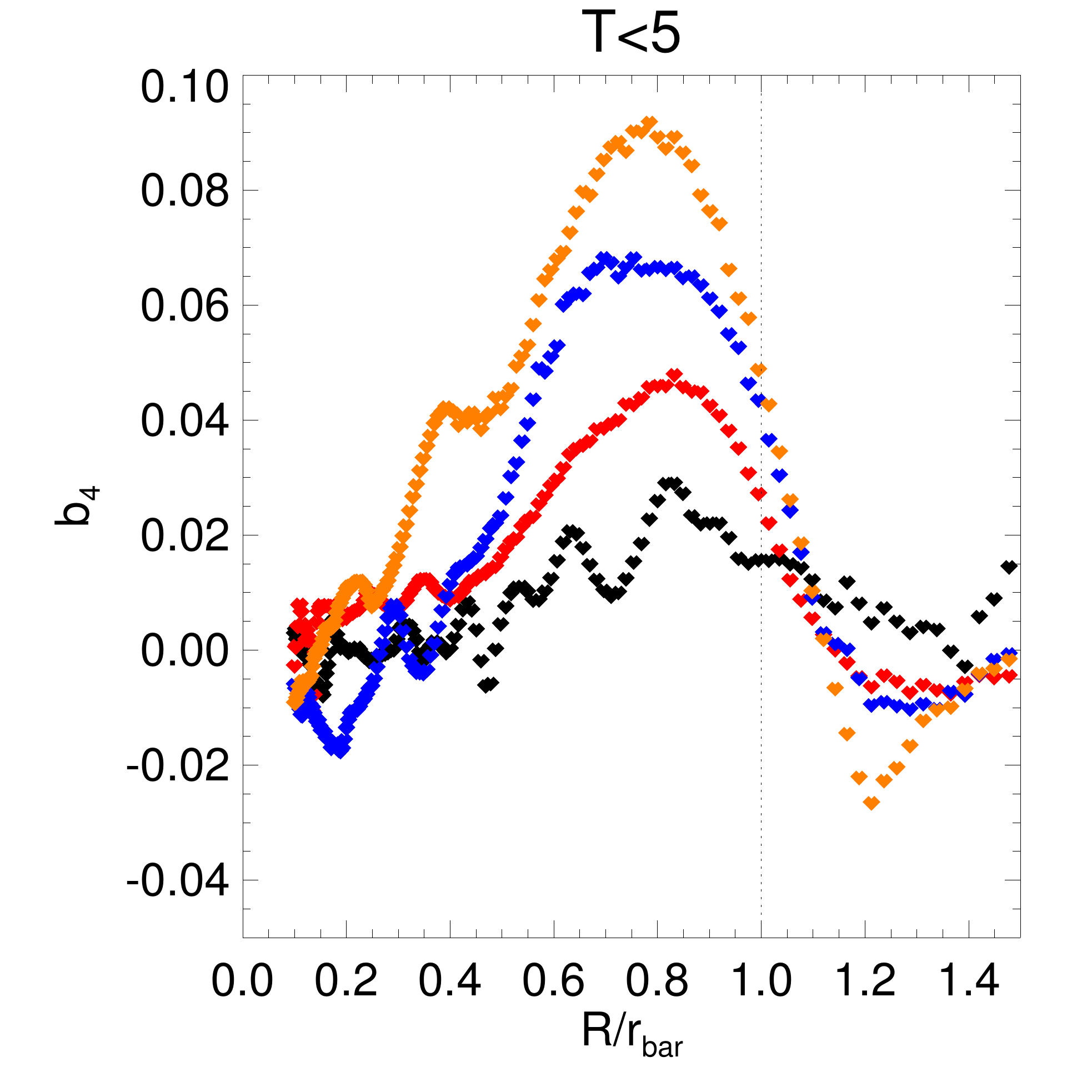}
     \includegraphics[width=0.25\textwidth]{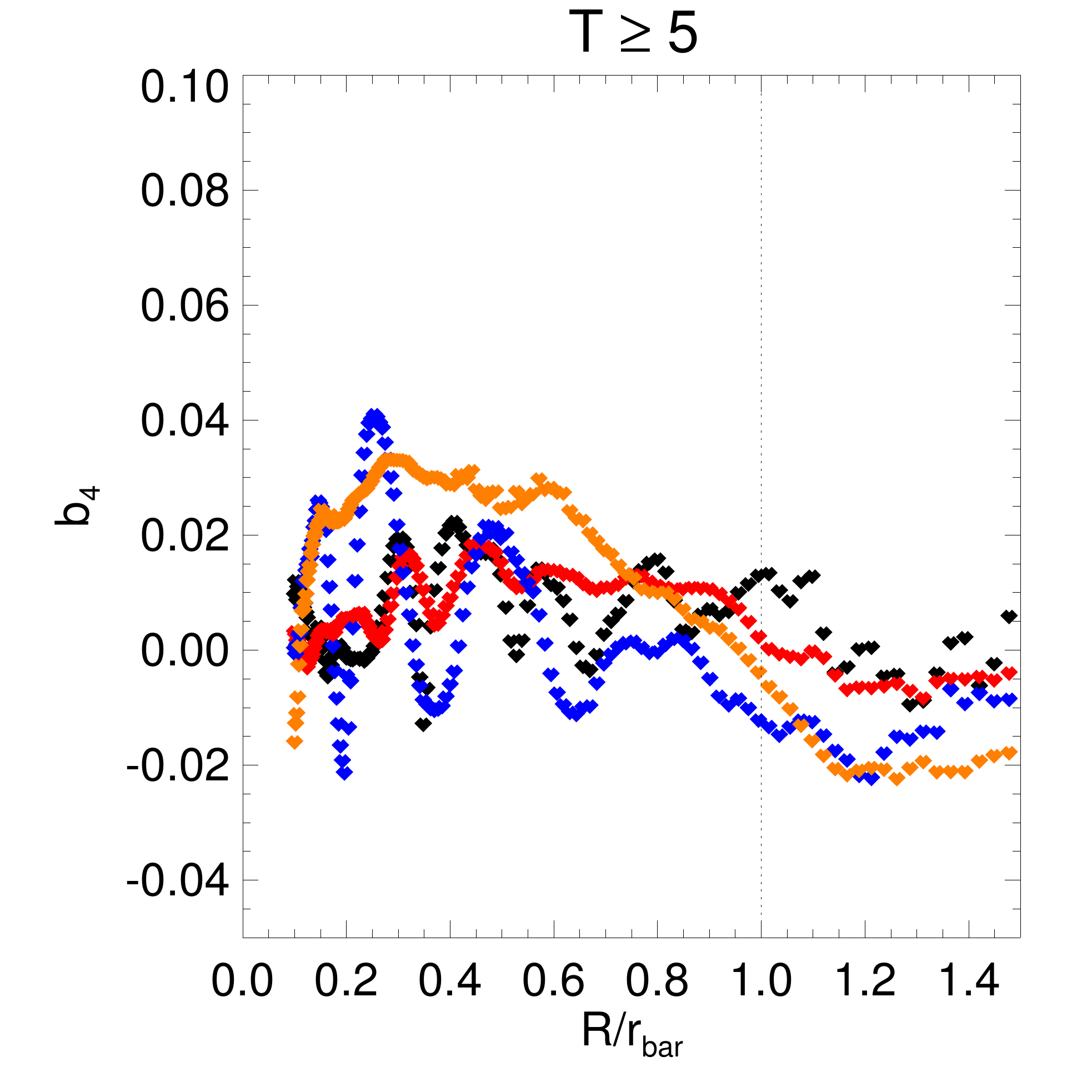}\\
\end{tabular}
\caption{
As in Fig.~\ref{stack_bars_ellip_b4}, but sampling the galaxies in terms of the galaxy family and separating early-type ($T<5$) and late-type ($T\ge5$) systems.
}
\label{stack_bars_ellip_b4_late_early_family}
\end{figure*}
%
%
\begin{figure*}
\centering   
\begin{tabular}{c c}
     \includegraphics[width=0.49\textwidth]{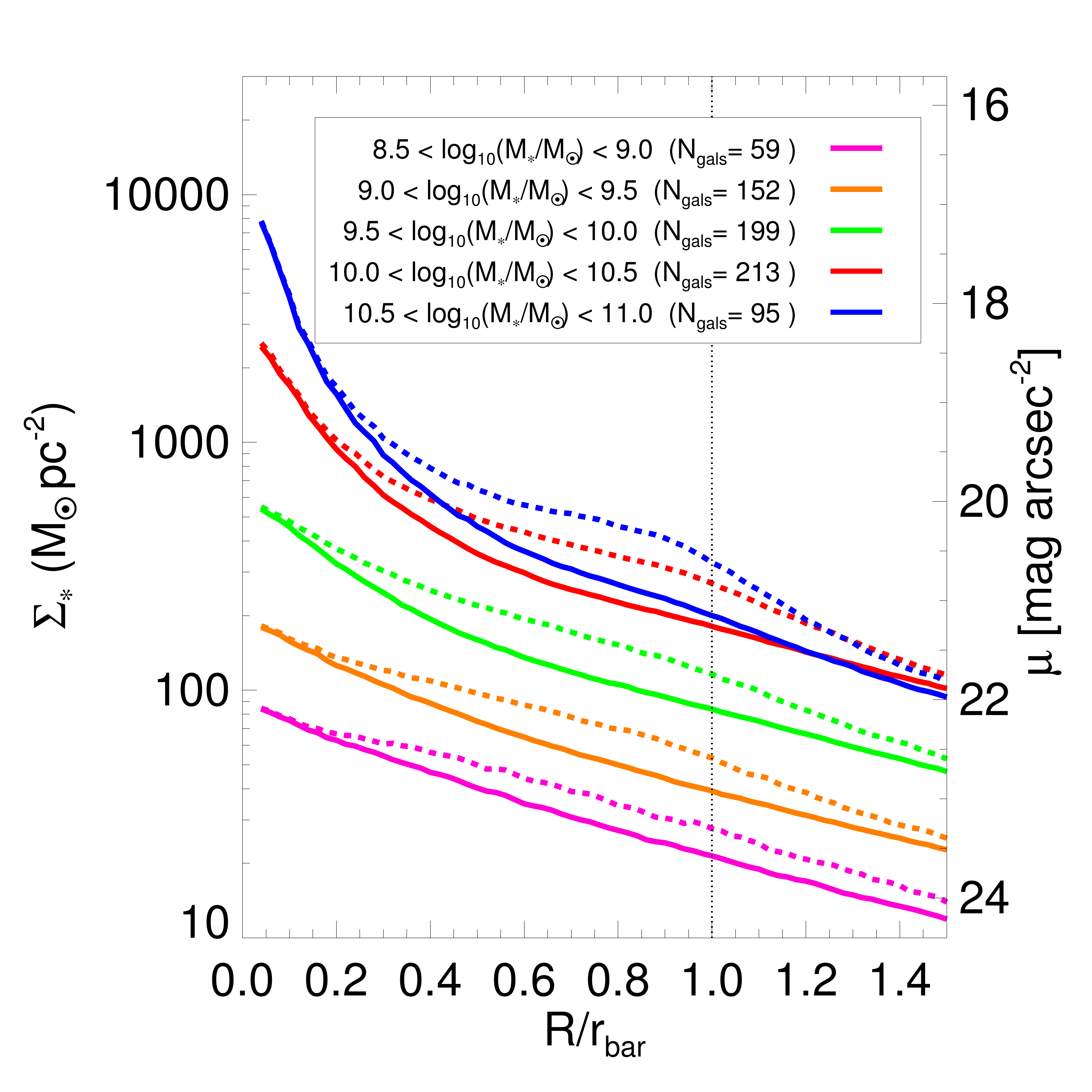}
     \includegraphics[width=0.49\textwidth]{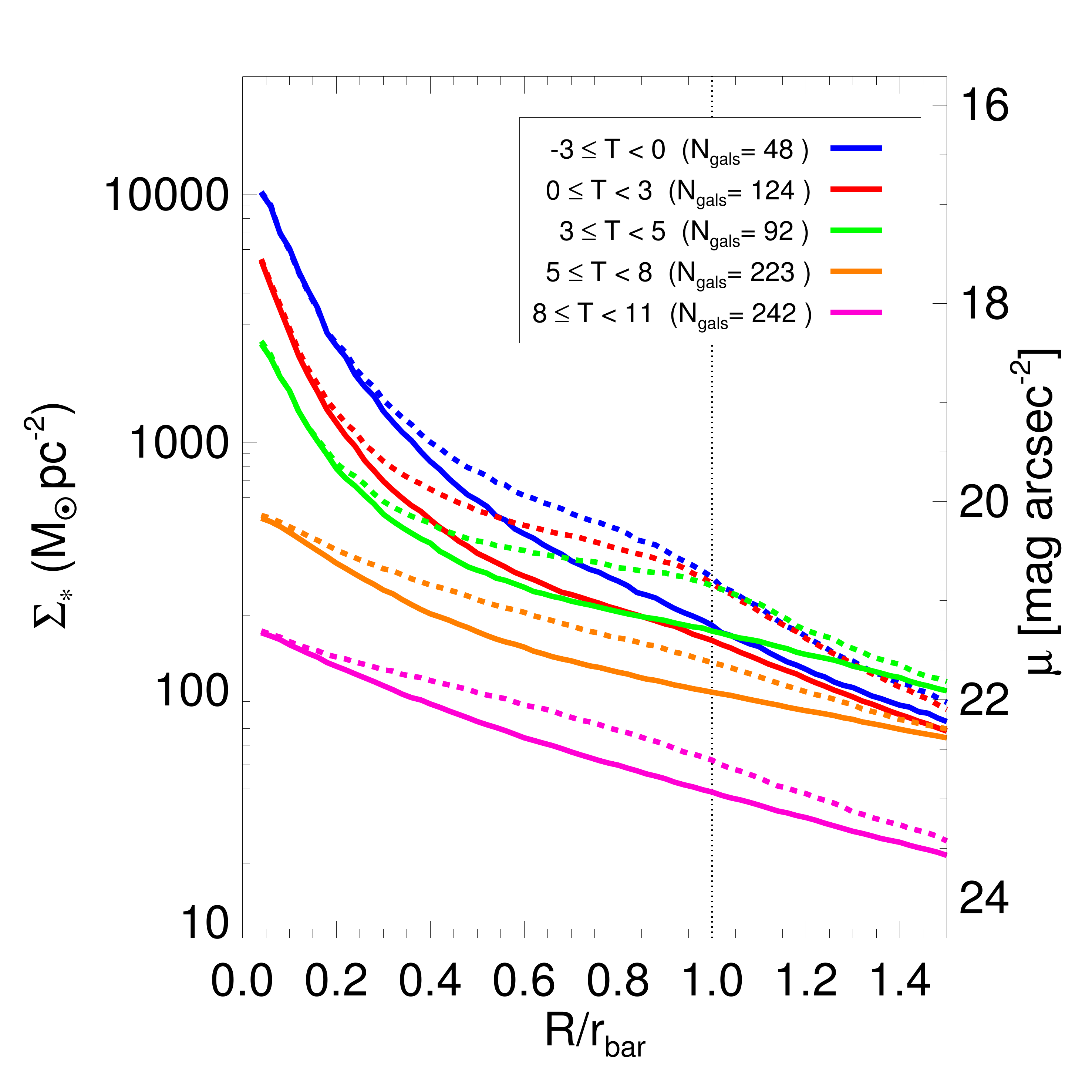}
\end{tabular}
\caption{
Azimuthally averaged mean 3.6~$\mu$m density profiles in bins of stellar mass and revised Hubble type obtained after resizing and reorienting the galaxy images with respect to the stellar bars (solid lines).
The dashed line corresponds the surface brightness cut along the bar major axis.
The vertical dotted line indicates the bar end.
}
\label{stack_bars_altogether}
\end{figure*}
%
%
In Fig.~\ref{stack3} we present the 2-D image of the resulting synthetic bars, binned in terms of $M_{\ast}$ and $T$. 
The morphology of the average bars is dependent on the mass and morphological type of the host galaxies: bars in lenticular galaxies are oval, while their form becomes more rectangular in spiral galaxies. 
Among early- and intermediate-type galaxies, we show that bulges and barlenses are clearly visible in the isocontours and make the inner bar isophotes rounder; 
this gives the whole bar  a peculiar diamond shape with rounded vertexes. The intrinsic boxiness of the  outermost bar isophotes is roughly the same for $10^9M_{\odot}<M_{\ast}<10^{11}M_{\odot}$. 
More massive early- and intermediate-type spirals seem to host somewhat narrower bars on average if one isolates the above-mentioned contribution of axisymmetric components. 
Irregular and Magellanic galaxies (fainter) seem to have on average bars which are slightly less eccentric. 
When $T>0$, the isocontours of the bar stacks at $r_{\rm bar}$ and beyond the bar end are somewhat bent towards the winding direction of the spiral arms 
(imposed to be clockwise before averaging the bars, as explained in Sect.~\ref{bar_disk_method}), which is most pronounced among intermediate- and late-type spirals.

In Fig.~\ref{stack_families} we show the shape of the average bars after splitting the sample based on the galaxy family. 
The bar ellipticities of the stacks associated with the different galaxy families gradually increase from $\rm S\underline{A}B$ towards SAB, $\rm SA\underline{B}$, and SB systems. 
Because early-type ($T<5$) and late-type ($T\ge5$) galaxies host two clearly distinct types of bars, 
we next study the above differences in the bar shape associated with galaxy family (see Fig.~\ref{stack_families_t5}). 
Despite the poorer statistics, the differences in bar ellipticity for the four families seem to be maintained for both early- and late-type systems.

In order to quantify the above described trends, we performed isophotal ellipse fitting over the bar stacks using the IRAF procedure \emph{ellipse} \citep{1987MNRAS.226..747J} and obtained radial profiles of the ellipticity and $b_4$ parameter 
shown in Fig.~\ref{stack_bars_ellip_b4}. The ellipticity is calculated as $1-b/a$, where $a$ and $b$ refer to the length of the major and minor axes, respectively, and 
$b_4$ measures the 4${\rm th}$ harmonic deviation from ellipse of the isophotal fit. Positive values of $b_4$ indicate \emph{disky} isophotes, whereas negative values correspond to \emph{boxy} isophotes \citep[see e.g.][]{2007ApJ...658L..91B}.

We show that the maximum ellipticity ($\epsilon$) is roughly the same for all stellar masses. 
In systems with $M_{\ast}>10^{10}M_{\odot}$ the radius of maximum ellipticity ($r_\epsilon$) lies very close to the bar end, 
while $r_\epsilon$ is smaller than $r_{\rm bar}$ in intermediate mass galaxies ($\sim80\%$ for $9<{\rm log}_{10}(M_{\ast}/M_{\odot})<10$). 
For the lowest mass systems ($M_{\ast}<10^{9}M_{\odot}$), the ellipticity profile is centrally peaked: it is largest at $\sim 0.3r_{\rm bar}$ and decays monotonically with increasing radius. 
In the upper central panel in Fig.~\ref{stack_bars_ellip_b4} the ellipticity profiles are studied as a function of Hubble type. 
Consistent with the behaviour  as a function of $M_{\ast}$, $r_\epsilon$ is smaller than $r_{\rm bar}$ among late-type spirals, Magellanic, and irregulars. 
For the galaxy family (see the upper right panel of Fig.~\ref{stack_bars_ellip_b4} and also the left panels in Fig.~\ref{stack_bars_ellip_b4_late_early_family}), 
weak and strong bars differ mainly in the amplitude of their ellipticity profiles.

For the bins with galaxies of stellar masses $M_{\ast}>10^{9.5}M_{\odot}$ and morphological types $T<5$, 
average bars typically show low values of $b_4$ in the inner parts ($R\lesssim 0.3 r_{\rm bar}$), 
then rise up to a maximum at $\sim 0.8r_{\rm bar}$ (larger values for early- and intermediate-type spirals), and drop in the outer part of the bar until $b_4$ reaches a minimum at $\sim 1.1-1.2 r_{\rm bar}$. 
These $b_4$ minima are negative for all the systems, except for the lenticular galaxies. 
For the less massive systems ($M_{\ast}<10^{9.5}M_{\odot}$ and $T\ge 5$), $b_4$ does not show any hump in the bar region and systematically becomes negative in the outer parts of the bar. 
Likewise, when $b_4$ is studied as a function of family (lower right panel of Fig.~\ref{stack_bars_ellip_b4}), 
we show that the radial profiles reach the minima slightly beyond the end of the bar for all the families except those corresponding to S$\underline{\rm A}$B galaxies.
We also find that the stronger the stacked bars, the lower the minima of the $b_4$ radial profiles. 
Shown in the last two panels of Fig.~\ref{stack_bars_ellip_b4_late_early_family} are the $b_4$ radial profiles of the bar stacks for late- and early-type systems, binned by family. 
Their shapes have the same bar imprint as described before. However, when $T\ge5$ the $b_4$ profiles are fairly noisy regardless of bar strength. 
The depths (amplitudes) of the $b_4$ minima (maxima) are larger for stronger bars for both early- and late-type galaxies, and when no $T$ selection is used.
%
%
\begin{figure*}
\centering   
\begin{tabular}{c c c c}
     \includegraphics[width=0.5\textwidth]{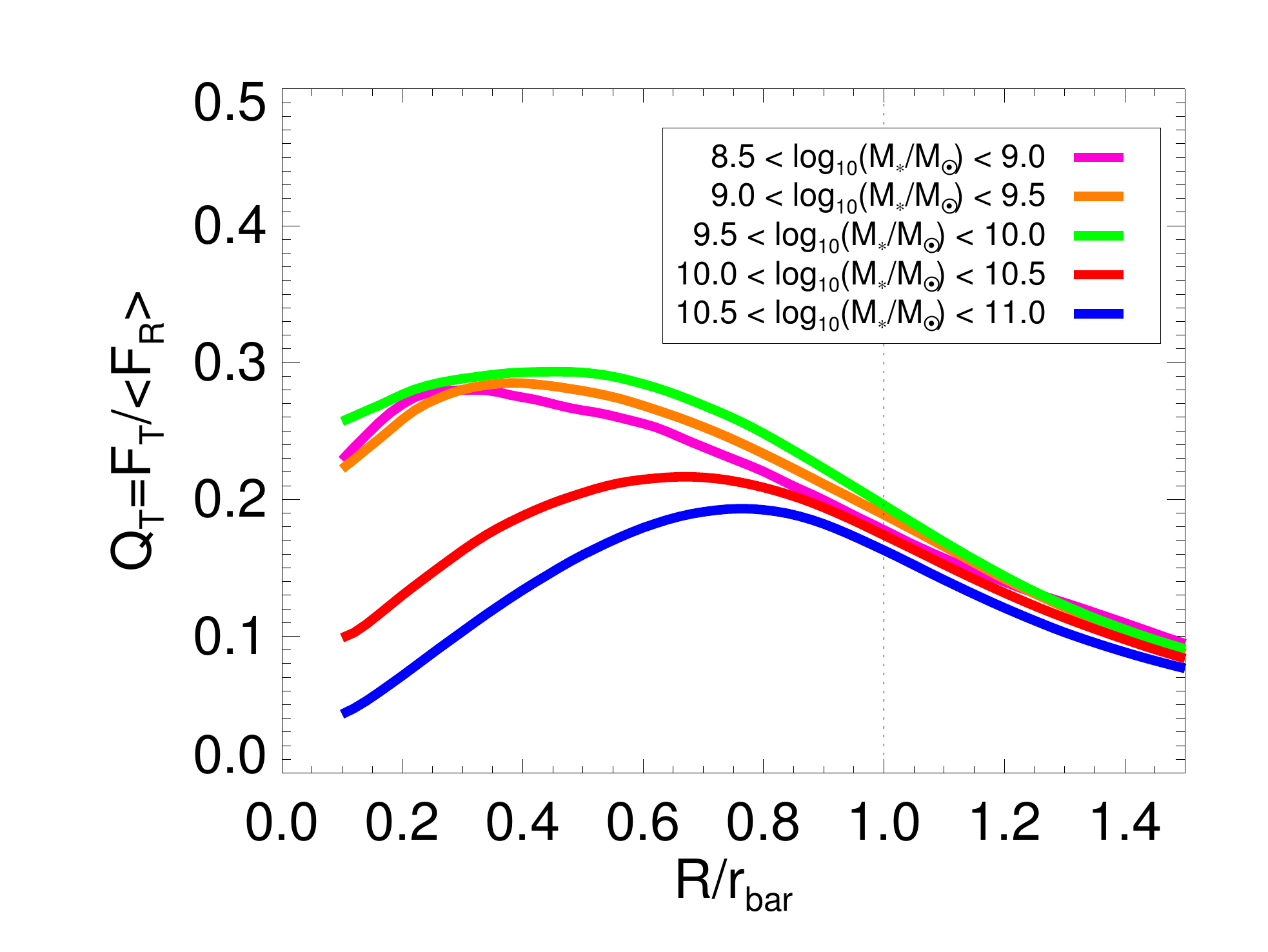}
     \includegraphics[width=0.5\textwidth]{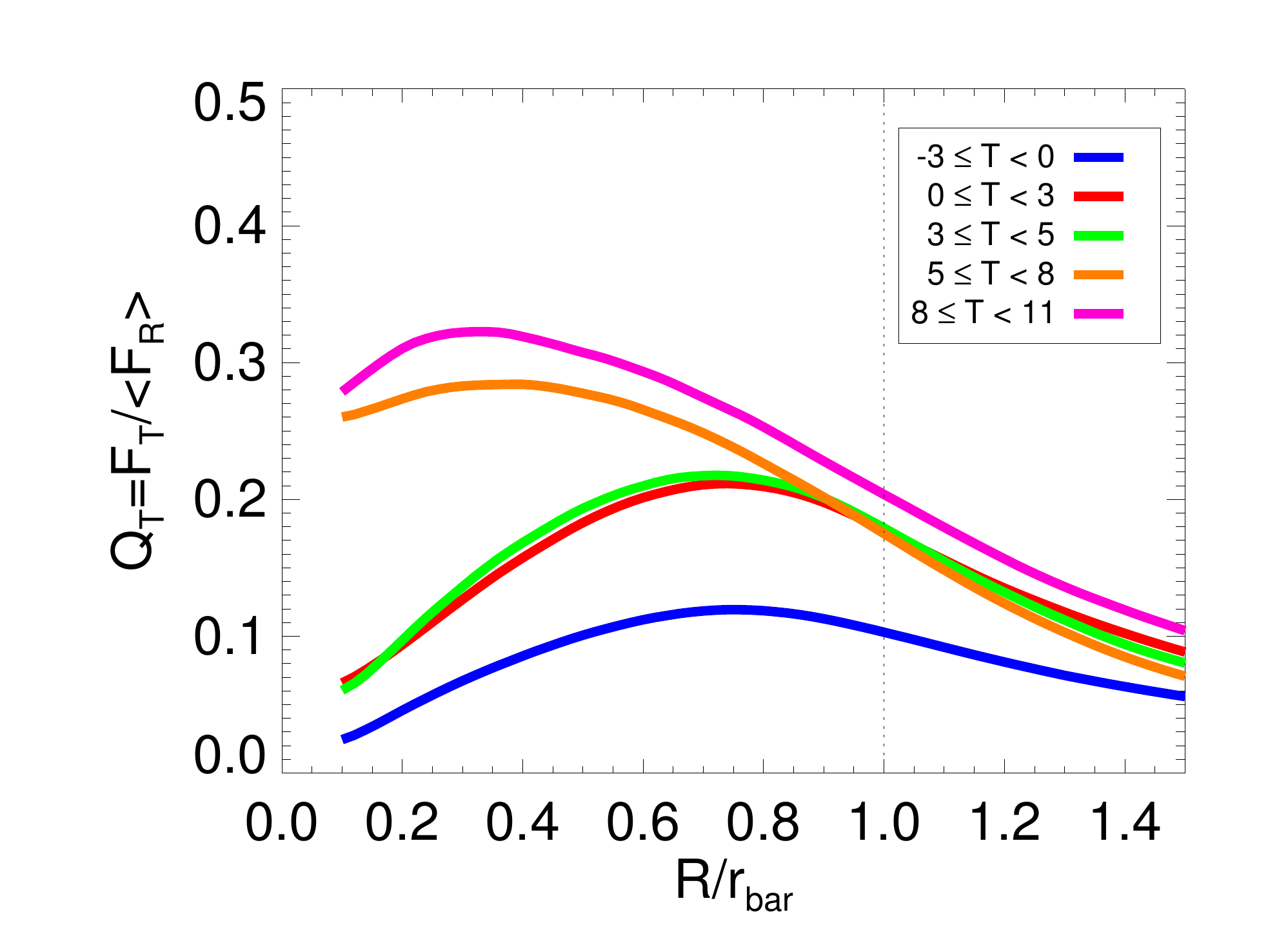}\\[-5ex]
     \includegraphics[width=0.5\textwidth]{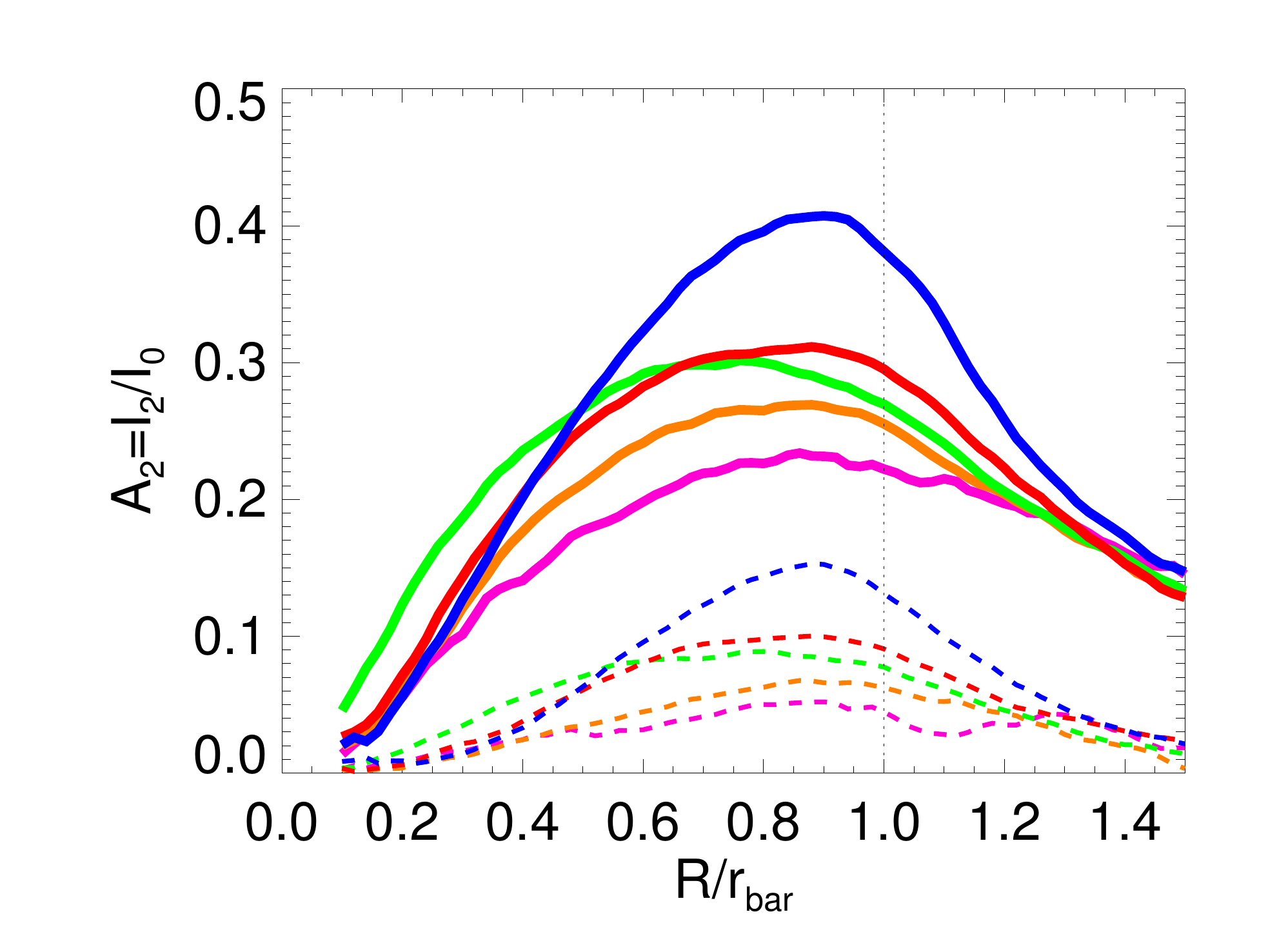}
     \includegraphics[width=0.5\textwidth]{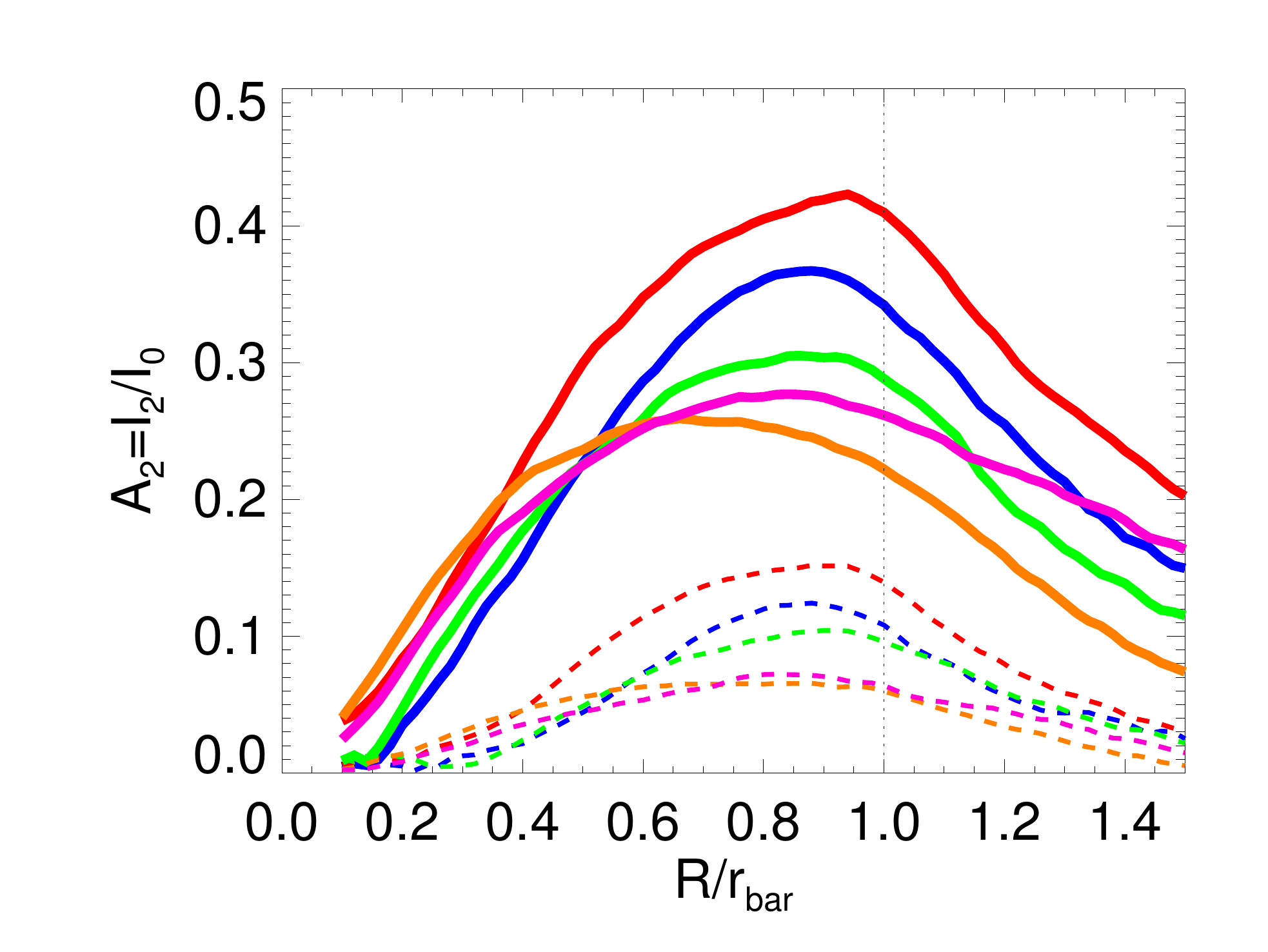}
\end{tabular}
\caption{
\emph{Upper row}: 
Tangential-to-radial force profiles associated with the average bars, binning the sample based on the total stellar mass (left) and the Hubble type (right) of the host galaxy. 
\emph{Lower row}: 
$m=2$ normalized Fourier amplitude of the average bars (solid lines) using the same sampling as in the upper panels. 
The dashed lines correspond to the $m=4$ normalized Fourier amplitudes. For all panels, the vertical dotted line corresponds to the end of the bar.
}
\label{qt_a2_mass_ttype_family}
\end{figure*}
\begin{figure*}
\centering   
\begin{tabular}{c c c c c c}
     \includegraphics[width=0.33\textwidth]{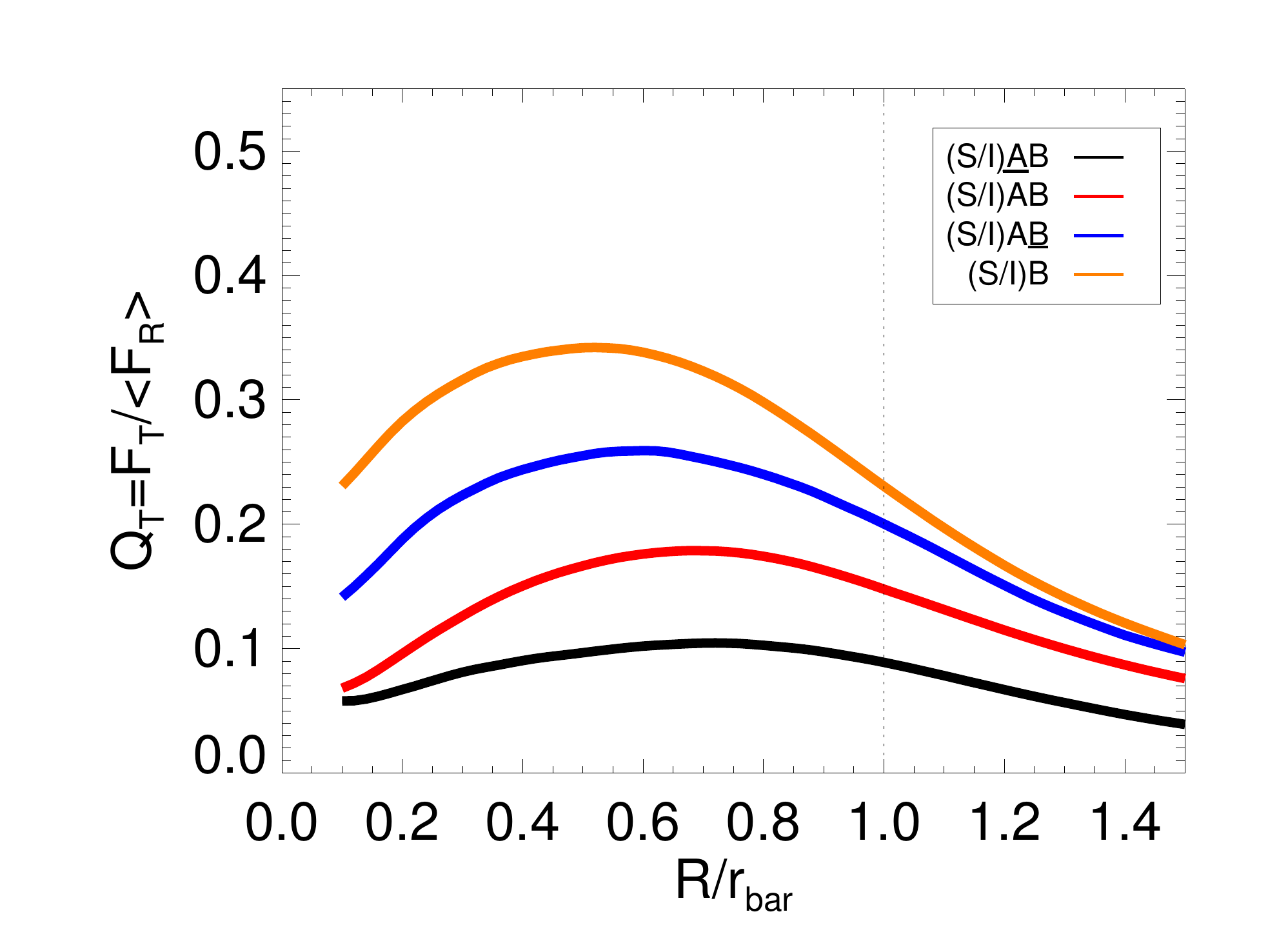}
     \includegraphics[width=0.33\textwidth]{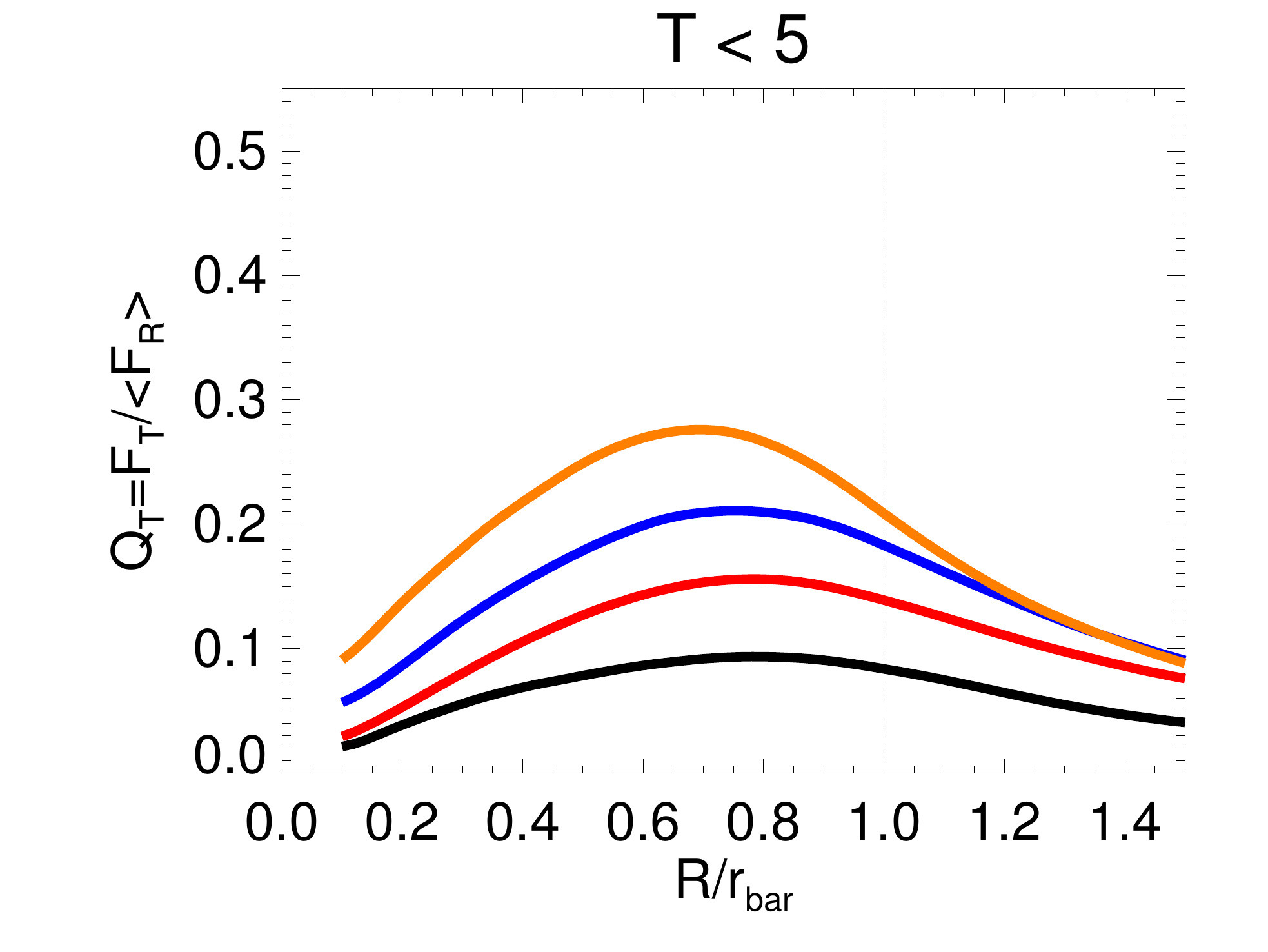}
     \includegraphics[width=0.33\textwidth]{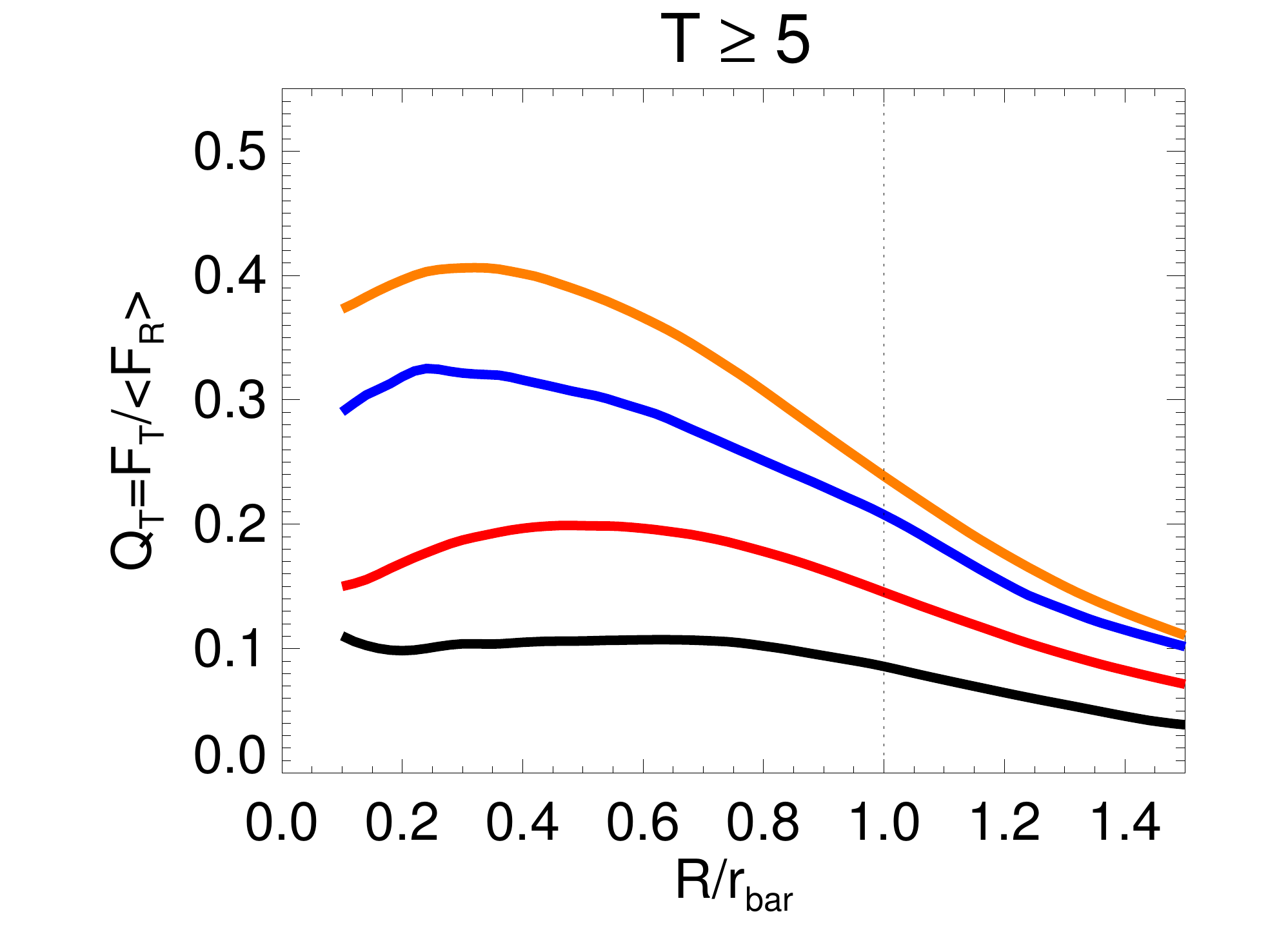}\\
     \includegraphics[width=0.33\textwidth]{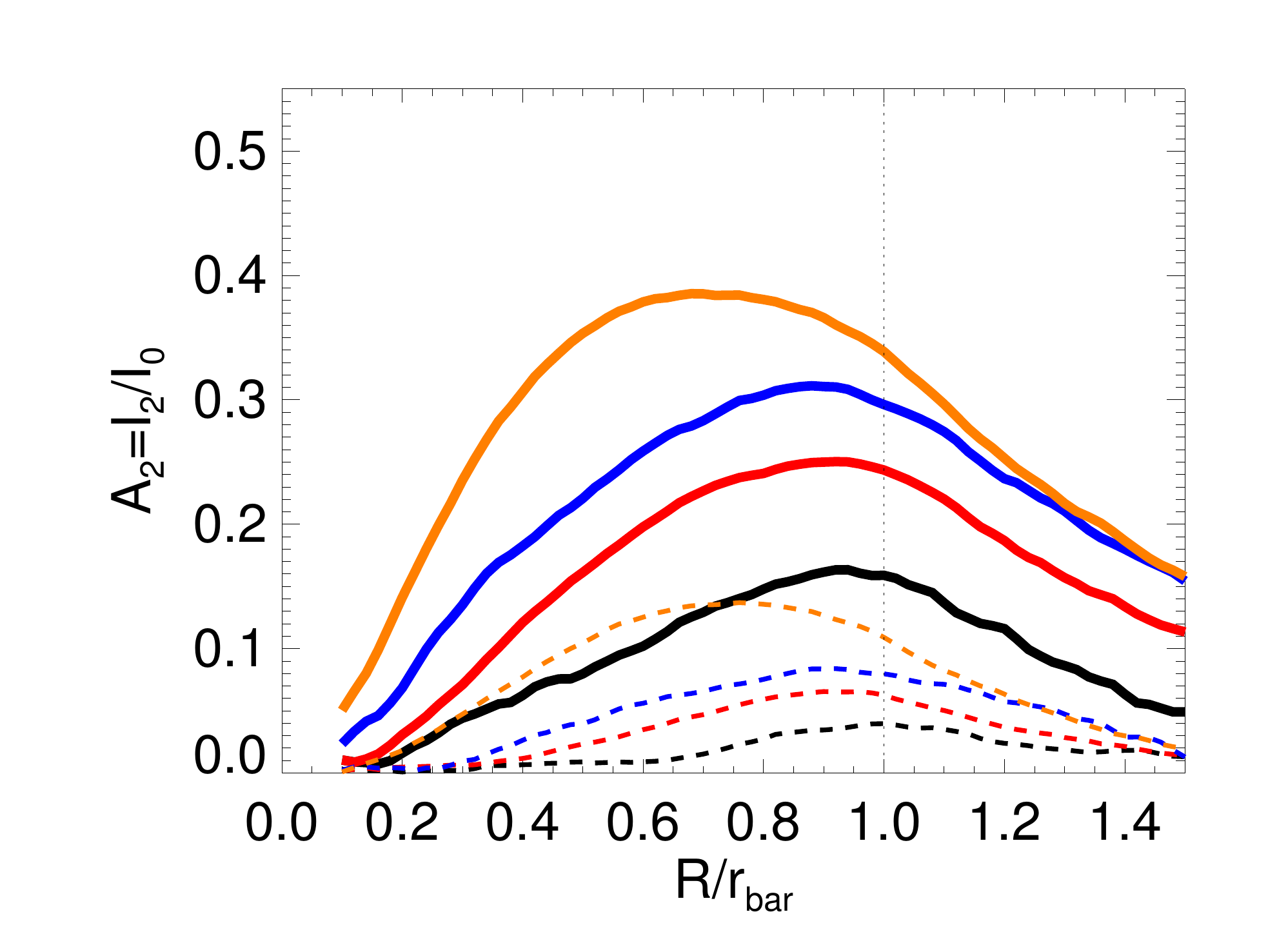}
     \includegraphics[width=0.33\textwidth]{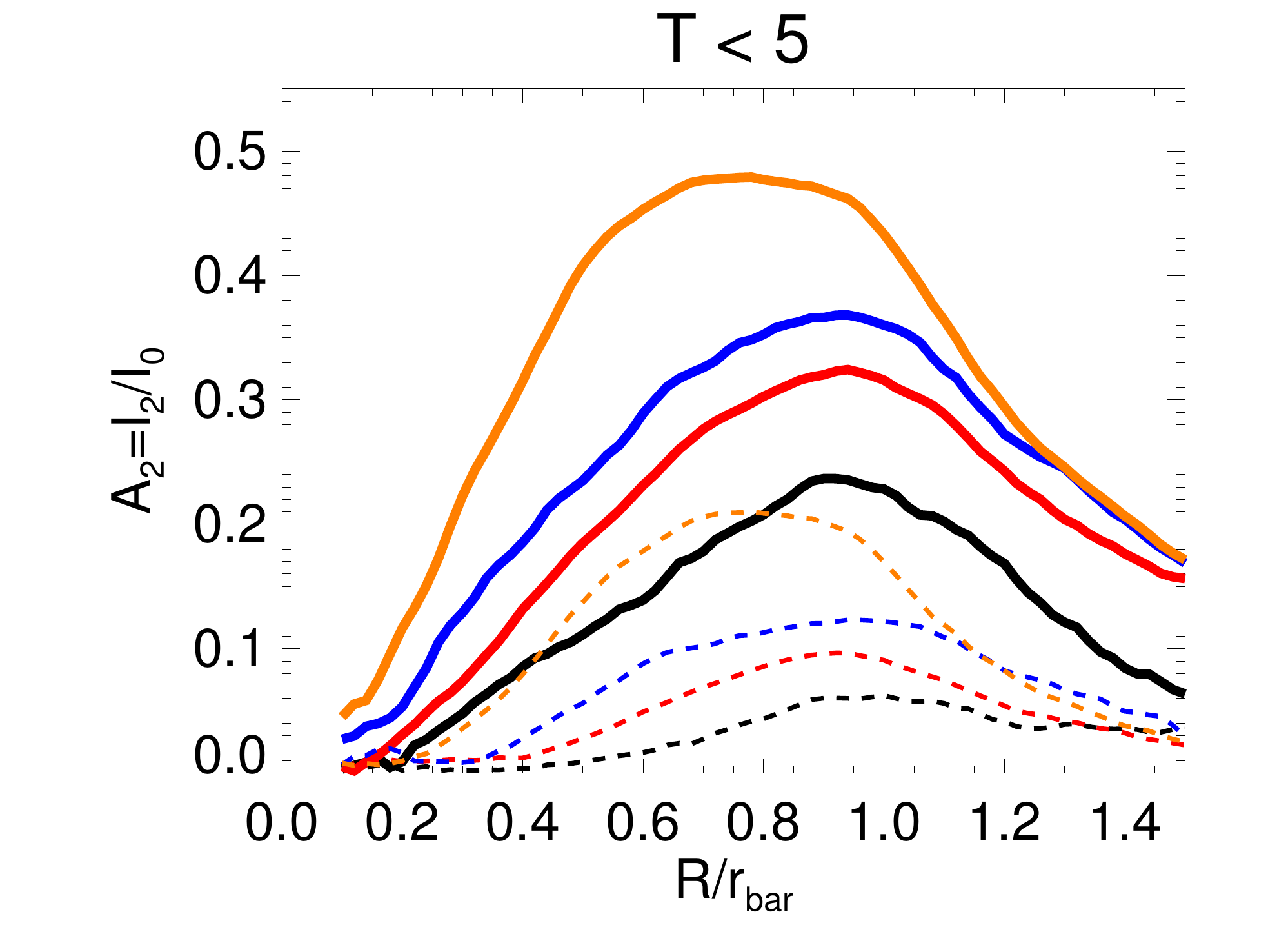}
     \includegraphics[width=0.33\textwidth]{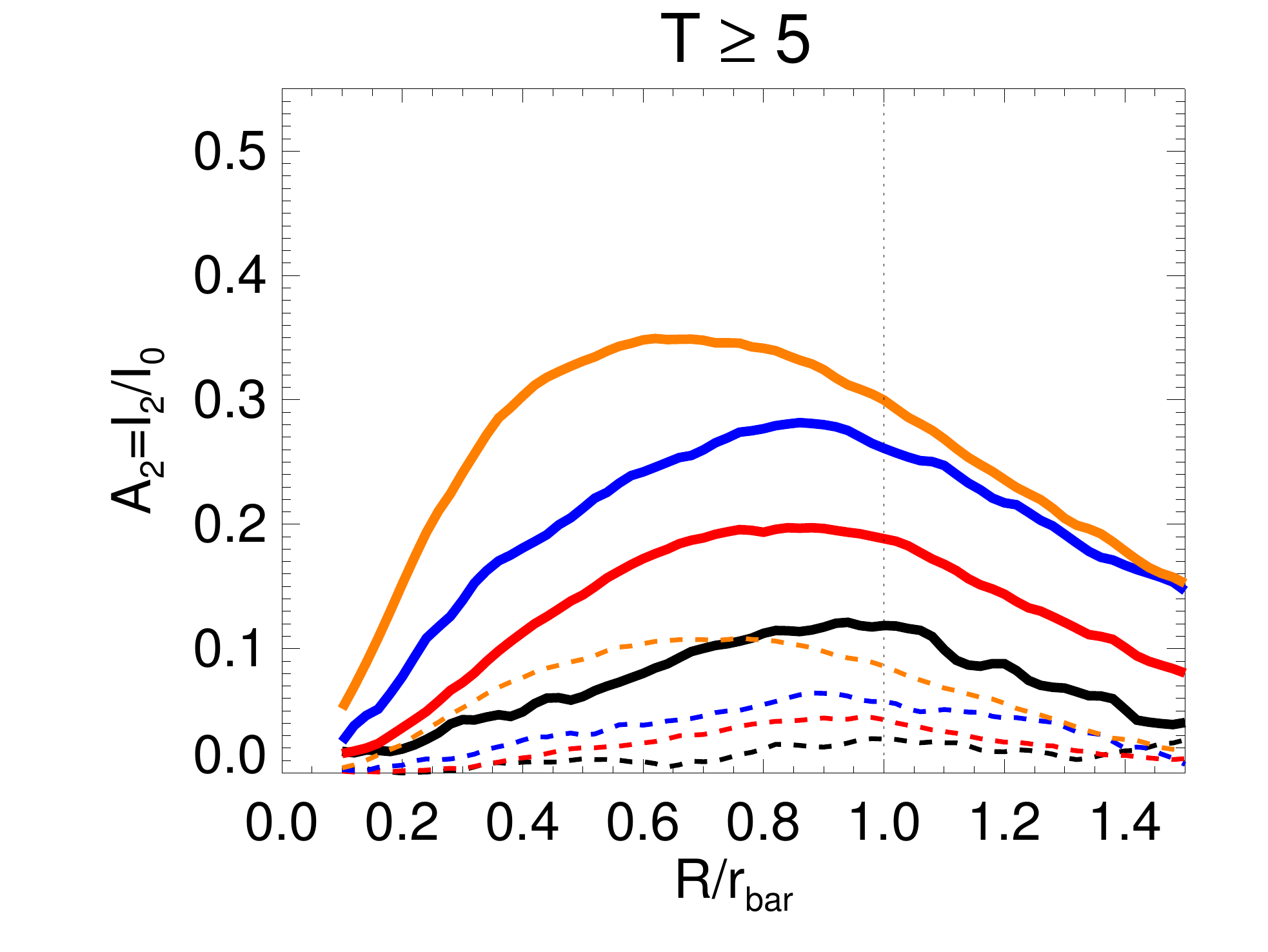}\\
\end{tabular}
\caption{
Radial force profiles (displayed as in Fig.~\ref{qt_a2_mass_ttype_family}) associated with the average bars as a function of the galaxy family. 
Early-type ($T<5$) and late-type ($T\ge5$) bars are shown separately (second and third columns, respectively).
}
\label{qt_a2_mass_ttype_family_early_late}
\end{figure*}
%
%
\begin{table*}
\centering
\tiny
\begin{tabular}{|| c|| c| c| c| c||}
\hline
\multicolumn{1}{||c}{\centering Bar parameters} & \multicolumn{1}{|c}{\centering $Q_{\rm b}$} & \multicolumn{1}{|c}{\centering $r_{\rm Qb}(\rm x \, r_{\rm bar})$} &
\multicolumn{1}{|c}{\centering $A_2$} & \multicolumn{1}{|c||}{\centering $r_{\rm A2}(\rm x \, r_{\rm bar})$} \tabularnewline
\hline
\hline
$8.5<\rm log_{10}M_{\ast}/M_{\odot}<9$&  0.28 (0.16)&  0.32 (0.23)&  0.23 (0.18)&  0.86 (0.22) \\
$9<\rm log_{10}M_{\ast}/M_{\odot}<9.5$&  0.29 (0.17)&  0.38 (0.21)&  0.27 (0.16)&  0.88 (0.20) \\
$9.5<\rm log_{10}M_{\ast}/M_{\odot}<10$&  0.29 (0.19)&  0.46 (0.24)&  0.30 (0.17)&  0.76 (0.32) \\
$10<\rm log_{10}M_{\ast}/M_{\odot}<10.5$&  0.22 (0.19)&  0.68 (0.24)&  0.31 (0.22)&  0.88 (0.23) \\
$10.5<\rm log_{10}M_{\ast}/M_{\odot}<11$&  0.19 (0.15)&  0.76 (0.20)&  0.41 (0.24)&  0.90 (0.20) \\
\hline
$-3\le \rm T<0$&  0.12 (0.08)&  0.76 (0.19)&  0.37 (0.21)&  0.88 (0.16) \\
$0\le \rm T<3$&  0.21 (0.14)&  0.74 (0.19)&  0.42 (0.24)&  0.94 (0.20) \\
$3\le \rm T<5$&  0.22 (0.13)&  0.72 (0.22)&  0.31 (0.20)&  0.86 (0.20) \\
$5\le \rm T<8$&  0.28 (0.20)&  0.40 (0.23)&  0.26 (0.19)&  0.66 (0.24) \\
$8\le \rm T<11$&  0.32 (0.16)&  0.32 (0.22)&  0.28 (0.16)&  0.84 (0.33) \\
\hline
(S/I)$\underline{\rm A}$B&  0.10 (0.13)&  0.72 (0.23)&  0.16 (0.14)&  0.94 (0.22) \\
(S/I)AB&  0.18 (0.13)&  0.68 (0.25)&  0.25 (0.21)&  0.92 (0.23) \\
(S/I)A$\underline{\rm B}$&  0.26 (0.14)&  0.60 (0.25)&  0.31 (0.17)&  0.88 (0.19) \\
(S/I)B&  0.34 (0.19)&  0.52 (0.25)&  0.39 (0.21)&  0.68 (0.25) \\
\hline
\multicolumn{1}{c}{$T<5$}\\
\hline
(S/I)$\underline{\rm A}$B&  0.09 (0.12)&  0.78 (0.22)&  0.24 (0.15)&  0.90 (0.23) \\
(S/I)AB&  0.16 (0.10)&  0.78 (0.21)&  0.32 (0.23)&  0.94 (0.20) \\
(S/I)A$\underline{\rm B}$&  0.21 (0.11)&  0.76 (0.20)&  0.37 (0.22)&  0.94 (0.17) \\
(S/I)B&  0.28 (0.13)&  0.70 (0.20)&  0.48 (0.23)&  0.78 (0.19) \\
\hline
\multicolumn{1}{c}{$T\ge5$}\\
\hline
(S/I)$\underline{\rm A}$B&  0.11 (0.13)&  0.62 (0.23)&  0.12 (0.05)&  0.94 (0.22) \\
(S/I)AB&  0.20 (0.14)&  0.48 (0.25)&  0.20 (0.13)&  0.84 (0.25) \\
(S/I)A$\underline{\rm B}$&  0.33 (0.12)&  0.24 (0.20)&  0.28 (0.12)&  0.86 (0.20) \\
(S/I)B&  0.41 (0.18)&  0.32 (0.22)&  0.35 (0.17)&  0.62 (0.29) \\
\hline
\end{tabular}
\caption{
Bar maximum gravitational torque ($Q_{\rm b}$) and normalized m=2 Fourier amplitude ($A_2^{\rm max}$), and radii ($r_{\rm Qb}$ and $r_{A2}$, relative to the bar length) for the different $M_{\ast}$-, T-, 
and galaxy family-dependent bins used for the bar stacks. In parentheses, we indicate the dispersion of the parameters based on the measurements on individual galaxies from \citet[][]{2016A&A...587A.160D}.
}
\label{bars_params}
\end{table*}

%
%
\subsection{ Luminosity profiles of average bars}\label{bars-lum}
%
%
In Fig.~\ref{stack_bars_altogether} we display the surface brightness profiles associated with the stacked images discussed above.
We computed the azimuthally averaged $\mu_{3.6\mu \rm m}$ profiles, and also the surface brightness along the bar major axis.
Only the most massive systems ($M_{\ast}\ge10^{10}M_{\odot}$) show a hint of flat bars when looking at the azimuthally averaged $\Sigma_{\ast}$ (left panel).
However, the bar prominence is easily identifiable when taking $T$-type bins and focusing on early- and intermediate-type spirals (right panel), 
which are the systems showing the highest bar-to-total flux ratio. The profile of S0s shows no trace of the outer part of the bar in the azimuthally averaged $\Sigma_{\ast}$. 
When looking at the cuts along the bar major axis, all systems show a hump in the bar end.
%
%
\subsection{Strength of average bars}\label{forces_sect}
%
%
From the bar stacks we also calculate the bar-induced perturbation strength, which is estimated from tangential-to-radial force profiles ($Q_{\rm T}$) and the $m=2$ Fourier amplitudes ($A_{2}$). 
For the potential calculation from the 2-D bar stack, the scaleheight is assumed to be the median $h_{\rm z}$ of the galaxies within the bin. 
Bar force parameters characterizing the stacks are listed in Table~\ref{bars_params}. 

As displayed in Fig.~\ref{qt_a2_mass_ttype_family}, late-type ($T\ge5$) and moderately faint systems ($M_{\ast}<10^{10}M_{\odot}$) typically have larger $Q_{\rm T}$ profiles in the bar region, 
whose maxima are reached close to the centre of the galaxy. 
As shown in Sect.~\ref{bars-lum}, these galaxies are almost bulge-less, therefore the radial force field in the inner parts is weaker, causing the tangential-to-radial force ratio to increase. 
However, early-type galaxies are characterized by lower gravitational torques as they are affected by the so-called bulge dilution \citep[][]{2004ApJ...607..103L}: 
central concentrations of mass enhance the radial force in the bar region, making the loci of maximum torque  move outwards and  diminish.

For the bar amplitudes, which are less affected by the prominence of the central components, $A_{2}^{\rm max}$ increases with increasing total mass and drops for intermediate- and late-type bar stacks ($T \ge 3$). 
The same trend is found with higher order components ($A_4$), which have much lower amplitudes.

In Fig.~\ref{qt_a2_mass_ttype_family_early_late}, we show the $A_{2}$ and $Q_{\rm T}$ profiles of the bar stacks, sampled based on the family of the galaxies. 
The amplitudes of these profiles increase from S$\underline{\rm A}$B towards SB galaxies (likewise the maximum $A_4$), in agreement with the analysis in DG2016. 
As shown in the same figure, the differences in the force profiles, depending on the morphological family, are qualitatively the same when splitting the sample into early-type ($T<5$) and late-type ($T\ge5$) systems.
%
%
%
\section{Bar properties in the Hubble sequence}\label{bar-hubble}
%
%
%
Based on the analysis of bar stacks, in this work we confirm many of the trends of the bar parameters in the Hubble sequence previously reported in the literature. 
Bars in early-type galaxies (and in more massive systems) typically show larger $m=2$ (and $m=4$) Fourier density amplitudes \citep[e.g.][]{2007MNRAS.381..401L} 
and lower gravitational torques \citep[e.g.][]{2004MNRAS.355.1251L,2010ApJ...721..259B} than in their later-type counterparts. 
The discrepancy in the statistical behaviour of these  bar strength estimates was discussed in \citet[][]{2004ApJ...607..103L} and DG2016, and attributed to the force dilution by bulges, causing the radius of maximum torque to move outwards, 
and to the weak underlying disks that is characteristic of late-type systems. 
Recent work by \citet[][]{2015MNRAS.451..936S} found that the kinematic perturbation related to the bar tightly correlates with the bar-induced tangential-to-radial forces derived from the gravitational potential, 
including cases with prominent bulges in the analysis. With regard to their form, bars are oval-shaped among lenticular galaxies, 
and look intrinsically narrower in early- and intermediate-type spirals ($0\le T < 5$) than in later types. 
However, among early-types the bar shape is rounded off by bulges and barlenses in the inner parts. 

In order to have a more quantitative analysis of the shape of bars of different masses and morphological types, we carried out ellipse fitting to the bar stacks and studied the radial profiles of ellipticity and $b_4$ parameter. 
The ellipticity profiles in barred galaxies typically show a monotonic growth until a maximum/plateau is reached at the end of the bar, after what it declines in the outer disk \citep[consistent with][]{1995AJ....109.2428M}. 
The contribution of the bulge and disk axisymmetric light distribution is known to diminish the bar ellipticity \citep[e.g.][]{1990MNRAS.245..130A,2008MNRAS.384..420G}, which in principle can be avoided by eliminating the bulge 
before measuring the isophotes of the bar. However, doing so is not trivial because in many cases the barlens would be removed, while it actually forms part of the bar \citep[][]{2007MNRAS.381..401L,2011MNRAS.418.1452L,2014MNRAS.444L..80L}. 
\citet[][]{1990MNRAS.245..130A} concluded that the bar isophotes of early-type strongly barred galaxies are rectangular. 
In addition, N-body simulations and observational studies indicate that the ellipticity profiles of barred galaxies peak somewhat before the end of the bar \citep[e.g.][]{1999A&A...348..737R,2002MNRAS.330...35A,1991A&AS...88..325W,2002MNRAS.337.1118L}.

S0s  show the smallest $\epsilon$ value in the bar stack ($\epsilon\approx0.35$), consistent with the drop in ellipticity among the earliest types of lenticulars reported by \citet[][]{2007MNRAS.381..401L}. 
All the spiral and irregular galaxies show on average similar bar ellipticities ($\epsilon \approx 0.5$), consistent with the analysis in \citet[][]{2007ApJ...659.1176M}. 
We show that the radii where the maximum ellipticities are reached lie very close to the bar end for all galaxies with stellar masses greater than $10^{10}M_{\odot}$. 
However, $r_{\epsilon}\approx0.6-0.8r_{\rm bar}$ when $10^{9}<M_{\ast}/M_{\odot}<10^{10}$, and even $r_{\epsilon}\approx0.2r_{\rm bar}$ when $M_{\ast}/M_{\odot}<10^{9}M_{\odot}$. 
In this last case the $M_{\ast}$-bins are likely to consist of late-type galaxies whose bars were not identified in DG2016 via ellipse fitting. 

The parameters obtained from bar stacks are consistent with the observations in \citet[][]{2007MNRAS.381..943G}, 
where the behaviour of the ellipticity and $b_4$ radial profiles were analysed for a sample of nine bright barred galaxies using J- and K$_{\rm s}$-band imaging. 
Their $b_4$ profiles generally had values close to zero in the nuclear parts, an enhancement due to the bar, and a fall-out after the radius of maximum ellipticity, $b_4$ eventually becoming negative. 
Likewise, in our bar stacks comprising early- and intermediate-type spirals, the $b_4$ show a similar bar hump and drop after $r_{\epsilon}$, turning negative slightly beyond the end of the bar. 
S0s are the only systems not showing negative $b_4$ values at any radii. 
For the late-type spirals, the $b_4$ profiles also decline at the bar end without any pronounced central increase (likely explained by their smaller central mass concentration). 
In the bin with the lowest stellar mass ($M_{\ast}<10^{9}M_{\odot}$), it is much harder to distinguish any well-defined radial pattern for $b_4$ associated with the bar stack. 

\citet[][]{1995AJ....109.2428M} and \citet[][]{2005MNRAS.364..283E} showed that bars in galaxies classified as SB and SAB differ in ellipticity. 
In our study the amplitude of $\epsilon$ also depends on the galaxy family, that is, the bar isophotes are more elliptical for SB and $\rm SA\underline{B}$ galaxies than for SAB and $\rm S\underline{A}B$ systems. 
Other quantitative estimates of the strength of the bar stacks (tangential-to-radial forces, $m=2$ Fourier amplitudes) also showed correlation with family. 

We found that the maxima of $b_4$ tend to be larger for SB galaxies, more clearly when $T<5$, implying more disky inner isophotes for stronger bars. 
The $b_4$ maxima systematically appear at $\sim 0.75r_{\rm bar}$. The depth of the $b_4$ minima was also found to be larger for SB galaxies, perhaps giving a hint of the link between the strength and boxiness of the bars. 
However, this minimum typically appears slightly beyond the mean bar end and therefore it may arise from the superposition of a bar and the spiral arm segments. 
In fact, in many of the radial profiles presented by \citet[][]{2007MNRAS.381..943G} the $b_4$ minimum is actually coincident with the $\epsilon$ minimum rather than the $\epsilon$ maximum. 

\citet[][]{1985ApJ...288..438E} found that bars in early-type galaxies are characterized by flat azimuthally averaged density profiles, i.e. 
the surface brightness is constant or slowly declining in the bar region and drops along the outer disk, 
whereas the surface brightness in later types decreases exponentially in the bar region at the same pace as in the rest of the disk. 
In Sect.~\ref{bars-lum} we confirmed that the bar stacks corresponding to early- and intermediate-type spirals ($0 \le T < 5$) are indeed rather \emph{flat}, while later type bars are \emph{exponential}. 
In the stack made out of bars hosted by lenticulars, we did not find any feature associated with the bar in the azimuthally averaged profiles. 
The bar hump is distinguishable in the cut along the bar major axis for all the bins. 
Our results are also consistent with \citet[][]{2015ApJ...799...99K}, who showed that bars in massive galaxies with bulges have a flat profile, whereas bars in less massive disk-dominated systems are characterized by exponential profiles.

Furthermore, in Sect.~\ref{a2_medi} we computed the mean $A_{2}$ profiles from the Fourier decomposition of individual galaxies carried out in DG2016. 
We confirmed that the amplitude of the $A_{2}$ hump associated with the bar depends on the family of the host galaxies. 
Interestingly, we found that among early- and intermediate-type spirals ($0 \le T < 5$), or when $M_{\ast}>10^{9.5}M_{\odot}$, the amplitude of the spiral arms (i.e. $A_2$ values beyond the typical bar region) is larger for barred systems. 
This  might be connected to the correlation between the local bar forcing and the local spiral amplitude reported in \citet[][]{2010ApJ...715L..56S}, presented as evidence for the role of bars driving spiral density waves. 
This interpretation is supported by the dependence of the spirals amplitude on the galaxy family, that is, larger $A_2$ amplitudes in the outer disk appear for $\rm SA\underline{B}$+SB systems than for $\rm S\underline{A}$B+SAB galaxies.

As a result of the angular momentum exchange, dynamical models predict that stellar bars become longer, narrower, and stronger and slow down their 
rotation speed within a Hubble time \citep[e.g.][]{1991MNRAS.250..161L,1998ApJ...493L...5D,2002MNRAS.330...35A}. 
Such evolution was addressed observationally in \citet[][]{2007MNRAS.381..401L} and DG2016, who found that early-type galaxies (more evolved systems) show on average larger density amplitudes and bar lengths (normalized to the disk size). 
Possible direct confirmation for the bar growth within a Hubble time is also provided in \citet[][]{2007ApJ...670L..97E}, \citet[][]{2011MNRAS.415.3308G}, and DG2016, 
who found a dependence between the sizes of bars and different estimates of their strengths. 
In this work we confirmed that bars hosted by early- and intermediate-type spirals are characterized by a larger m=2 Fourier density amplitudes and intrinsically narrower bars. 
In summary, our observations are consistent with a scenario in which late-type spirals move in the Hubble sequence towards earlier-types \citep[][and references therein]{2013seg..book....1K} 
as they deplenish their gas and their bars trap particles from the disk and become narrower and longer.

Based on the distinct bar properties, two types of bars were reported in DG2016, those hosted by early- and intermediate-type galaxies ($T<5$), and those hosted by later type systems. 
This twofold nature of bars was argued to be linked to the higher halo-to-stellar mass ratio within the optical radius 
found among galaxies with morphological types later than Sc, which is known to affect the disk stability properties. 
This duality is also manifested in the differences in bar shapes and density amplitudes of the stacks discussed in this work. 
%
%
%
\section{The assembly of galactic disks}\label{disk-assemb}
%
%
%
\citet[][]{2013ApJ...771L..35V} concluded that although galaxies with present-day stellar masses similar to that of the Milky Way built $\sim90\%$ of their stellar mass since $z = 2.5$,
most of the star formation took place already before $z = 1$, 
at which time these systems typically have their structures assembled to a large extent. 

Using $u$- and $g$-band imaging from SDSS \citep[][]{2006AJ....131.2332G}, 3D-HST \citep[][]{2012ApJS..200...13B}, and CANDELS \citep[][]{2011ApJS..197...36K} surveys, 
\citet[][]{2013ApJ...771L..35V} binned their sample based on galaxy redshift $0 \le z \le 2.5$. They obtained averaged density profiles by applying the stacking technique from \citet[][]{2010ApJ...709.1018V}, 
which yields very similar profiles at $z\approx0$ to those of  our method  (shown in Sect.~\ref{coaddbars}). 
They showed that the mass growth of the MW-type systems was not limited to large radii (i.e. bulges seem to grow at the same time as the rest of the disk) 
and that their buildup could be fully explained by the SFRs of the galaxies without invoking mergers. 
This track for the galaxy assembly seems to be in contrast to the behaviour of more massive galaxies 
($M_{\ast}\approx3\cdot10^{11}M_{\odot}$ today, comprising elliptical galaxies and massive S0s), 
which are characterized by inside-out growth according to \citet[][]{2010ApJ...709.1018V}. 
The central regions of these massive galaxies were formed earlier than $z=2$, having a factor of $\sim$~3 increase in stellar mass between $z=3$ and $z=0$ \citep[][]{2013ApJ...766...15P}, which cannot be uniquely explained by star formation.

Since the early studies by \citet[][]{1970ApJ...160..811F}, it has been known that spiral galaxies have disks whose surface brightness tends to decay exponentially as a function of galactocentric radius. 
The whole picture increased in complexity with the advent of deeper surveys in optical wavelengths, which revealed breaks in the disk luminosity profiles \citep[][]{2005ApJ...626L..81E,2006A&A...454..759P,2008AJ....135...20E}. 
Disks with an exponential decay in luminosity and truncations (double-exponentials) had been observed in the local universe \citep[e.g.][]{2002A&A...392..807P} and at redshifts $0.6 < z < 1.0$ \citep[][]{2004A&A...427L..17P}. 
Using 3.6~$\mu$m imaging, \citet[][]{2014MNRAS.441.1992L} shed light on the nature of disk breaks by linking the disk profile types to structural components of galaxies such as rings, lenses, and spirals (type II), 
and to tidal interactions (type III). \citet[][]{2013ApJ...771...59M} associated the loci of breaks with the resonance radii of bars and the dynamical coupling between the bar and the spiral pattern. 
Recent work by \citet[][]{2016MNRAS.456L..35R} indicates that the stellar population distributions in disk galaxies are decoupled from the shape of the surface brightness profiles.

In Sect.~\ref{disk_kpc} we showed that the mean $\Sigma_{\ast}$ profiles have larger disk scalelengths ($h_{\rm R}$) and fainter extrapolated central surface brightnesses ($\mu_{\circ}$) for the bins with larger $M_{\ast}$. 
The roughly parallel trends (in log-log plot) reported by \citet[][]{2013ApJ...771L..35V} for the surface brightnesses of the Milky Way progenitors at different redshifts (see profiles in their Fig.~3)
resemble the tendency of our $M_{\ast}$-binned averaged stellar density profiles, especially beyond $\sim$~2 kpc. 
Nevertheless, without the aid of simulation models in a cosmological framework that enable comparisons with our stellar profiles at $z\approx0$, 
any conclusion for the large-scale assembly of stellar mass in a cosmic time exclusively based on our observations would be overly speculative. We encourage such comparison to be done elsewhere. 
Together with mean profiles, we also provide dispersion measurements. Fundamental physics determining the formation of disk galaxies \citep[see e.g.][]{1997ApJ...482..659D} 
are encoded in these dispersion measurements (e.g. distribution of the spin parameter, acquisition of angular momentum by baryons, or secular evolution). 
Thus, the scatter itself, which is typically larger among non-barred systems, may also be a valuable tool for the comparison with models. 
We also consider that the average central mass concentration as a function of $M_{\ast}$ obtained in this work, which is independent of any decomposition method, might also be a strong constraint for galaxy formation models.

We also showed that systems with $M_{\ast}>10^{10}M_{\odot}$ had steeper mean $\Sigma_{\ast}$ in the central parts. 
The prominence of this central mass concentration among the more massive systems was found to scale with the Hubble type of the host galaxy. 
It mostly results from the combined contribution to the mean $\Sigma_{\ast}$ by different types of bulges \citep[classical, disky pseudobulges, and boxy/peanut bulges in the notation of][]{2005MNRAS.358.1477A}. 
To some degree, our central mass concentrations are produced by central stellar structures such as inner and nuclear rings, spirals, or bars. 
We show that for galaxies with Hubble types $T\ge8$ and stellar masses $M_{\ast}<10^{9} M_{\odot}$ the average $\Sigma_{\ast}$ is bulge-less, 
while intermediate systems with $10^{9} \le M_{\ast}<10^{10} M_{\odot}$ and $5 \le T < 8$ are characterized by fairly small central mass concentrations. 
Our results are consistent with the analysis of \citet[][]{2015ApJS..219....4S}, where for Hubble types $T=5-7$ the bulges in disk galaxies gradually disappear and for the latest types most of the systems are pure disks. 
This is expected because for the stacked profiles we used the same galaxy images as in \citet[][]{2015ApJS..219....4S}.

In Sect.~\ref{disk_kpc} and Sect.~\ref{ttypesprofs} we also showed that for all $M_{\ast}$- and $T$-bins, the mean $\Sigma_{\ast}$ follow an exponential slope down to at least $\sim 10M_{\odot}/$pc$^{2}$. 
Beyond this depth, the sample coverage in the radial direction compromises the robustness of our statistics (e.g. possible appearance of artificial up-bending profiles), as explained in Sect.~\ref{coaddbars}. 
Because our methodology inhibits us from extending the analysis to fainter regions of the average $\Sigma_{\ast}$, in this work we do not study disk breaks. 
In addition, the reported average exponential trends must not lead to the conclusion that, typically, the disks in our sample do not present breaks 
as up-bending and down-bending sections of the binned profiles may cancel out in the final mean $\Sigma_{\ast}$. 

Although there are  several proposed theories \citep[e.g.][]{1970ApJ...160..811F,1980MNRAS.193..189F,1987MNRAS.225..607L,1991A&A...252...75P}, the way in which exponential disks assemble is still unclear. 
In contrast to early cosmological simulations that failed to produce bulge-less exponential disks \citep[e.g.][]{2001MNRAS.327.1334V}, 
\citet[][]{2009MNRAS.396..121D} was successful at modelling exponential or quasi-exponential stellar disks in a $\Lambda$CDM framework
over a wide range of total stellar masses without any {a priori} assumption about the disk shape. 
This model followed the accretion, cooling, and ejection of baryonic mass within dark matter halos; the exponential disks resulted from the combined effect of supernovae driven galactic outflows, 
variations in the distributions of the specific angular momentum of the baryons, and the star formation inefficiency at large radii. 
Exponential disks had been produced earlier in models with star formation feedback \citep[e.g.][]{2003ApJ...597...21A,2004ApJ...606...32R}. 
Using galaxy imaging in the \emph{Hubble Space Telescope} Ultra Deep Field \citep[UDF;][]{2006AJ....132.1729B}, 
\citet[][]{2005ApJ...634..101E} showed that blue star-forming clumps in high-redshift galaxies appear to be distributed following an exponential profile; 
these clumps, formed by gravitational instabilities, are candidates to disperse into the  exponential uniform disks that we observe today. 
In addition, models by \citet[][]{2008ApJ...688...67E} have shown that bulges (with high Sérsic indexes) can form by the coalescence of those clumps that resisted the SF disruption, lost angular momentum, and migrated to the galaxy centre. 
Classical bulges, which look indistinguishable from elliptical galaxies \citep[][]{1994cag..book.....S},
have  traditionally been thought to form through mergers \citep[][]{1977egsp.conf..401T} followed by a rapid relaxation \citep[for a review, see][]{2016ASSL..418..317B}, which is the way ellipticals possibly formed \citep[][]{1967MNRAS.136..101L}. 
On the other hand, pseudobulges result from the secular evolution of disks, for instance via bar instabilities \citep[see][and the discussion in   Sect.~\ref{disk-hubble}]{2004ARA&A..42..603K}.
Recent cosmological simulations of galaxy formation by \citet[][]{2012MNRAS.423.1544S} have shown that the accretion of cold gas from filaments,
combined with a considerable misalignment in the spin of the galaxy components can also result in the formation of slowly rotating spheroidal central masses.

In Sect.~\ref{rotcurs} we provided the mean stellar contribution to the circular velocity as inferred from near-IR wavelengths
whose maximum amplitude and inner slope was found to scale with the total stellar mass and bulge prominence of the galaxies.
These curves are an additional constraint for testing galaxy formation models,
and may contribute to more accurate estimates of the amount of dark matter that is present in the inner parts of present-day galaxies based on kinematic data. 
Combining the disk rotation curves inferred from 3.6~$\mu$m photometry with H{\sc\,i} velocity measurements from the literature, DG2016 obtained a first-order estimate of the halo-to-stellar mass ratio ($M_{\rm halo}/M_{\ast}$) within the optical radius. 
Systems with $T\gtrsim 5$ or $M_{\ast}\lesssim10^{10} M_{\odot}$ were found to be more dark matter dominated inside the optical disk than the early-type massive galaxies \citep[see also][]{2015ApJ...801L..20C}. 
Under the assumption that the halo within the optical disk contributes approximately a constant fraction of the total halo mass ($\sim4\%$), 
we found that the trend of the $M_{\rm halo}/M_{\ast}$-$M_{\ast}$ relation agreed with models at $z\approx0$ in $\Lambda$CDM cosmological simulations fitted based on abundance matching and halo occupation distribution methods, 
with constraints from weak lensing analysis \citep[e.g.][]{2010ApJ...710..903M,2010ApJ...717..379B,2010MNRAS.404.1111G,2012ApJ...746...95L}. 
A possible direction for studies of galaxy formation motivated by this result is to obtain the stellar-to-halo mass ratio within the central parts of galaxies in cosmological simulations. 
Another interest scaling relation showing the coupling between the baryonic and dark matter in the central parts of disk galaxies was presented by \citet[][]{2013MNRAS.433L..30L}, 
who found a linear correlation between the inner gradient of the circular velocity curve (i.e. inner shape of the potential well) and the central surface brightness. 
This relation was confirmed with S$^4$G galaxies by \citet[][]{2016MNRAS.458.1199E} using $H_{\alpha}$ kinematics from \citet[][]{2015MNRAS.451.1004E}. 
%
%
%
\section{Role of bars in the evolution of galactic disks}\label{disk-hubble}
%
%
%
The effect of bars in the evolution of their host galaxies is manifested in the redistribution of gaseous and stellar material within the disk. 
Specifically, the material beyond the bar corotation is pushed outwards, while the material within the CR 
is pushed inwards because of the torques exerted by the bar \citep[][and references therein]{2013seg..book..305A}.

The gas funneled by bars towards the central regions of the galaxies might eventually be used in starbursts when it reaches high densities; 
in fact, many barred galaxies have been observed to have dense central concentrations of gas and star formation \citep[][and references therein]{2013seg..book....1K}. 
This gas can cause the growth of the inner disk \citep[e.g.][]{1992MNRAS.259..328A,1992MNRAS.258...82W,1993A&A...268...65F,2004ARA&A..42..603K,2005MNRAS.358.1477A}, 
and perhaps fuel the active galactic nucleus \citep[e.g.][]{1989Natur.338...45S} \citep[but see][]{2013ApJ...776...50C}. 

Based on the analysis of disk stacks normalized to $h_{\rm R}$, in Sect.~\ref{ttypesprofs} we showed that both early- and late-type systems have larger central mass concentrations when a stellar bar is present, confirming the above listed expectations. 
Furthermore, slightly higher (lower) concentrations were detected for strongly (weakly) barred galaxies. 
As discussed in Sect.~\ref{bar-hubble}, this could be explained by the higher efficiency of strong bars fueling matter towards the central regions and perhaps also by the larger prominence of barlenses in strongly barred systems. 
This is also consistent with bars being robust structures \citep[e.g.][]{2004ApJ...604..614S,2010ApJ...719.1470V,2013MNRAS.429.1949A,2015A&A...584A..90G,2016MNRAS.460.3784S} in spite of the exchange of angular momentum with the driven gas. 
The mean $\Sigma_{\ast}$ of unbarred galaxies showed a substantial central mass concentration ($\sim 6$ and $\sim 2$ times the extrapolated central stellar density of the underlying disk for $T<5$ and $T\ge5$, respectively), 
which is in principle independent of any bar-driven secular evolution or presence of B/P bulges. 
However, for $T<5$ we found that barred galaxies have a  mean central concentration that is a factor of 2 higher than non-barred galaxies, which we attribute to secular evolution.

The redistribution of material beyond the CR driven by the bar torques can enhance substantially the outer disk stellar mass density, increasing the disk scalelength \citep[e.g.][]{1971ApJ...168..343H,2002MNRAS.330...35A,2003MNRAS.346..251O}. 
Particularly, \citet[][]{2003MNRAS.345..406V} found that for a simulation model with a substantial amount of dark matter the bar formation increases the exponential scalelength of the disk by a factor of $1.2-1.5$. 
In the simulations of \citet[][]{2006ApJ...645..209D} the angular momentum redistribution within a galaxy leads to an increase in $h_{\rm R}$ even when the disk angular momentum is conserved (i.e. with a rigid halo). 
 In addition, they show that the distribution of dark matter halo angular momenta does not unambiguously determine the increase in the disk scalelength 
since the evolution of the density profile linked to the bar formation is also sensitive to initial disk kinematics: the larger the Toomre parameter $Q$ (hotter disk), the smaller the change.

There can be multiple physical processes leading to the radial migration that causes the outer disk growth in barred galaxies. 
Mixing due to resonant scattering by transient spiral structure \citep[][]{2002MNRAS.336..785S,2008ApJ...684L..79R} has been 
proposed to account for a radial flattening of metallicity gradients in the disk, but this process does not seem to alter the surface density profile of the disk. 
A different mechanism for the redistribution of angular momentum in the disk, proposed in \citet[][]{2010ApJ...722..112M}, 
results from the overlap of spiral and bar resonances. Non-linear couplings between bar and spirals with different patterns are known to take place in galactic disks 
\citep[][]{1985MNRAS.217..127S,1987ApJ...318L..43T,1988MNRAS.231P..25S,1988MNRAS.232..733S,1997A&A...322..442M,1999A&A...348..737R}. 
Using self-consistent Tree-SPH simulations and high-resolution N-body simulations, 
\citet[][]{2011A&A...527A.147M} demonstrated that such radial migration induces a strong extension of the disk, even in low-mass galaxies. 
This bar+spirals coupling mechanism is expected to be ubiquitous for all barred galaxies as bars are likely to be drivers of spiral density waves 
\citep[evidence for this can be found in][and also in Sect.~\ref{bar-hubble} of this work]{2010ApJ...715L..56S}. 
The disk growth in barred galaxies can also be induced by flux-tube manifold spirals \citep[][]{2006A&A...453...39R,2006MNRAS.369L..56P,2006MNRAS.373..280V,2007A&A...472...63R,2009MNRAS.394...67A}, 
as the disk size can increase as much as 50~$\%$ after a few episodes of spiral arm formation \citep[][]{2012MNRAS.426L..46A}.

In Sect.~\ref{disk_kpc} we found a significant difference in the density profiles of barred and non-barred galaxies in  the inner parts and in the outskirts of their disks (Fig.~\ref{stack_kpc_disk_mass}).
For a fixed total stellar mass, the average $\Sigma_{\ast}$ of unbarred galaxies intersected the average $\Sigma_{\ast}$ of barred galaxies 
slightly beyond the mean bar radii of the latter sample that presented a shallower surface brightness in the outer kpc.
Because stellar bars are known to push outwards the material which is beyond corotation, we speculate that such crossing point might be the mean corotation resonance radius of the binned barred galaxies. 
This constitutes an observational confirmation of the expectations from the simulations described above. 
In addition, this imprint of the bar-driven rearrangement of material in the disk was shown to be clear for  early-type ($T\ge5$) and late-type ($T>5$) systems (Fig.~\ref{stack_kpc_disk_ttype}), 
which are known to host two distinct types of mass distributions (DG2016). 

Among the faintest systems analysed in this study ($10^{8.5}\lesssim M_{\ast}/M_{\odot}\lesssim10^{9}$), barred and non-barred galaxies were found to have almost identical mean stellar mass density profiles. 
This  probably occurs because  the systems in this bin, mostly constituted by irregular and Magellanic galaxies, 
are characterized by bars and spiral arms of low amplitudes, as shown in Sect.~\ref{forces_sect}, which might be less efficient redistributing angular momentum. 
Another explanation would come from their typically larger halo-to-stellar mass ratio within the optical disk, shown in Fig.~6 of DG2016, 
causing the halo resonances to act as the main angular momentum sinks in detriment to the stellar disk (i.e. less angular momentum absorbed by particles near the outer disk resonances). 
This is  oversimplified, however, since the maximum exchange of angular momentum requires an optimum equilibrium between emitters (material surrounding the bar inner Lindblad resonance) 
and absorbers (material at halo and outer disk resonances), as discussed in \citet[][]{2003MNRAS.341.1179A} and \citet[][]{2013seg..book..305A}. 
In addition, for systems with $10^{9}\lesssim M_{\ast}/M_{\odot}\lesssim10^{9.5}$, which are dark matter dominated, the evidence for the disk spread being caused by bars is fairly clear.

Observational evidence for the bar-driven secular evolution analysing disk structural parameters was also provided by \citet[][]{2013MNRAS.432L..56S},
who studied the difference in the disk central surface brightnesses and scalelengths between barred and non-barred galaxies of stellar masses $M_{\ast} > 10^{10} M_{\odot}$ at redshifts $0.02 \le z \le 0.07$. 
Using a sample of $\sim700$ SDSS-DR2 galaxies and parameters from the $i$-band 2-D bulge/disk/bar decompositions performed in \citet[][]{2009MNRAS.393.1531G},
they found that barred disks tend to have fainter extrapolated central surface brightnesses ($\Delta \mu_{\circ} \approx 0.25$) and larger disk scalelengths ($\Delta h_{\rm R} \approx 15\%$), 
which is more or less consistent with our results, in the same $M_{\ast}$-bins, based on mean stellar mass density profiles. 
In addition to using deeper data which are less affected by dust obscuration, we also apply a different methodological approach that is not sensitive to any decomposition technique. 
In fact, in 2-D decompositions bars, lenses, and barlenses need to be modelled in order to recover intrinsic structural parameters of disks and, 
more severely, bulges \citep[e.g.][]{2005MNRAS.362.1319L,2010MNRAS.405.1089L,2014MNRAS.444L..80L}; this is not an issue in the qualitative study presented here. 
In addition, we have extended the analysis to bluer and fainter systems, among which the the role of bars in the disk growth is even more evident. 

\citet[][]{2013MNRAS.432L..56S} also discussed (and discarded) an alternative scenario that would eventually explain the statistical differences in
disk structural parameters between barred and non-barred galaxies, namely that bars form preferentially
in disks with faint disk central surface brightnesses and large scalelengths. 
Using S$^4$G data, DG2016 showed that galaxies with $T>5$ and $M_{\ast}<10^{10}M_{\odot}$, typically bluer and richer in cold gas,
are characterized by higher halo-to-stellar mass ratios. 
In addition, in \citet[][]{2015ApJ...807..111C} (see also DG2016) the bar fraction was lower for systems with a larger $M_{\rm halo}/M_{\ast}(<R_{\rm opt})$, 
providing possible observational evidence for the role of halos stabilizing the disk against bar formation 
shown in simulation models \citep[e.g.][]{1976AJ.....81...30H,1997ApJ...477L..79M,2004MNRAS.347..277M}. 
This can be an excessively simplified analysis though 
as live halos (i.e. halos absorbing angular momentum) did not prevent the long-term formation of bars in the models developed by \citet[][]{2002ApJ...569L..83A}, and eventually these bars became stronger than when weaker halos were used. 
However, recent simulations with live disks and halos in a cosmological context by \citet[][]{2012MNRAS.426..983D} favour the stabilizing role of dark matter halos. 
In summary, based on observations, we expect that disks with lower surface brightness (more dark matter dominated) are less prone to the bar instability. 
In addition, it is unlikely that the average profiles for barred and non-barred galaxies converge close to the mean bar corotation by coincidence. 
Altogether, we reject any other scenario but the bar-driven secular evolution of disks to explain some of the results presented in this paper.

In conclusion, although hierarchical clustering is still taking place at $z=0$, there has been enough time for secular processes to have an effect in the stellar mass distribution of disk galaxies.

%
%
%
%
\section{Summary and conclusions}\label{summarysection}
%
%
In this work we provide observational constraints for galaxy formation models by using 3.6~$\mu$m imaging for 1154 non-highly inclined disk galaxies ($i<65^{\circ}$) in the local Universe, 
of which $\sim 2/3$ are barred, with total stellar masses $10^{8.5} \lesssim M_{\ast}/M_{\odot} \lesssim 10^{11}$ 
and Hubble types $-3 \le T \le 10$, drawn from the S$^{4}$G sample \citep[][]{2010PASP..122.1397S}. 

We obtained average disks (1-D) and bars (2-D) by stacking galaxy images which were rescaled to a common frame determined by their stellar disk and 
bar sizes and orientations (Fig.~\ref{stack1}, Fig.~\ref{stack_MW_dokum_methods}, and Fig.~\ref{stack1-bars}). 
We describe the stellar density profiles ($\Sigma_{\ast}$) of disks and bars as a function of $M_{\ast}$ and morphology. 
Based on the Fourier decomposition of galaxies and the calculation of gravitational potentials done in DG2016, 
we also characterize the mean disk(+bulge) contributions to the circular velocity ($V_{3.6\mu \rm m}$) and $m=2$ Fourier amplitude ($A_{2}$) radial profiles in disk galaxies. 
We calculated the statistical dispersion as a function of radius.

The main results of this paper based on the 1-D study of average disks are the following:
\begin{itemize}
\item For all the $M_{\ast}$- and $T$-bins, the mean stellar density profiles follow an exponential slope (Fig.~\ref{stack_kpc_disk_mass}, Fig.~\ref{stack_kpc_disk_ttype} and Fig.~\ref{stack_1}) 
down to at least $\sim 10 M_{\odot} \rm pc^{-2}$. Beyond this image depth the robustness of our statistical methodology is compromised by the sample coverage in the radial direction. 
The disk extrapolated central surface brightness and scalelength of the mean $\Sigma_{\ast}$ increase with increasing $M_{\ast}$.

\item The mean $\Sigma_{\ast}$ profiles associated with galaxies with Hubble types $T\ge8$ and stellar masses $M_{\ast}<10^{9} M_{\odot}$ are bulge-less. 
Intermediate systems with $10^{9} \le M_{\ast}<10^{10} M_{\odot}$ and $5 \le T < 8$ present fairly small central mass concentrations, 
while more massive systems have significantly larger concentrations whose prominence scales with $T$.

\item We present observational evidence for the bar-induced secular evolution of galactic disks. 
For a given $M_{\ast}$-bin ($\ge 10^{9}M_{\odot}$), (i) mean stellar mass density profiles computed for barred systems present larger scalelengths and 
fainter extrapolated central surface brightnesses than for their non-barred counterparts; 
(ii) the average $\Sigma_{\ast}$ of barred and non-barred galaxies intersect each other slightly beyond the mean bar length of the former subsample, 
most likely at the corotation radius; and (iii) the central mass concentration of barred galaxies is larger (almost a factor~2 when $T<5$) than in the non-barred systems. 
To our knowledge, the last trend is observationally confirmed for the first time based on a large unbiased sample in near-IR wavelengths. 
The imprint of the bar-driven rearrangement of material is found for both early- and late-type galaxies. 

\item For early- and intermediate-type spirals ($0 \le T < 5$), we find that the mean $A_{2}$ is substantially larger within and outside the typical
bar region among barred galaxies, compared to the non-barred systems (Fig.~\ref{a2_prof_ind}). 
For all stellar masses, $\rm SA\underline{B}$+SB galaxies tend to have larger $A_{2}$ amplitudes than $\rm S\underline{A}$B+SAB systems at all radii.
This provides possible observational evidence for the role of bars driving spiral density waves. 

\item The shape of the average stellar component of the circular velocity (maximum velocity and inner slope) is strongly dependent 
on the total stellar mass and bulge prominence (i.e. morphological type) (Fig.~\ref{stack_vdisk_comp}, Fig.~\ref{stellar-halo-ratio}, Table~\ref{vcirc_mass} and Table~\ref{vcirc_ttype}). 
Early-type systems ($T<5$) present similar $V_{3.6\mu \rm m}$ maximum amplitudes ($\sim 120-130$ km s$^{-1}$) and inner slopes that increase with decreasing $T$. 
\end{itemize}

From the analysis of 2-D bar stacks we confirm previous studies of bar properties in the local universe, 
some of them consistent with a framework in which bars trap particles from the disk and become narrower and stronger, while the host galaxies evolve secularly and move in the Hubble sequence: 
\begin{itemize}
\item Bar stacks comprising early-type systems present larger $m=2$ and $m=4$ Fourier density amplitudes and lower non-centrally peaking gravitational torques (Fig.~\ref{qt_a2_mass_ttype_family} and Table~\ref{bars_params}).

\item Early- and intermediate-type spirals host intrinsically narrower mean bars than later types (Fig.~\ref{stack3} and Fig.~\ref{stack_bars_ellip_b4}). 
In contrast, lenticulars and faint galaxies ($M_{\ast}<10^9 M_{\odot}$) have oval-shaped average bars.

\item Qualitatively, bar stacks corresponding to early- and intermediate-type spiral galaxy bins are typically flat, 
while later-types show exponential profiles, in agreement with the classic result by \citet[][]{1985ApJ...288..438E} (Fig.~\ref{stack_bars_altogether}). 
The azimuthally averaged density profile of lenticulars does not show any feature associated with the stellar bar. 
The bar hump is very clear in the cut along the bar major axis for all systems.

\item By grouping galaxies based on the galaxy family from \citet[][]{2015ApJS..217...32B}, we show a clear correspondence between visual 
($\rm \underline{A}B$/AB/$\rm A\underline{B}$/B) and qualitative and quantitative estimates of the bar strength (tangential-to-radial forces, $m=2$ Fourier amplitudes, ellipticity), 
for both early-type ($T<5$) and late-type ($T\ge5$) systems (Fig.~\ref{stack_families}, Fig.~\ref{stack_bars_ellip_b4}, Fig.~\ref{stack_bars_ellip_b4_late_early_family}, and Fig.~\ref{qt_a2_mass_ttype_family_early_late}). 
\end{itemize}
%
%
The FITS files of the synthetic images and the tabulated radial profiles of average stellar mass density, luminosity, Fourier amplitudes, gravitational force, and the stellar contribution to the circular velocity 
are also available on-line at www.oulu.fi/astronomy/S4G$\_$STACKS/. The sample dispersions of the 1-D profiles within each radial bin are also provided. 
A brief description of the content of the website can be found at www.oulu.fi/astronomy/S4G$\_$STACKS/readme.txt. 
In this paper we also provide simple fitting formulae to the 1-D average stellar density profiles (Table~\ref{disk_fits_params}) and disk component of the rotation curves (Table~\ref{rcur_fits_params}).
%
%
\begin{acknowledgements}
We thank the anonymous referee for comments that improved this paper. We acknowledge financial support to the DAGAL network from the People Programme (Marie Curie Actions)
of the European Union's Seventh Framework Programme FP7/2007- 2013/ under REA grant agreement number PITN-GA-2011-289313. 
We thank Sebastien Comerón, Jarkko Laine, Ryan Leaman, Martín Herrera Endoqui, Marie Martig, Reynier Peletier, and Glenn van de Ven for very useful conversations. 
We thank the S$^{4}$G team for their work on the different pipelines.
{\it Facilities}: Spitzer (IRAC).
%
%
\end{acknowledgements}
\bibliographystyle{aa}
\bibliography{bibliography}
\clearpage
%
%
\end{document}